\documentclass[aps,prx,reprint,twocolumn,superscriptaddress]{revtex4}

\usepackage{graphicx}
\usepackage{subfigure}
\usepackage[utf8]{inputenc}
\usepackage{amsmath}
\usepackage{amssymb}
\usepackage{appendix}
\usepackage{color}

\usepackage{hyperref}
\usepackage{cleveref}

\crefname{equation}{Eq.}{Eqs.}
\Crefname{equation}{Equation}{Equations}
\crefname{figure}{Fig.}{Figs.}
\Crefname{figure}{Figure}{Figures}
\crefname{section}{Sec.}{Secs.}
\Crefname{section}{Section}{Sections}
\crefname{appendix}{Appendix}{Appendices}
\Crefname{appendix}{Appendix}{Appendices}

\newcommand{\bra}[1]{\langle #1|}
\newcommand{\ket}[1]{|#1\rangle}
\newcommand{\braket}[1]{\langle #1 \rangle}

\newcommand{\tr}{\text{tr}}
\newcommand{\e}{e}

\newcommand{\bigO}{\mathcal{O}}
\newcommand{\dd}{d}

\newcommand{\half}{\frac{1}{2}}

\newcommand{\hphi}{\hat{\phi}}

\newcommand{\dphi}{\dot{\phi}}
\newcommand{\hpi}{\hat{\pi}}
\newcommand{\ha}{\hat{a}}
\newcommand{\ta}{\tilde{a}}
\newcommand{\tb}{\tilde{b}}
\newcommand{\haR}{\hat{a}_{\rm R}}

\newcommand{\hb}{\hat{b}}

\newcommand{\hX}{\hat{X}}
\newcommand{\hY}{\hat{Y}}

\newcommand{\hU}{\hat{U}}
\newcommand{\hc}{\hat{c}}
\newcommand{\hx}{\hat{x}}
\newcommand{\hy}{\hat{y}}
\newcommand{\hH}{\hat{H}}
\newcommand{\hK}{\hat{K}}

\newcommand{\hV}{\hat{V}}
\newcommand{\hB}{\hat{B}}
\newcommand{\hs}{\hat{\sigma}}

\newcommand{\aR}[1]{\ha_{R #1}^{\rm out}}
\newcommand{\aRd}[1]{\ha_{R #1}^{{\rm out} \dagger}}

\newcommand{\fq}{\frac{2\pi}{\Phi_0}}

\newcommand{\CL}{\mathcal{L}}

\renewcommand{\Re}{\text{Re}}
\renewcommand{\Im}{\text{Im}}

\begin{document}

\title{
Squeezing and quantum state engineering with Josephson traveling wave amplifiers} 

\author{Arne L. Grimsmo}\email{arne.loehre.grimsmo@usherbrooke.ca}
\affiliation{Institut Quantique and D\'epartment de Physique, Universit\'e de Sherbrooke, 2500 boulevard de l'Universit\'e, Sherbrooke, Qu\'ebec J1K 2R1, Canada}
\author{Alexandre Blais}
\affiliation{Institut Quantique and D\'epartment de Physique, Universit\'e de Sherbrooke, 2500 boulevard de l'Universit\'e, Sherbrooke, Qu\'ebec J1K 2R1, Canada}
\affiliation{Canadian Institute for Advanced Research, Toronto, Canada}

\date{\today}

\begin{abstract}
  We develop a quantum theory describing the input-output properties of Josephson traveling wave parametric amplifiers. This allows us to show how such a device can be used as a source of nonclassical radiation, and how dispersion engineering can be used to tailor gain profiles and squeezing spectra with attractive properties, ranging from genuinely broadband spectra to ``squeezing combs'' consisting of a number of discrete entangled quasimodes. The device's output field can be used to generate a multi-mode squeezed bath---a powerful resource for dissipative quantum state preparation. In particular, we show how it can be used to generate continuous variable cluster states that are universal for measurement based quantum computing.
  The favourable scaling properties of the preparation scheme makes it a promising path towards continuous variable quantum computing in the microwave regime.
\end{abstract}


\maketitle

\section{Introduction}

Superconducting microwave circuits can be used to behave as artificial atoms in engineered electromagnetic environments, where strong light-matter interaction is achieved by confining the electromagnetic field in microwave resonators~\cite{Wallraff09} or one-dimensional waveguides~\cite{Van13}. This is in close analogy, respectively, with cavity and waveguide quantum electrodynamics with real atoms~\cite{Haroche06,Vetsch10,Goban14}. The flexibility offered by microwave engineering also allows experimentalists to go beyond the limits of conventional quantum optics in many ways. Examples include realizing light-matter coupling strengths that are unachievable with real atoms~\cite{Niemczyk10}, and using nonlinear microwave resonators to simulate relativistic quantum effects~\cite{Wilson11}.

A recent advancement in microwave quantum optics is the bottom-up design of nonlinear, one-dimensional metamaterials with strong photon-photon interactions and engineered dispersion relations~\cite{OBrien14,Macklin15,White15}. The nonlinearity is provided by Josephson junctions embedded in a transmission line, with photon-photon interactions activated by a strong pump tone through a parametric four-wave mixing process. These devices have been dubbed Josephson traveling Wave Parametric Amplifiers (JTWPAs)~\cite{Macklin15}, and are analogous to one-dimensional $\chi^{(3)}$ nonlinear crystals~\cite{Hillery04}.

The development of JTWPAs is motivated by their potential use as amplifiers for readout of superconducting qubits. The extremely high measurement fidelity necessary for fault tolerant quantum computing requires phase preserving amplifiers with added noise near the fundamental quantum limit~\cite{Caves82,Jeffrey14}. 
The JTWPA design is in this respect a very promising candidate. Early experimental realizations have shown impressive performance, and are already expected to have sufficient dynamic range and bandwidth to read out several tens of superconducting qubits with a single device~\cite{Macklin15,White15}. The key advantage to the JTWPA design is the operational bandwidth which is in the GHz range. This is in contrast to other near-quantum limited microwave amplifiers based on resonant cavity interactions, which typically have bandwidths of tens of MHz~\cite{Castellanos07,Bergeal10,Hatridge11}.

An amplifier operating near the quantum limit is, however, very different from a classical amplifier. Quantum-limited phase preserving amplification implies the presence of entanglement between the amplified signal and an ``idler'' signal in a two-mode squeezed state~\cite{Caves82,Caves12}. This motivates an alternative viewpoint on the JTWPA: Besides using the device to amplify a signal of interest, one can also view it as a broadband source of nonclassical radiation. 

In this paper we show how the inherent flexibility in the bottom-up JTWPA construction allows for designing broadband squeezing spectra with attractive properties. 
In particular, we show how to tailor the squeezing spectrum such that some frequency ranges are unaffected by the nonlinear interaction.
This is useful for example to avoid unwanted quantum heating~\cite{Ong13} of systems placed at the JTWPA output.

We subsequently demonstrate how the JTWPA can be used as a resource for dissipative quantum state preparation, including resource states for universal measurement based quantum computing~\cite{Raussendorf01,Menicucci06}. Dissipative quantum state preparation has over the last years emerged as an exciting alternative to preparation of entangled states using coherent Hamiltonian~\cite{Aharonov08} or gate-based methods~\cite{Nielsen10}. It has been shown that universal quantum computing can be achieved through dissipative processes alone~\cite{Verstraete09}, and in a similar vein that highly correlated states such as stabilizer states and projected entangled pair states can be created as stable steady states of dissipative processes~\cite{Verstraete09,Kraus08}. The early theoretical proposals in Refs.~\cite{Verstraete09,Kraus08}, however, involve many-body dissipative interactions that are hard to realize in practice. As a consequence, searching for simpler schemes for dissipative state preparation that can be implemented in present day experiments has become an active area of research~\cite{Diehl08,Krauter11,Lin13,Shankar13,Stannigel12,Bardyn12}.

We show that broadband squeezed radiation, such as the radiation emitted by a JTWPA, is a particularly potent resource for dissipative quantum state preparation. The emitted radiation generates a broadband squeezed bath which can be used to cool quantum systems placed at the source's output into entangled states. This contrasts the amplifier mode of operation, where the systems of interest are located at the amplifier's input, and do not see the nonclassical radiation emitted by the device. We show that by engineering such a squeezed bath one can produce pairs of entangled qubits, as well as continuous variable cluster states that are universal for measurement based quantum computing. The preparation schemes are simple, requiring no Hamiltonian interactions or complicated reservoir engineering. For the case of JTWPAs as squeezing sources, the large bandwidth furthermore makes the process very hardware efficient, making this an attractive avenue for measurement based quantum computing with microwaves.

The purely dissipative nature of the preparation process distinguishes our proposal from similar approaches for generating cluster states in the optical regime~\cite{Menicucci08,Flammia09,Menicucci11b,Wang14,Chen14,Alexander15}.
A distinct advantage of a dissipative scheme is that it relaxes constraints on locality, which might allow for a more modular architecture that avoids spurious interactions and increases scalability~\cite{Brecht:16a}.

Although we focus on JTPWAs as squeezing sources in this work, due to their design flexibility and large bandwidth, we emphasize that the dissipative quantum state preparation schemes we develop are relevant for any type of broadband squeezing source that can be integrated with coherent quantum systems, such as other types of traveling wave amplifiers~\cite{Eom12,Bockstiegel14}, impedance engineered Josephson parametric amplifiers~\cite{Roy15}, squeezing sources based on reservoir engineering~\cite{Metelmann14}, or even the nonclassical radiation emitted by an ac-biased tunnel junction~\cite{Forgues15,Grimsmo16}.

\section{\label{sect:inout}Asymptotic Input-output Theory}
\begin{figure}
  \centering
  \includegraphics{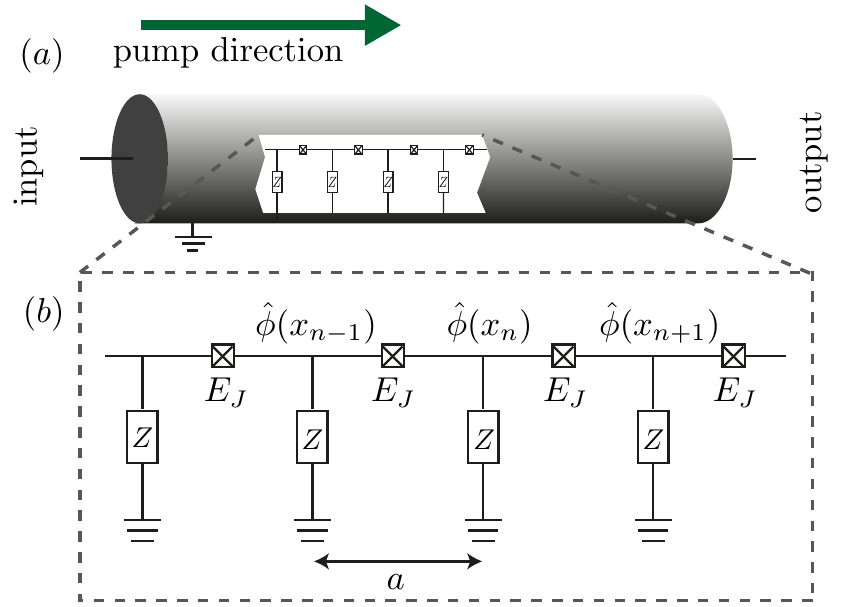}
  \caption{\label{fig:TWPA}Josephson traveling wave parametric amplifier. A chain of identical coupled Josephson junctions, with Josephson energy $E_J$ and plasma frequency $\omega_P$, are coupled in series. Each junction is furthermore coupled to ground by a passive, dissipationless element described by an impedance $Z(\omega)$. By designing this impedance one can engineer the dispersion relation for waves traveling through the device. A strong right-moving pump actives a four-wave mixing process through the Josephson potential, which can be used to amplify a signal and generate squeezed light.}
\end{figure}

To describe the JTWPA's squeezing properties, we first need a quantized theory of its dynamics. Classical treatments of a JTWPA are presented in Refs.~\cite{Yaakobi13,OBrien14,White15}.
In the following we give a quantized Hamiltonian treatment of the nonlinear dynamics, taking into account dispersion and the continuum nature of the electromagnetic field in the waveguide.

The resulting theory is in general difficult to treat analytically due to time-ordering effects~\cite{Quesada14,Quesada15}, and we therefore take a perturbative approach treating the non-linearity to first order. An input-outpu relation linking the field entering the JTWPA to the emitted output field is derived in the usual scattering limit where the initial and final times of the problem are taken to minus and plus infinity, respectively~\cite{Breit54,Liscidini12,Drummond14,Quesada14,Quesada15}.

Equations of motion similar to those we derive here have been used previously by Caves and Crouch in a study of wideband traveling wave amplifier~\cite{Caves87}, where they were taken as operator versions of macroscopic Maxwell's equations for a nonlinear, homogeneous and dispersionless medium~\cite{Hillery84}. We here rigorously justify these equations by deriving them from a microscopic theory, taking into account the finite extent of the nonlinearity as well as dispersion, the latter stemming from both junctions' plasma oscillations, non-linear phase-modulation and engineered bandgaps in the medium.

The device we consider in this paper is depicted in \cref{fig:TWPA}. It consists of a series of identical coupled Josephson junctions with Josephson energies $E_{J}$ and junction capacitances $C_{J}$. Each junction is coupled to ground by a passive, dissipationless element with impedance $Z(\omega)$, which is left arbitrary for now. By engineering $Z(\omega)$ one can modify the dispersion relation of waves propagating through the device as shown in Ref.~\cite{OBrien14}. We show in~\cref{sect:spectrumengineering} how this can be used to tailor the squeezing properties of the output field leaving the device. 

\begin{figure}
  \centering
  \includegraphics{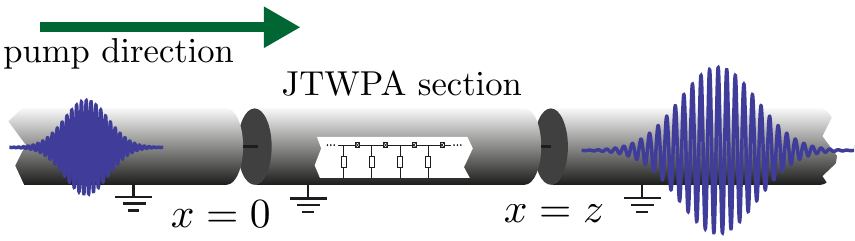}
  \caption{\label{fig:waveguide}An infinite transmission line with a JTWPA section of length $z$ sandwiched by two identical semi-infinite linear transmission lines. The wave-packet illustrated a signal at the input, which is amplified at the JTWPA's output.}
\end{figure}

Realistic JTWPAs have several thousand junctions with a unit cell distance much smaller than the relevant wavelengths~\cite{Macklin15,White15}. One can therefore approximate the device with a continuum description (formally taking the unit cell distance, $a$, to zero). We furthermore assume that the JTWPA is coupled to identical, semi-infinite and impedance matched transmission lines to the left and the right, as illustrated in~\cref{fig:waveguide}. Note that other variants of the JTWPA device where the Josephson junctions are replaced by SQUIDs have recently been discussed~\cite{Bell15,Zorin16}. We do not consider such modifications here, but the general approach we develop below can be used to formulate a quantum theory also in these cases.

As shown in~\cref{app:Hamiltonian}, the position-dependent flux, $\hphi(x)$ (in the Schrödinger picture), along a transmission line with a JTWPA section extending from $x=0$ to $x=z$ can in the continuum limit be expanded in terms of a set of left- and right-moving modes,
\begin{equation}\label{eq:phianzats}
  \begin{aligned}
    \hphi(x) = \sum_{\nu={\rm L,R}}& \int_{0}^\infty \dd\omega \sqrt{\frac{\hbar}{2 c(x) \omega}} g_{\nu\omega}(x)
    \ha_{\nu\omega}\\
    &+ \text{H.c.},
  \end{aligned}
\end{equation}
where $[\ha_{\nu\omega},\ha^\dagger_{\mu\omega'}]=\delta_{\nu\mu}\delta(\omega-\omega')$ and the mode functions are given by
\begin{align}\label{eq:modeanzats}
  g_{\nu\omega}(x) ={}& \sqrt{\frac{1}{2\pi \eta_{\omega}(x)v(x)}} \e^{\pm i k_{\omega}(x) x}.
\end{align}
Here, $+$ ($-$) corresponds to $\nu={\rm R}$ ($\nu={\rm L}$),
$k_\omega(x) = \eta_\omega(x) \omega/v(x)$ is the wavevector, with $\eta_\omega(x)$ the refractive index,
and $v(x) = 1/\sqrt{c(x)l(x)}$.
The $x$-dependent parameters are defined such that they take one (constant) value inside the JTWPA section, and another value outside this section. $c(x)$ is the capacitance to ground per unit cell, $l(x)$ is the linear inductance of the transmission line.

The only difference from the usual prescription for the quantized flux in a linear, homogeneous and dispersion free transmission line~\cite{Yurke84b,yurke:2004a} is the $x$-dependent wavevector, which now takes a different form inside and outside the nonlinear section. Explicitly, the dispersion relation is found to be (see~\cref{app:Hamiltonian})~\cite{OBrien14}
\begin{align}\label{eq:disprel}
  k_\omega(x) = \left\{
  \begin{array}{ll}
    \sqrt{\frac{-i\omega z^{-1}(\omega) l(x)}{1-\omega^2/\omega_P^2}}
    &\text{for } 0 < x < z \\
    \frac{\omega}{v(x)}
    &\text{otherwise},
  \end{array}\right.
\end{align}
where $z^{-1}(\omega) = Z^{-1}(\omega)/a$ is the admittance to ground per unit cell in the JTWPA section, and $\omega_P$ is the junctions' plasma frequency.

\begin{figure}
  \centering
  \includegraphics{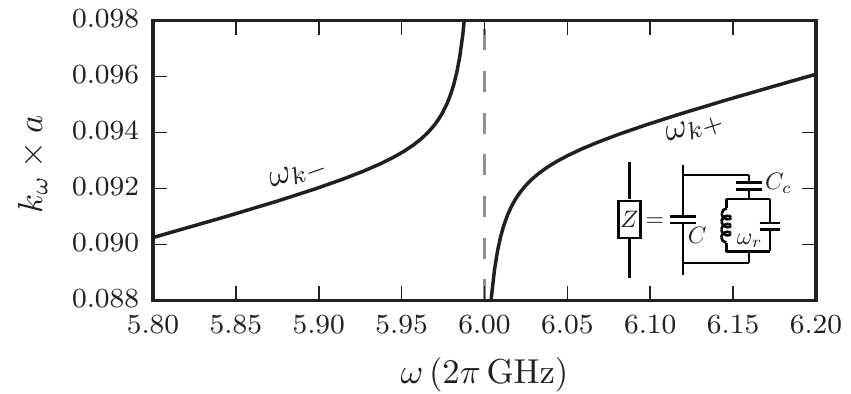}
  \caption{\label{fig:bandgap}Illustration of the disperion relation when $Z(\omega)$ (illustrated in the inset) describes a single resonant mode at a frequency $\omega_{r}$ linearly coupled to the flux field in every unit cell. A bandgap opens up around the resonance frequency, close to 6 GHz in this example. The width of the bandgap is set by the coupling capacitance $C_c$ shown in the inset.}
\end{figure}

Implicit in the continuum description is that we are considering sufficiently low frequencies, such that the wavelengths are large compared to the unit cell distance, $a$.
Furthermore, plane wave solutions only exists when the right hand side of~\cref{eq:disprel} is real. In practice, we are interested in frequencies $\omega^2 \ll \omega_P^2$ such that the dispersion due to the plasma oscillations of the junctions is relatively small. If, however, the admittance $z^{-1}(\omega)$ describes a linear element with a resonant mode, a bandgap opens around the resonance frequency for which no plane wave solutions exists. Physically such a resonant mode behaves as a ``matter field'' in the continuum limit, and the excitations of the systems resemble light-matter polaritons~\cite{Huttner91,Huttner92,Drummond14}. As long as we are away from any bandgap, however, these ``matter fields'' slave the photonic field, $\hphi(x,t)$, and only modifies the dielectric properties of the medium, manifest in the dispersion relation~\cref{eq:disprel}. The behavior of the dispersion relation close to a bandgap is illustrated in~\cref{fig:bandgap}.

As shown in detail in~\cref{app:Hamiltonian} a continuum limit Hamiltonian for the system can be written
\begin{equation}
  \hH = \hH_0 + \hH_1,
\end{equation}
where $\hH_0$ is a linear contribution containing all terms up to second order in the fields,
and 
$\hH_1$
is a nonlinear contribution due to the Josephson junction potential.
The linear Hamiltonian can be diagonalized in terms of the frequency modes introduced in~\cref{eq:phianzats},
\begin{equation}
  \begin{aligned}
    \hH_0 ={}& \sum_{\nu={\rm L,R}} \int_{0}^\infty \dd \omega \hbar \omega
    \ha_{\nu\omega}^\dagger\ha_{\nu\omega},
  \end{aligned}
\end{equation}
where we have omitted the zero-point energy.

For the nonlinear Hamiltonian, we systematically perform a series of approximations that are ultimately equivalent to those used in the classical treatment given in Refs.~\cite{Yaakobi13,OBrien14,White15}. A quantized analog of the classical equation of motion found in previous work is shown to be a limiting case of a more general theory.

We assume the presence of a strong right-moving classical pump centered at a frequency $\Omega_p$ with corresponding wavevector $k_p$, and replace $\ha_{R\omega} \to \ha_{R\omega} + b(\omega)$, with $b(\omega)$ a complex valued function centered at $\Omega_p$. The fields, $\ha_{\nu\omega}$, are assumed to be sufficiently weak so that we can drop in $\hH_1$ terms that are smaller than second order in the pump. Also dropping fast rotating terms and the highly phase mismatched left moving field, this leads to the approximate Hamiltonian
\begin{equation}\label{eq:H_1}
  \hH_1 = \hH_{\rm CPM} + \hH_{\rm SQ},
\end{equation}
where
\begin{equation}\label{eq:H_CPM}
  \begin{aligned}
    \hH_{\rm CPM} ={}& -\frac{\hbar}{2\pi} \int_0^\infty \dd \omega \dd \omega' \dd\Omega \dd\Omega' \sqrt{k_\omega k_{\omega'}} \\
      &\times \beta^*(\Omega)\beta(\Omega')\Phi(\omega,\omega',\Omega,\Omega')
      \ha^\dagger_{R\omega} \ha_{R \omega'}\\
    &+ \text{H.c.},
  \end{aligned}
\end{equation}
describes cross-phase modulation due to the pump, and
\begin{equation}\label{eq:H_SQ}
  \begin{aligned}
    \hH_{\rm SQ} = &-\frac{\hbar}{2\pi} \int_0^\infty \dd \omega \dd \omega' \dd\Omega \dd\Omega' \sqrt{k_\omega k_{\omega'}}  \\
      &\times \beta(\Omega)\beta(\Omega')\Phi(\omega,\Omega,\omega',\Omega')
      \ha^\dagger_{R\omega} \ha^\dagger_{R\omega'} \\
    &+ \text{H.c.},
  \end{aligned}
\end{equation}
describes broadband squeezing. The dynamics of the classical pump is governed by a classical Hamiltonian which includes self-phase modulation, given in~\cref{eq:appH:H_SPM}.
For notational convenience, we have defined the phase matching function~\cite{Quesada14,Quesada15}
\begin{equation}
  \begin{aligned}
    &\Phi(\omega_1,\omega_2,\omega_3,\omega_4) = \\
    &\int_{0}^{z} \dd x \e^{-i\left[ k_{\omega_1}(x)-k_{\omega_2}(x)+k_{\omega_3}(x)-k_{\omega_4}(x) \right] x},
  \end{aligned}
\end{equation}
and a dimensionless pump amplitude, $\beta(\Omega)$, which as shown in~\cref{app:Hamiltonian}, can be written
in terms of the ratio of the pump current to the Josephson junction critical current,
\begin{equation}
  \beta(\Omega) = \frac{I_p(\Omega)}{4 I_c},
\end{equation}
where $I_c = (2\pi/\Phi_0) E_J$.

Treating $\hH_1$ as a perturbation, it is natural to go to an interaction picture with respect to $\hH_0$. The time-evolution operator for the problem in this picture is
\begin{equation}\label{eq:timeevol}
  \hU(t_0,t_1) = \mathcal{T} \e^{-\frac{i}{\hbar} \int_{t_0}^{t_1} \dd t \hH_1(t)},
\end{equation}
where $\hH_1(t) = \exp\left[i\hH_0t\right] \hH_1 \exp\left[-i\hH_0t\right]$ and $\mathcal{T}$ is the time-ordering operator. 

Solving the time-dynamics according to~\cref{eq:timeevol} is difficult in general~\cite{Quesada14,Quesada15}. A greatly simplified approximate theory can be derived, however, by 1) treating $\hH_1$ as a perturbation to first order only, in which case the time-ordering in~\cref{eq:timeevol} can be dropped, and 2) taking the initial and final times to $t_0=-\infty$ and $t_1=\infty$, respectively. The time-integral then gives rise to delta-functions in frequency space, and we are left with approximate asymptotic evolution operator, or scattering matrix~\cite{Drummond14},
\begin{equation}\label{eq:timeevol2}
  \hU \equiv \hU(-\infty,\infty) = \e^{-\frac{i}{\hbar} \hK_1 },
\end{equation}
where
\begin{equation}\label{eq:K_1}
  \hK_1 = \hK_{\rm CPM} + \hK_{\rm SQ}.
\end{equation}
Explicit expressions for $\hK_1$ for a general classical pump are given in~\cref{app:Hamiltonian}. We from now on focus on the monochromatic pump limit, taking $b(\omega) \to b_p \delta(\omega-\Omega_p)$, with $b_p$ a c-number. In this limit we have
\begin{equation}\label{eq:K_CPM}
  \begin{aligned}
    \hK_{\rm CPM} ={}& -2 \hbar z \int_0^\infty \dd \omega |\beta|^2 k_\omega 
    \ha^\dagger_{R\omega} \ha_{R\omega},
  \end{aligned}
\end{equation}
and
\begin{equation}\label{eq:K_SQ}
  \begin{aligned}
    \hK_{\rm SQ} =
    -\hbar \int_0^\infty {}&\dd \omega
    \lambda(\omega)\,\Phi[-\Delta k_L(\omega)] \\
    &\times \ha^\dagger_{R\omega} \ha^\dagger_{R(2\Omega_p-\omega)}
    + \text{H.c.},
  \end{aligned}
\end{equation}
where
we have defined $\beta = \beta(\Omega_p)$ and
\begin{align}
  \lambda(\omega) ={}& \beta^2\,\sqrt{k_\omega k_{2\Omega_p-\omega}},\\
  \Delta k_L(\omega) ={}& 2k_p - k_{1\omega}-k_{1(2\Omega_p-\omega)}.
\end{align}
$\Delta k_L(\omega)$ here quantifies a phase-mismatch due to the linear dispersion in the JTWPA section. As we show below there is also an additional nonlinear contribution to the phase-mismatch that must be taken into account.

Defining Heisenberg picture output fields, $\ha_{R\omega}^{\rm out} = \hU^\dagger\ha_{R\omega}\hU$, we find the following input-output relation
\begin{equation}\label{eq:modesol}
  \begin{aligned}
    &\ha_{R\omega}^{\rm out} = \e^{i\left[2|\beta|^2 k_\omega + \Delta k(\omega)/2\right] z}\\
    &\times \bigg[ u(\omega,z) \ha_{R\omega}
    + i v(\omega,z) \ha_{R(2\Omega_p-\omega)}^\dagger  \bigg],
  \end{aligned}
\end{equation}
where the functions $u(\omega,z)$ and $v(\omega,z)$, defined in~\cref{app:inout}, satisfy $|u(\omega,z)|^2-|v(\omega,z)|^2 = 1$, and 
\begin{equation}\label{eq:phasemismatch}
\Delta k(\omega) = \Delta k_L(\omega) + 2|\beta|^2(k_p - k_{2\Omega_p-\omega} - k_\omega),
\end{equation}
is the phase mismatch, including a nonlinear correction due to to the cross- and self-phase modulation of the pump. 

\cref{eq:modesol} is formally identical, up to a frequency-dependent normalization of the wave amplitudes, to the classical solution derived in Refs.~\cite{Yaakobi13,OBrien14,White15}. To summarize, this limiting equation is valid for weak nonlinearity and weak fields, only treating the nonlinear Hamiltonian $\hH_1$ to first order, a strong monochromatic classical pump at a frequency $\Omega_p$, and in an asymptotic large time limit where $t_0=-\infty$ and $t_1=\infty$.

How can we interpret the asymptotic limit where the initial and final times are taken to minus and plus infinity, respectively? If we consider a situation where an initial wave packet is localized far away at $x\ll 0$ at an early time $t_0 \ll 0$, this can be interpreted as a ``scattering'' limit, where we let the wave packet propagate through the nonlinearity and consider the asymptotic field at $x\gg z$ for a late time $t_1 \gg 0$~\cite{Liscidini12,Drummond14}. However, since the initial evolution before the wave packet enters the nonlinear section is governed by $\hH_0$, it is trivial to propagate the wave packet forward towards the nonlinearity. The late evolution after the wave packet has left the nonlinear section is similarly trivial. We can therefore think of $\ha_{R\omega}$ as a frequency domain input field entering the JTWPA and $\ha_{R\omega}^{\rm out}$ as the corresponding output field leaving the device. This is similar to the definition of input and output fields used in the description of damped quantum optical systems~\cite{Gardiner85,Liscidini12}.
One should keep in mind, however, that the validity of this interpretation depends on the problem one is trying to solve: it is clearly not appropriate if, for example, the initial state of the field is delocalized over the nonlinear section.

\section{\label{sect:spectrumengineering}Engineering Nonclassical Radiation}

The quantum input-output theory developed above allows us predict features of the JTPWA's output field, such as the device's gain profile and output field squeezing spectrum. In this section we show how output spectra can be tailored through dispersion engineering. We focus first on an ideal device and discuss the effect of loss in~\cref{sect:nonideal}.

\subsection{Ideal Squeezing Spectra}

From~\cref{eq:modesol}, the JTWPA's amplitude gain is given by $u(\omega,z)$, and we define the power gain as $G(\omega,z) = |u(\omega,z)|^2$~\cite{Caves82,Yaakobi13,OBrien14}. This function grows exponentially with $z$ for small phase mismatch, $\Delta k(\omega) \simeq 0$, but is only of order one if the phase mismatch is large (see~\cref{app:inout}). 

The squeezing of the JTWPA's output field is manifest in correlations between frequencies $\omega$ and $2\Omega_p-\omega$, symmetric around the pump frequency.
It is convenient to define for the right moving field the thermal photon number
\begin{equation}\label{eq:N}
    N_R(\omega,z) = \int_{0}^\infty \dd\omega' \Big\{\braket{\aRd{\omega}\aR{\omega'}}
    - \braket{\aRd{\omega}}\braket{\aR{\omega'}}\Big\},
\end{equation}
the squeezing parameter
\begin{equation}\label{eq:M}
    M_R(\omega,z) = \int_{0}^\infty \dd\omega' \Big\{\braket{\aR{\omega}\aR{\omega'}}
    - \braket{\aR{\omega}}\braket{\aR{\omega'}}\Big\},
\end{equation}
and the squeezing spectrum~\cite{Wustmann13}
\begin{equation}\label{eq:S}
  \begin{aligned}
    S_{R}(\omega,z) ={}& \int_{0}^\infty \dd\omega' \braket{\Delta \hY^{\theta}_{R\omega}\Delta \hY^{\theta}_{R\omega'}} \\
  ={}& 2N_R(\omega,z)+1-2|M_R(\omega,z)|,
  \end{aligned}
\end{equation}
where we have defined quadratures 
\begin{equation}
  \hY^\theta_{R\omega} =i\left[\e^{i\theta/2}\aRd{\omega} - \e^{-i\theta/2}\aR{\omega}\right],
\end{equation}
with fluctuations $\Delta \hY_{R\omega} = \hY_{R\omega} - \braket{\hY_{R\omega}}$, 
and $\theta$ is the squeezing angle, given through $M_R(\omega,z) = |M_R(\omega,z)|\e^{i\theta}$.
We emphasize that~\cref{eq:N,eq:M,eq:S} are defined exclusively in terms of the right-moving field. The left-moving field also contributes vacuum noise and might add to the total photon number, but will have zero squeezing parameter in the absence of left-moving pump fields. The squeezing spectrum is typically probed in experiments by heterodyne measurement of filtered field quadratures~\cite{Eichler11,Flurin12,Eichler14,Forgues15}. We discuss this in some more detail in~\cref{app:inout}.

For a vacuum input field, where $\braket{\haR(\omega,0)\haR^\dagger(\omega',0)} = \delta(\omega-\omega')$ and all other second order moments vanish, it follows that $N_R(\omega,z) = G(\omega,z)-1 = |v(\omega,z)|^2$ and $M_R(\omega,z) = iu(\omega,z)v(\omega,z)\e^{i\Delta k(\omega) z}$. This assumes no internal loss in the JTWPA device. These expressions satisfy $|M_R(\omega,z)|^2 = N_R(\omega,z)[N_R(\omega,z)-1]$, the maximum value allowed by the Heisenberg uncertainty relation and also imply quantum-limited amplification~\cite{Caves82}.

The gain and the squeezing at the output depends strongly on the phase mismatch $\Delta k(\omega)$. The phase mismatch can however be compensated for by tuning $Z(\Omega_p)$, as this allows for tuning the pump wavevector $k_p = k_{\Omega_p}$ according to~\cref{eq:disprel}. As was proposed theoretically in Ref.~\cite{OBrien14} and demonstrated experimentally in Refs.~\cite{Macklin15,White15}, it is possible to tune the phase mismatch to zero at the pump frequency, $\Delta k (\Omega_p) \simeq 0$, and greatly reduce it across the whole JTWPA bandwidth. 
This is done by placing LC (or transmission line) resonators with resonance frequency $\omega_{r0} \simeq \Omega_p$ regularly along the JTWPA transmission line. This technique is referred to as resonant phase matching~(RPM)~\cite{OBrien14}. 

\begin{figure}
  \centering
  \includegraphics{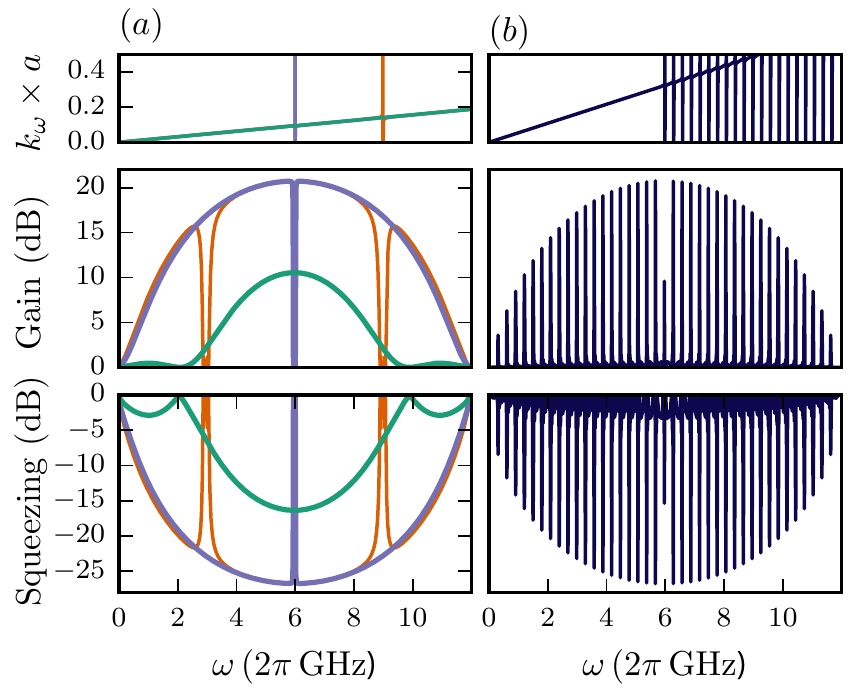}
  \caption{\label{fig:spectra}
    Gain profile and squeezing spectra for an ideal JTWPA with 2000 unit cells and parameters given in the text. $(a)$ The green lines are for a device without RPM. The blue lines are for a device with identical parameters, but where RPM has been used to tune $\Delta k(\Omega_p) \simeq 0$.
    The orange lines show a device where in addition to RPM, a second resonance has been placed at 9 GHz, punching two symmetric holes in the gain and squeezing spectra.
    $(b)$ A JTWPA with nineteen additional resonances a ``squeezing comb.''}
\end{figure}
\begin{figure}
  \centering
  \includegraphics{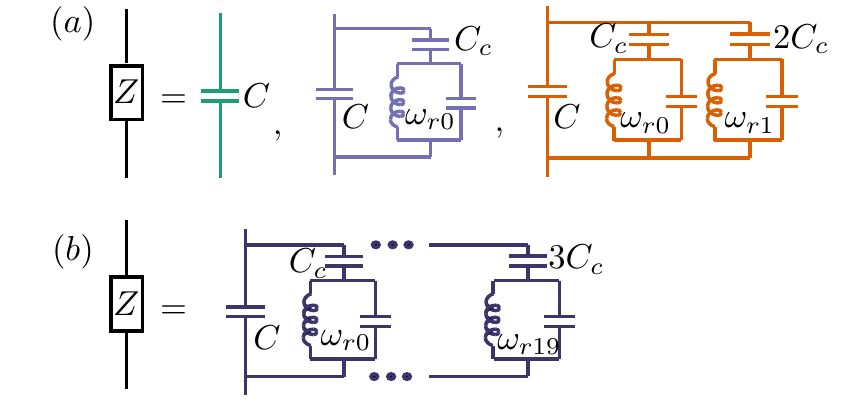}
  \caption{\label{fig:impedances}Illustration of the choice of impedances used to dispersion engineer the gain profiles and squeezing spectra in~\cref{fig:spectra}. The color codes correspond to those in~\cref{fig:spectra}.}
\end{figure}

The effect of RPM on the gain and squeezing spectra is illustrated in \cref{fig:spectra} for a simulated device similar to what has been realized experimentally in Refs.~\cite{Macklin15,White15}: The device length was chosen to be 2000 unit cells with characteristic impedance $Z_0 = \sqrt{l/c} = 50\,\Omega$, critical current $I_c = (2\pi/\Phi_0) E_J = 2.75\,\mu$A, dimensionless pump strength $\beta = 0.125$, junctions' plasma frequency $\Omega_p/2\pi=5.97$ GHz, and pump frequency $\Omega_p^2/\omega_P^2 = 6.7\times 10^{-3}$. The green lines in~\cref{fig:spectra} $(a)$ show the gain profile and squeezing spectrum of the output field for the device without RPM, while the blue lines show results for an identical device where RPM has been used to tune $\Delta k(\Omega_p) = 0$.
The circuit parameters for the LC resonator are $C_c = 10$ fF, $C_r = 7.0$ pF, $L_r = 100$ pH, giving a resonance frequency of $\omega_{r0}/2\pi = 6.0$ GHz.
The corresponding impedances to ground in each unit cell, $Z(\omega)$, are illustrated schematically in~\cref{fig:impedances} $(a)$.

Two-mode squeezing has applications for entanglement generation~\cite{Palma89,Gomez15}, quantum teleportation~\cite{Furusawa98}, interferometry~\cite{Yurke86}, creation of so-called quantum mechanics free subsystems~\cite{Tsang12}, high-fidelity qubit readout~\cite{Barzanjeh14,Didier15} and logical operations~\cite{Royer16}, amongst others.
A broadband squeezing source such as the JTWPA might have a great advantage for scalability, as tasks can be parallelized with many pairs of far-separated two-mode squeezed frequencies using a single device. It is, however, not necessarily desirable to have squeezing at \emph{all} frequencies over the operational bandwidth as this might lead, \emph{e.g.}, to unwanted quantum heating~\cite{Ong13}.
This is the case, for example, for the qubit measurement scheme with Heisenberg limited scaling of the signal-to-noise ratio proposed in Ref.~\cite{Didier15}.
It is important in this scheme to \emph{not} have high degrees of squeezing at the \emph{qubit} frequency. The reason being that the qubit sees thermal noise with photon number $N_R(\omega_q,z)$, $\omega_q$ being the qubit frequency---and even if the qubit is not directly coupled to the squeezing source, this leads to increased Purcell decay via the cavities (see Supplemental Material in Ref.~\cite{Didier15}). This problem can be avoided with the JTWPA through dispersion engineering. In the following we show how to shape the squeezing spectrum to prohibit squeezing in certain frequency bands, and create spectra with a comb-like structure.

Building on the RPM technique, we consider placing additional resonances in each unit cell with resonance frequencies $\omega_{rk}$ away from $\Omega_p$. This leads to a bandgap and a divergence in $k(\omega)$ close to each resonance $\omega_{rk}$, as illustrated in~\cref{fig:bandgap}. The huge phase mismatch close to these resonances prohibits gain at $\omega \simeq \omega_{rk}$ and $\omega \simeq 2\Omega_p-\omega_{rk}$, effectively punching two symmetric holes in the gain and squeezing spectra.
This is illustrated by the orange lines in \cref{fig:spectra}~$(a)$, where a single additional resonace has been placed at $\omega_{r1} = 9.0\times 2\pi$ GHz. The parameters are otherwise as before, except that the second LC resonator is chosen to have twice the coupling capacitance, $2 C_c$. This choice serves to illustrate how the width of the hole in the spectrum is determined by the coupling capacitance to the resonator, as is clearly seen when comparing the holes at $\omega_{r0}$ and $\omega_{r1}$.

In~\cref{fig:spectra} $(b)$ we
demonstrate how this technique can be used to engineer a ``squeezing comb'' where there is considerable gain and squeezing only for a discrete set of narrow quasimodes. With a larger number of closely spaced resonance frequencies---either using individual lumped LC circuits or the resonances of a multi-mode transmission line resonator---it is possible to have phase matching only over narrow frequency bandwiths. In \cref{fig:spectra} $(b)$ we show the gain profile and squeezing spectrum where nineteen additional resonances at $\omega_{rk} = \omega_{r0} + k\times \omega_{r0}/20$, $k=1,2,\dots,19$ has been used to create a squeezing comb with 38 quasimodes.
Slightly different parameters were chosen for this device, to get similar gain and squeezing profiles as before: $Z_0 = 14\,\Omega$, $I_0 = 2.75\,\mu$A, $\beta = 0.069$, while the additional coupling capacitances were chosen to be $3.0 C_c$. The corresponding impedance to ground is illustrated in~\cref{fig:impedances}~$(b)$.

\begin{figure}
  \includegraphics{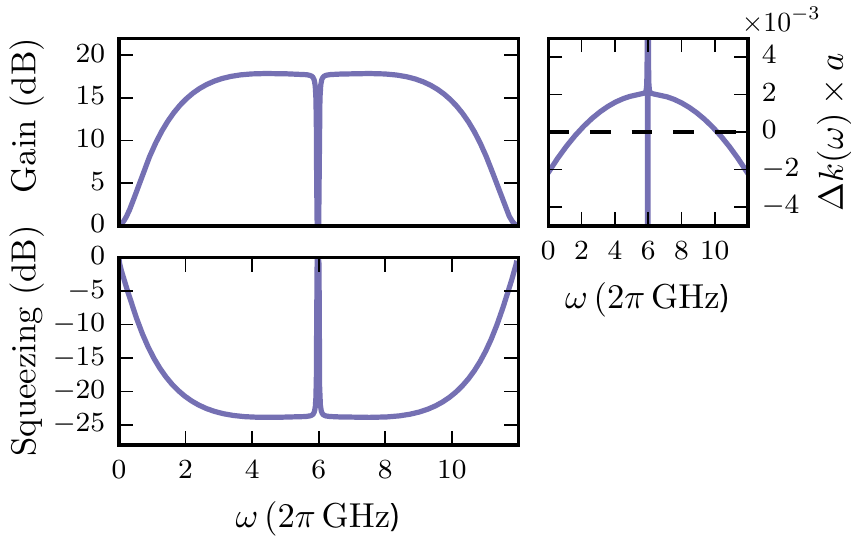}
  \caption{\label{fig:flatspectrum}A device similar to those in~\cref{fig:spectra}, but where RPM has been used to tune $\Delta k(\omega) = 0$ for $\omega/(2\pi) \simeq 1.8$ GHz. The larger phase mismatch around $\omega\simeq \Omega_p$ gives a flatter profile for both the gain and squeezing spectra.}
\end{figure}
For certain applications it might also be of interest to have a squeezing spectrum with a flatter profile than what is shown in~\cref{fig:spectra}. This can be achieved by suitably engineering the phase mismatch. In~\cref{fig:flatspectrum} we show a device where RPM has been used to tune $\Delta k(\omega) = 0$ for $\omega/2\pi \simeq 1.8$ GHz, with the pump frequency close to to the resonance frequency at $\omega_{r0}/2\pi = 6$ GHz. This leads to larger phase mismatch in the center region of the spectrum, close to the pump, giving the flatter profile shown in the figure. The simulated device otherwise has parameters $Z_0 = 60\,\Omega$, $I_0 = 1.75\,\mu$A, $\beta = 0.113$.

\subsection{\label{sect:nonideal}Reduction in Squeezing Due to Loss}

Internal loss in the JTWPA is likely to be a source of reduction in squeezing from the ideal results shown in~\cref{fig:spectra}. A simplified model for losses is a beam splitter with transmittance $\sqrt{\eta(\omega)}$ placed after the JTPWA, with vacuum noise incident on the beam splitter's second input port~\cite{Mallet11}. This leads to a reduction in photon number, $N_R(\omega,z) \to |\eta(\omega)| N_R(\omega,z)$, and squeezing parameter, $M_R(\omega,z) \to \sqrt{\eta(\omega)}\sqrt{\eta(2\Omega_p-\omega)} M_R(\omega,z)$. Taking $\eta = \eta(\omega)$ frequency independent for simplicity, this gives a reduction in squeezing, $S_R(\omega,z) \to 2|\eta| N_R(\omega,z) + 1  - 2|\eta||M_R(\omega,z)|$. Note that distributed loss throughout the JTWPA can be taken into account through a simple phenomenological model~\cite{Caves87}, but this is beyond the scope of the present discussion.
\begin{figure}
  \includegraphics{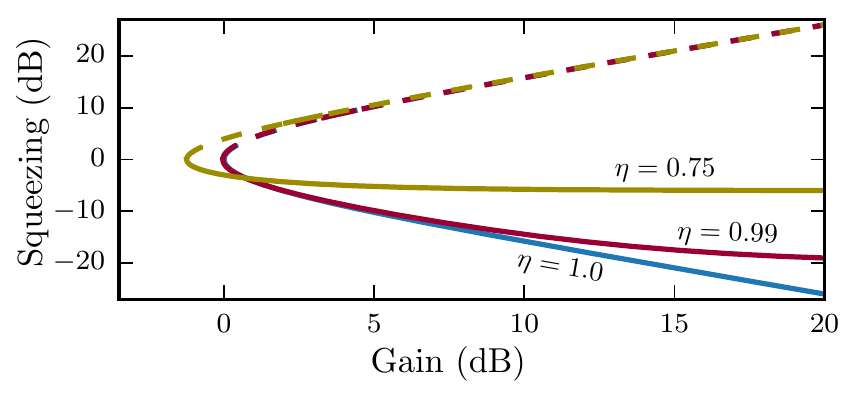}
  \caption{\label{fig:loss}Squeezing as a function of gain, $G(\omega,z) = \eta |u(\omega,z)|^2$, in the presence of loss, modelled as a beam splitter with transmittance $\eta$ placed at the JTWPA output. The solid lines show the maximally squeezed quadrature for three different values of $\eta$, while the dashed lines show the corresponding anti-squeezed quadrature.}
\end{figure}

\cref{fig:loss} shows the maximum squeezing level as a function of gain as the pump strength is ramped up. The parameters are otherwise identical to those used for the blue lines displayed in~\cref{fig:spectra} $(a)$. The solid lines show the maximally squeezed quadrature, while the dashed lines show the corresponding anti-squeezed quadrature, for three different values $\eta=0.75$ (yellow), $0.99$ (dark red) and $1.00$ (blue). Note that the gain is also reduced by the loss, $G(\omega,z) = \eta|u(\omega,z)|^2$, such that we have attenuation at zero pump power.

For a non-unity $\eta$, the squeezing level saturates with gain, while the anti-squeezed quadrature keeps growing proportionally. The maximal squeezing depends sensitively on $\eta$: while a quantum-limited device with $\eta=1$ would produce more than 25 dB of squeezing at 20 dB of gain, a device with $\eta=0.75$ only gives about 6.5 dB of squeezing for the same gain.

\section{\label{sect:twoqubits}Probing the output}
The examples discussed above demonstrate how the flexible JTWPA design allows for generating nonclassical light with interesting and useful squeezing spectra.

The squeezing spectrum 
can be found experimentally by measuring the variance of filtered two-mode quadratures (see~\cref{app:inout} and, e.g.,~\cite{Eichler11,Flurin12,Eichler14,Forgues15}). However, this necessarily includes insertion loss and noise from subsequent parts of the amplification chain Ref.~\cite{Macklin15}.
For a more direct probing of the JTWPA's performance we propose placing two superconducting qubits capacitively coupled to the transmission line at the output port. 

For two off-resonant qubits with respective frequencies $\omega_1 \ne \omega_2$, and $\omega_1+\omega_2 \not\simeq 2\Omega_p$, the qubits will be in uncorrelated thermally populated states. If, however, $\omega_1+\omega_2 = 2\Omega_p$, the qubits become entangled and information about the JTWPA's squeezing spectrum is encoded in the joint two-qubit density matrix. This information can then be extracted by measuring qubit-qubit correlation functions.

Assuming for simplicity that the qubits are both located at the JTWPA output, $x_0>z$, their reduced dynamics after tracing out the bath is governed by a Markovian master equation, $\dot{\rho} = \mathcal{L}\rho$.
The form of $\mathcal{L}$ for the general case is given in~\cref{app:bath}, while we here focus on the most interesting situation when the two qubits are tuned in with the squeezing interaction, such that $\omega_1+\omega_2=2\Omega_p$, where $\omega_m$ is the frequency of the $m$th qubit. We can then write the Lindbladian in the interaction picture
\begin{equation}\label{eq:twoqubitme}
  \begin{aligned}
    \mathcal{L} =
    \sum_{\substack{\nu={\rm L,R} \\ m=1,2}}&\left\{
    \frac{\gamma_m}{2}(N_{m,\nu}+1) \mathcal{D}[\hs_-^{(m)}] + \frac{\gamma_m}{2} N_{m,\nu} \mathcal{D}[\hs_+^{(m)}] \right\} \\
    & -\frac{\sqrt{\gamma_1\gamma_2}}{2} \mathcal{S}_{M_\nu}[\hs_+^{(1)},\hs_+^{(2)}],
  \end{aligned}
\end{equation}
where
\begin{equation}\label{eq:Sdef}
  \mathcal{S}_{M}[A,B]\rho = M \left(A\rho B + B\rho A - \{AB,\rho\}\right) + \text{H.c.},
\end{equation}
describes a dissipative squeezing interaction, 
and $\mathcal{D}[A]\rho = A\rho A^\dagger - \{A^\dagger A,\rho\}/2$ is the usual dissipator. 
$\gamma_m$ is the decay rate of qubit $m$ and $\hs_- = \ket{g}\bra{e}$ ($\hs_+ = \ket{e}\bra{g}$) is the qubit lowering (raising) operator.
The Lindbladian has two contributions coming from the left- and the right-moving field respectively. In general both fields can have non-zero thermal photon number $N_{m,\nu}=N_\nu(\omega_m)$ and squeezing parameter $M_{\nu}=\left[M_\nu(\omega_1)+M_\nu(\omega_2)\right]/2$.
If, on the other hand, the qubits are tuned out of resonance with the squeezing interaction, $\omega_1+\omega_2\not\simeq 2\Omega_p$, the last line in~\cref{eq:twoqubitme} will be fast rotating and can be dropped in a rotating wave approximation (see~\cref{app:bath} for more details).

Assuming for simplicity a single right-moving pump and a left-moving field in the vacuum state, we have that
for $\omega_1+\omega_2\not\simeq 2\Omega_p$, 
the steady state 
of the two qubits 
is the product state $\rho = \rho_1 \otimes \rho_2$, where $\rho_m$ is a thermal state with thermal population $N_{R}(\omega_m)/2$ and inversion
$\braket{\hs_z^{(m)}} = -1/(N_{R}(\omega_m)+1)$.
On the other hand, for $\omega_1+\omega_2 = 2\Omega_p$ the qubits become entangled. 
Under the simplifying symmetric assumptions $N_{R}(\omega_m) \equiv N$ and $\gamma_m \equiv \gamma$ we find that
\begin{equation}\label{eq:twoqubits:xx_corr}
  \begin{aligned}
    &\braket{\hs_x^{(1)}\hs_x^{(2)}} = \frac{\Re[M]}{(N+1)\left[(N+1)^2-|M|^2\right]} \\
  &= -\braket{\hs_x^{(1)}\hs_x^{(2)}},
  \end{aligned}
\end{equation}
and
\begin{equation}\label{eq:twoqubits:xy_corr}
  \braket{\hs_x^{(1)}\hs_y^{(2)}} = -\frac{\Im[M]}{(N+1)\left[(N+1)^2-|M|^2\right]}, 
\end{equation}
in steady state, where $M(\omega_i)\equiv M$. More general expressions are given in~\cref{app:bath}. Hence, by measuring qubit-qubit correlation functions and single-qubit inversion using standard qubit readout protocols~\cite{Blais04,Vijay11,Evan14}, one can map out the squeezing spectrum and the quantum efficiency of the JTWPA 
(the latter also requires knowledge of the thermal noise at the input, which could be probed in a similar way by a single qubit located at the input port, see~\cite{Macklin15} for a similar experiment).

\begin{figure}
  \includegraphics{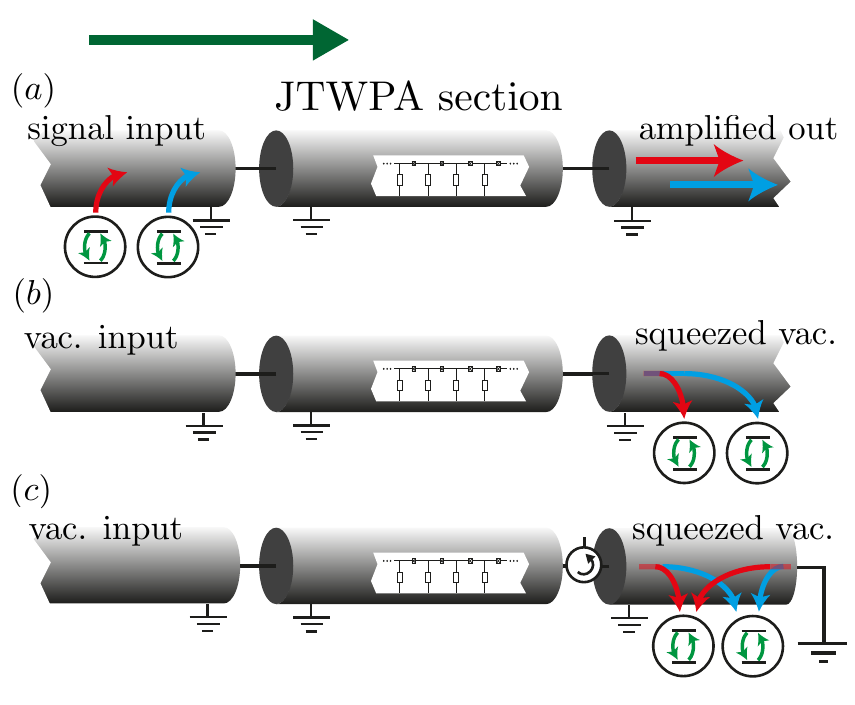}
  \caption{\label{fig:opmodes}Three different modes of operation for a JTPWA. $(a)$ Amplification mode: Quantum systems (here depicted as two-level systems to illustrate) are placed at the device input. $(b)$ Probing mode: Quantum systems placed at the output absorbs correlated photons from the JTWPA's output field and become entangled. $(c)$ Reflection mode: Higher degrees of entanglement can be reached by avoiding the left-moving vacuum noise. A circulator can be added to avoid back scattering into the JTWPA.}
\end{figure}

We can also turn this around and, rather than view the two qubits as a probe of the JTWPA's performance, view the JTWPA as a source of entanglement for the qubits. To achieve maximal degree of entanglement between the qubits, it is desirable to avoid the vacuum noise of the left-moving field. This can be achieved by squeezing the left-moving field with a separate JTWPA section, or more simply by operating the device in reflection mode, as illustrated in \cref{fig:opmodes}~$(c)$.

Assuming ideal conditions where the qubits couple symmetrically to equally squeezed left- and right-moving fields,
$N_{L}(\omega_i)=N_{R}(\omega_i)\equiv N/2$,
$M_L(\omega_i)=M_R(\omega_i)\equiv M/2$,
and ideal lossless squeezing,
the steady state of the two qubits is the pure state (see~\cref{app:bath} for more information)
\begin{equation}\label{eq:twoqubits:Psi}
  \ket{\Psi^\theta} = \frac{1}{\sqrt{2N+1}}\left(\sqrt{N+1}\ket{gg} + \e^{i\theta}\sqrt{N}\ket{ee} \right),
\end{equation}
where $\theta$ is the squeezing angle.
For large $N$, this pure state approaches a maximally entangled state with entanglement entropy 
$E(\ket{\Psi^\theta}) = -\tr[\rho_1\log_2(\rho_1)] \simeq 1 - 1/4N^2$, 
where $\rho_1 = \tr_2 \left[\ket{\Psi^\theta}\bra{\Psi^\theta}\right]$. 

\begin{figure}
  \centering
  \includegraphics{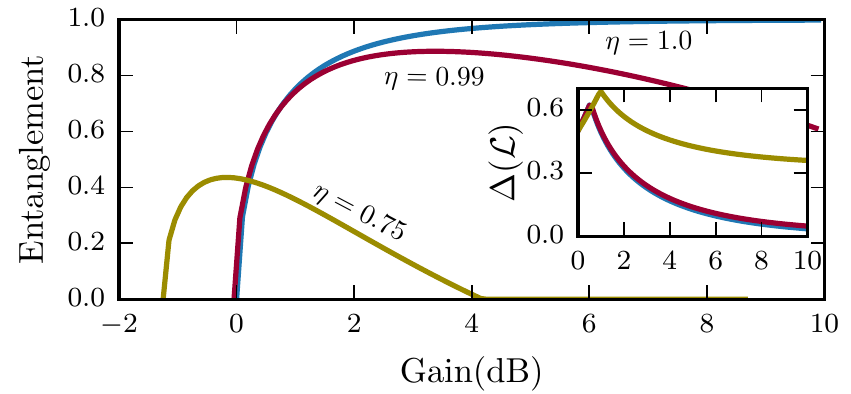}
  \caption{\label{fig:ent}Concurrence of two qubits in a two-mode squeezed bath as a function of the gain of the squeezing source, $G(\omega,z) = \eta |u(\omega,z)|^2$, for three different source loss levels $\eta=0.75,0.99,1.00$. No thermal noise at the squeezing source input is assumed.}
\end{figure}

Of practical importance is the steady state entanglement's dependence on the degree of loss,
and the behaviour of the spectral gap of the Lindbladian in~\cref{eq:twoqubitme}. The latter is important because it sets the time-scale for approaching the steady state. It is defined as $\Delta(\mathcal{L}) = |\Re\,\lambda_1|$, where $\lambda_1$ is the non-zero right-eigenvalue of $\mathcal{L}$ with real part closest to zero. 
In \cref{fig:ent} we plot the steady state entanglement, quantified by the concurrence~\cite{Wootters01}, and the spectral gap
as a function of gain for different values of $\eta$ (as defined in~\cref{sect:nonideal}). These results show that the achievable entanglement is very sensitive to loss, but an upshot is that relatively modest gains are needed to achieve high degree of entanglement, which might facilitate creating devices with higher $\eta$.
Furthermore, note that multiple pairs of qubits can be entangled using a single squeezing source. Due to the large bandwidth of the JTWPA several tens of entangled qubit pairs can likely be generated in this way using a single device.

Our scheme for probing the squeezing spectrum is similar to previous proposals for extracting information about single-mode squeezing through the resonance fluorescence emitted by a single atom~\cite{Gardiner86,Carmichael87}. The predictions of Refs.~\cite{Gardiner86,Carmichael87} were recently confirmed experimentally using a superconducting artificial atom coupled to the squeezed output field of a Josephson parametric amplifier~\cite{Toyli16}. Our scheme extends this to probing correlation between different frequency components of the squeezed radiation by going from a single to two qubits.

\section{\label{sect:cluster}Continuous variable cluster states}

The two-qubit dynamics considered above demonstrates the JTWPA's potential for entanglement generation. By adding multiple pump tones, a single frequency can become entangled with multiple other ``idler'' frequencies in a multi-mode squeezed state, and complex patters of entanglement can emerge. 
Together with its broadband nature and the potential for dispersion engineering, this turns the JTWPA into a powerful resource for dissipative quantum state engineering. 

As a demonstration of the JTWPA's potential as a source of nonclassical radiation, we show below how continuous variable (CV) cluster states can be generated through a dissipative and deterministic process, using the output field of multiple JTWPAs. Th cluster states are a powerful class of entangled many-body quantum states that are resource states for measurement based quantum computing. Given a universal cluster state, an algorithm is executed using only single-site measurements and classical feed forward on the state~\cite{Raussendorf01,Raussendorf03,Menicucci06,Briegel09,Gu09}.

A CV cluster state is defined with respect to a (weighted) simple graph $G = (V,E)$, with $V$ the set of vertices and $E$ the set of edges. A CV quantum systems with quadratures $\hx_v = (\hc_v + \hc_v^\dagger)/\sqrt{2}$ and $\hy_v = -i(\hc_v - \hc_v^\dagger)/\sqrt{2}$, where $\hc_v$ ($\hc_v^\dagger$) is a bosonic annihilation (creation) operator, is associated to each vertex $v$. The ideal CV cluster state (with respect to $G$) is defined as the unique state $\ket{\phi_G}$ satisfying~\cite{Menicucci06,Gu09,Menicucci11}
\begin{equation}\label{eq:csdef1}
  \Big( \hy_v - \sum_{w\in \mathcal{N}(v)} a_{vw} \hx_w \Big)\ket{\phi_G} = 0 \quad\forall v \in V,
\end{equation}
where $\mathcal{N}(v)$ is the neighborhood of $v$, \emph{i.e.}, all the vertices connected to $v$ by an edge in $E$ and $a_{vw} = a_{wv} \in [-1,1]$ is the weight of the edge $\{v,w\}$~\footnote{Note that $\ket{\phi_G}$ is an infinitely squeezed state, and thus not physical. In practice one has to work with Gaussian states that approaches $\ket{\phi_G}$ in a limit of infinite squeezing. We still refer to $\ket{\phi_G}$ as a ``state'' in this work, with the implicit understanding that it should be taken as a limit.}.
The operators $L_v \equiv \hy_v - \sum_{w\in \mathcal{N}(v)} a_{v,w} \hx_w$ are referred to as the nullifiers of $\ket{\phi_G}$.
We can define an adjacency matrix $A = [a_{vw}]$ for the graph, where $a_{vw} = 0$ if there is no edge $\{v,w\} \in E$. Since the adjacency matrix uniquely defines the graph, and vice versa, we use the symbol $G$ to interchangeably refer to both the graph and its adjacency matrix in the following.

We focus here on a class of graphs, first studied in Refs.~\cite{Menicucci08,Flammia09}, satisfying two simplifying criteria: 1) The graph is bicolorable. This means that every vertex can be given one out of two colors, in such a way that every edge connects vertices of different colors (the square lattice is an example). 2) The graph's adjacency matrix is self-inverse, $G=G^{-1}$. The latter constraint has a simple geometric interpretation described in Ref.~\cite{Flammia09}.
We show in~\cref{app:cluster} that for a graph $G$ satisfying these critera, the Lindblad equation $\dot{\rho} = \mathcal{L}_G\rho$, with Lindbladian
\begin{equation}\label{eq:L_G}
  \begin{aligned}
    \mathcal{L}_G ={}&
    \sum_{v \in V} \Big\{ 
      \kappa (N+1)\mathcal{D} [\hc_v]
      + \kappa N \mathcal{D} [\hc_v^\dagger]  \Big\} \\
      - &\sum_{\{v,w\} \in E} \kappa a_{vw} \mathcal{S}_{iM}[\hc_v^\dagger,\hc_w^\dagger],
  \end{aligned}
\end{equation}
where $M=\sqrt{N(N+1)}$ and $\mathcal{S}_{iM}[A,B]$ is defined in~\cref{eq:Sdef},
has a unique steady state $\ket{\phi_G(M)}$ that approaches $\ket{\phi_G}$ as $M\to\infty$.
The existence of graphs satisfying all the listed criteria, with associated cluster states, $\ket{\phi_G}$, that are universal for quantum computing, was shown in Refs.~\cite{Menicucci08,Flammia09}.

In Ref.~\cite{Wang14} Wang and coworkers showed how cluster states with graphs of the type considered here could be generated through Hamiltonian interactions between the modes of optical parametric oscillators (OPOs), followed by an interferometer combining modes from distinct OPOs. We adopt this scheme in the following, using JTWPAs (other types of broadband squeezing sources can also be used) in place of OPOs. The main difference between our proposal and that of Ref~\cite{Wang14} and previous proposals~\cite{Menicucci08,Flammia09} is that our scheme is purely dissipative: the CV modes of the cluster state never interact directly, but rather become entangled through absorption and stimulated emission of correlated photons from their environment. We focus primarily on a situation where the modes are embodied in multimode resonators, which is a particularly hardware efficient implementation. We emphasize, however, that due to the dissipative nature of the scheme, this is not a necessary constraint. The modes could in principle all be embodied in physically distinct and remote resonators, removing any constraints on locality. This is a distinct advantage of such a dissipative scheme.

\begin{figure}
  \includegraphics{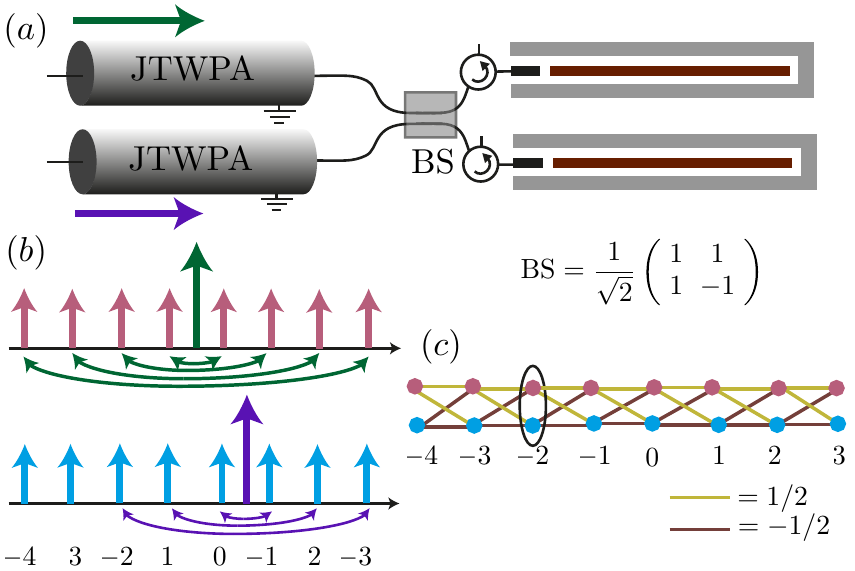}
  \caption{\label{fig:cluster_D1}Dissipative generation of a linear cluster state. $(a)$ Two JTWPAs are used as squeezing sources. The output field of the two devices are interfered on a 50-50 beam splitter enacting a Hadamard transformation, before impinging on two identical multi-mode resonators. $(b)$ Each JTWPA is pumped by a single pump tone, generating entanglement (curved arrows) between pairs of frequencies satisfying $\omega_n+\omega_m=2\Omega_i$. We focus here on center frequencies corresponding to the frequencies of the resonator modes, illustrated by the pink and blue arrows. The numbers show the macronode index of each frequency. $(c)$ Linear graph defining the steady state cluster state of the resonator modes. The horizontal edges are generated by the two pumps, while the diagonal edges are generated by the Hadamard transformation (see~\cref{app:cluster} for details). The numbers show the macronode index, and the circle shows macronode $\mathcal{M}=-2$ in the graph.}
\end{figure}

Following Ref.~\cite{Wang14}, the modes of the cluster states are resonator modes with equally spaced frequencies $\omega_m = \omega_0+m\Delta$, where $m$ is an integer, $\omega_0$ is some frequency offset and $\Delta$ the frequency separation. We require a number of degenerate modes for each frequency $\omega_m$: to create a $D$-dimensional cluster state requires a $2\times D$-fold degeneracy per frequency. This can be achieved using $2\times D$ identical multi-mode resonators, as illustrated for $D=1$ in~\cref{fig:cluster_D1}. Each mode is a vertex in the cluster state graph, and as will become clear below, a set of degenerate modes can be thought of as a graph ``macronode''~\cite{Wang14}. It is convenient to relabel the frequencies with a ``macronode index'' $\mathcal{M} = (-1)^m m$.

We show in~\cref{app:cluster} that a master equation with Lindbladian of the form~\cref{eq:L_G} is realized for a single resonator interacting with a bath generated by the output field of a JTWPA with a single pump frequency
$\Omega_p = \omega_0 + p\Delta/2$ where $p=m+n$ for some choice of frequencies $\omega_{m}\neq\omega_{n}$.
The graph is in this case a trivial graph consisting of a set of disjoint pairs of vertices connected by an edge, \emph{i.e.}, a set of two-mode cluster states which can be represented as
$G_0 =$ \includegraphics{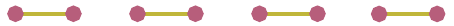} $\dots$ 
The edges have weight $+1$, under the assumption of a quantum limited, flat squeezing spectrum $M(\omega) = iM = i\sqrt{N(N+1)}$ with $N(\omega) = N$ over the relevant bandwidth.

\begin{figure}
  \centering
  \includegraphics{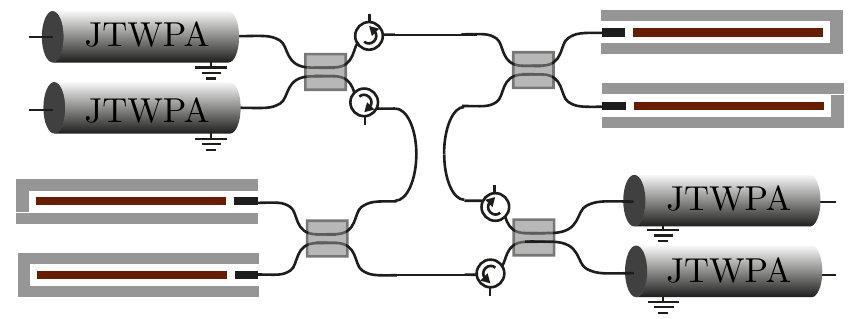}
  \caption{\label{fig:cluster_D2}Schematic setup for a universal microwave quantum computer. Four JTWPAs are used as squeezing sources to dissipatively prepare the modes of four identical multi-mode resonators in a two-dimensional cluster state. The quantum computation is subsequently performed through Gaussian and non-Gaussian (\emph{e.g.}, photon-number resolving~\cite{Schuster07}) single-mode measurements on the resonators~\cite{Gu09}.}
\end{figure}

More complex and useful graphs can be constructed using these two-mode cluster states as basic building blocks~\cite{Wang14}. Taking a number of JTWPAs, each labelled by $i$ and acting as a squeezing source independently generating a disjoint graph $G_i = $ \includegraphics{trivialgraph.pdf} as above, universal cluster states can be created by combining the output fields of the different sources on an interferometer.
The action of the interferometer can be written as a graph transformation $G = \bigoplus_i G_i$ to $G \to R G R^T$, where $R = \bigoplus_{\mathcal{M}} H_D^{(\mathcal{M})}$ represents an interferometer acting independently on each macronode $\mathcal{M}$, \emph{i.e.}, each set of $2\times D$-fold degenerate modes.
$R$ has to be orthogonal for the transformed graph to be self-inverse, $G=G^{-1}$, which we recall is one of the criteria for~\cref{eq:L_G} to generate the corresponding cluster state.
As shown in Ref.~\cite{Wang14} this is the case if the $2D\times 2D$ matrix $H_D$ is a Hadamard transformation $H_D = H^{\otimes D}$ built from $2\times 2$ Hadamard matrices
\begin{equation}\label{eq:Hadamard}
  H = \frac{1}{\sqrt{2}}
  \left(\begin{array}{cc}
      1 & 1 \\
      1 & -1
  \end{array}\right).
\end{equation}
Physically such a transformation can be realized by pairwise interfering the output fields of the JTWPAs on 50-50 beam splitters with beam splitter matrix as in~\cref{eq:Hadamard}. The network of beam splitters needed for the case $D=1$ is illustrated in~\cref{fig:cluster_D1}, for $D=2$ in~\cref{fig:cluster_D2}, and for higher dimensions in Ref.~\cite{Wang14}.

In Ref.~\cite{Wang14} it was shown that graphs $G$ constructed in this way can give rise to $D$-dimensional cluster states that are universal for measurement based quantum computing. Let us consider an example with $D=1$ in some more detail to illustrate the basic principles, while referring the reader to~\cite{Wang14} for more details. First, take two JTWPAs pumped individually with respective pump frequencies $\Omega_i$ and $\Omega_j$, with $i=-j=\Delta\mathcal{M}$. On the macronode level, this gives exactly one edge between macronodes separated by $|\Delta \mathcal{M}|$, as illustrated by the horizontal edges in~\cref{fig:cluster_D1} for $\Delta \mathcal{M} = 1$. By interfering the output fields of the two JTWPAs on a beam splitter defined by \cref{eq:Hadamard},
every node in each macronode becomes entangled with every node in the neighboring macronode, as illustrated by the diagonal edges in the figure. This gives a graph $G$ with a linear structure,
corresponding to a one-dimensional cluster state that is universal for single-mode quantum computation~\cite{Wang14,Flammia09,Menicucci11b}.

The scheme can straight-forwardly be scaled up to arbitrary $D$-dimensional cluster states using $2\times D$ JTWPAs and the same number of beam splitter transformations as shown in Ref.~\cite{Wang14}. 
$D=2$ is sufficient for universal quantum computation; a possible setup of JTWPAs and resonators is illustrated in~\cref{fig:cluster_D2}. As emphasized in Ref.~\cite{Wang14}, the relative ease of creating even higher dimensional cluster states is a very attractive property of the scheme. $D=3$ might allow for error correction with high thresholds based on surface-code encodings~\cite{Zhang08}, and $D\ge 4$ might allow for simulating systems with topological self-correcting properties~\cite{Dennis02}.

\section{Conclusions}
We have shown how the recently developed JTWPA devices are powerful sources of nonclassical radiation. The design flexibility and broadband nature of the radiation allows us engineer the output field's squeezing spectrum. We have shown how one can create multi-mode squeezed baths that can be used for dissipative quantum state preparation. By placing quantum systems at the output of a broadband squeezing source, the systems are cooled into non-trivial entangled states by interacting with a multi-mode squeezed vacuum.
In particular we have shown how to prepare pairs of entangled qubits and continuous variable cluster states that are universal for quantum computing. In both cases the large bandwidth of the JTWPA makes the state preparation highly hardware efficient.

The ability to prepare cluster states demonstrates the universal power of broadband squeezing as a resource. We hope this motivates experimental efforts to demonstrate high degrees of squeezing over large bandwidths. It should also motivate a broader theoretical study of how squeezing sources such as the JTWPA can be used to generate quantum radiation with complex entanglement structures geared towards particular applications in quantum technology and information processing.

\begin{acknowledgments}
A.~L.~G thanks N. Quesada and J. Sipe for helpful discussions on quantization in dispersive and inhomogeneous media, and N. Menicucci and O. Pfister for helpful comments regarding continuous variable cluster states.
We also thank A. Clerk, L. Govia and A. Kamal for useful discussions.
This work was supported by the Army Research Office under Grant No. W911NF-14-1-0078 and NSERC.
This research was undertaken thanks in part to funding from the Canada First Research Excellence Fund.
\end{acknowledgments}

\appendix

\section{\label{app:Hamiltonian}Hamiltonian Treatment of a JTWPA}

In this Appendix we give a more detailed Hamiltonian treatment of the JTWPA. The device we consider consists of $N$ coupled Josephson junctions, as illustrated in~\cref{fig:TWPA}. Each junction has Josephson energy $E_J$, junction capacitance $C_J$, and is coupled to ground by an impedance $Z(\omega)$, describing a reactive and dissipationless element.  We first treat the case of a single capacitance to ground, $Z(\omega) = 1/i\omega C$, before considering the more general case where the impedance contains resonances.

\subsection{Without Resonances}

We write the Lagrangian of the JTWPA in terms of the node fluxes, $\phi_n$, following the standard lumped element approach~\cite{Devoret95},
\begin{equation}
  \begin{aligned}
    L = \sum_{n=0}^{N-1}& \bigg\{
      \frac{C}{2} \dphi_n^2 + 
      \frac{C_{J}}{2}\Delta \dot{\phi}_{n}^2
      - \frac{E_{J}}{2}\left(\fq \right)^2\Delta\phi_{n}^2\\
    &+ \frac{E_{J}}{4!} \left(\fq \right)^4\Delta\phi_{n}^4 \bigg\},
  \end{aligned}
\end{equation}
where $\Delta \phi_n = \phi_{n+1}-\phi_n$, $\Phi_0 = h/2e$ is the magnetic flux quantum, and we have expanded the Josephson junction cosine potential to fourth order. 

In experimental realizations, the number of junctions is of the order of a few thousand, and the unit cell distance, $a$, is much smaller than the relevant microwave wavelengths. This justifies a continuum limit treatment of the device. Taking
\begin{align}
  &x_n = na,\\
  &\phi_{n}(t) \to \phi(x_n,t),\\
  &\Delta \phi_{n} \to a\partial_x \phi(x_n,t),
\end{align}
and defining the continuum parameters
\begin{subequations}\label{eq:continuumparams}
\begin{align}
  &C = c a,\\
  &E_{J} \left(\frac{2\pi}{\Phi_0}\right)^2 = \frac{1}{l a},\\
  &C_{J} = \frac{1}{\omega_P^2 l a},\\
  &\frac{E_{J}}{12} \left(\frac{2\pi}{\Phi_0}\right)^4 = \frac{\gamma}{a^3},
\end{align}
\end{subequations}
where $\omega_P = (2\pi/\Phi_0) \sqrt{E_J/C_J}$ is the junctions' plasma frequency,
we can formally take the continuum limit $N\to\infty$, $a\to 0$ such that the length of the device $N a \equiv z$ is kept constant. A continuum Lagrangian can then be introduced
\begin{equation}\label{eq:appH:L}
  \begin{aligned}
    &L[\phi,\partial_t\phi] ={} \half \int_{-\infty}^\infty \dd x \bigg\{ c(x)[\partial_t\phi(x,t)]^2 \\
      &- \frac{1}{l(x)}[\partial_x \phi(x,t)]^2
      + \frac{1}{\omega_P^2(x) l(x)} [\partial_x\partial_t \phi(x,t)]^2 \\
      &+ \gamma(x) [\partial_x \phi(x,t)]^4 \bigg\}.
  \end{aligned}
\end{equation}
The $x$-dependent parameters in the above expression are defined such that they take the values in~\cref{eq:continuumparams} for $0 < x < z$, while outside this region we take $c(x) = c_0$, $l(x) = l_0$, $\omega_P(x) = \infty$ and $\gamma=0$. \cref{eq:appH:L} thus represents a JTWPA section extending from $x=0$ to $x=z$, sandwiched between two identical semi-infinite linear transmission line sections extending from $x=-\infty$ to $x=0$ and $x=z$ to $x=\infty$, respectively. We furthermore assume that the sections are impedance matched, $Z_0 = \sqrt{l_0/c_0} = \sqrt{l/c}$, ensuring that there are no reflections at the boundaries.

The Euler-Lagrange equation found from~\cref{eq:appH:L} is
\begin{equation}\label{eq:appH:EL}
\left(c\partial_t^2 - \frac{1}{l}\partial_x^2 - \frac{1}{\omega_P^2 l} \partial_x^2\partial_t^2 \right) \phi = 2\gamma \partial_x [\partial_x \phi]^3,
\end{equation}
where we have left out the $x$ and $t$ dependence of the fields and the parameters for notational simplicity.
It is useful to find stationary solutions of the form $\phi(x,t) = \phi(x)\e^{-i\omega t}$ to the linear part of~\cref{eq:appH:EL}, \emph{i.e.}, the left hand side only of this equation. These classical solutions serve to determine the spatial dependence of the field in the absence of nonlinearity and are useful for constructing quantized solutions to the full problem~\cite{Bhat06}. We find that solutions of the form $\phi(x,t) = A \e^{-i\omega t + i k_\omega x}$ satisfies the linear part of~\cref{eq:appH:EL} with a dispersion relation
\begin{subequations}\label{eq:appH:disprel}
\begin{align}
  &\left.
  \begin{array}{l}
    k_\omega^2 = \frac{\omega^2}{v_1^2} \frac{1}{1-\omega^2/\omega_P^2}
  \end{array}\right.
  \qquad \text{for } 0 < x < z \label{eq:appH:disprel_inside}, \\
  &\left.
  \begin{array}{l}
    k_\omega^2 = \frac{\omega^2}{v_0^2}
  \end{array}\right.
  \qquad \text{otherwise},
\end{align}
\end{subequations}
where $v_1 = 1/\sqrt{cl}$ and $v_0=1/\sqrt{c_0l_0}$. Of course, the right hand side of~\cref{eq:appH:disprel_inside} has to be positive for traveling wave solutions to exist. In practice, we are interested in frequencies $\omega^2 \ll \omega_P^2$, such that the dispersion due to the junctions' plasma oscillations is relatively small.

We now wish to transform to a Hamiltonian. The canonical momentum to $\phi(x,t)$ is found from~\cref{eq:appH:L} in the usual way,
\begin{equation}\label{eq:appH:pi}
  \begin{aligned}
    \pi ={}& \frac{\delta L}{\delta [\partial_t \phi]} = \frac{\partial\mathcal{L}}{\partial[\partial_t \phi]} - \partial_x \frac{\partial\mathcal{L}}{\partial[\partial_x\partial_t \phi]} \\
  ={}& c \partial_t \phi - \frac{1}{\omega_P^2 l} \partial_x^2\partial_t \phi,
  \end{aligned}
\end{equation}
where $\CL$ is the Lagrangian density.
Note that this differs from the usual prescription, $\pi = \partial \CL/\partial(\partial_t\phi)$, due to the term proportional to $1/\omega_P^2$~\cite{Drummond14}.
From this we can define a Hamiltonian $H=H[\phi,\pi]$,
\begin{equation}
  H = \int_{-\infty}^\infty \dd x \pi \partial_t\phi - L,
\end{equation}
which after an integration by parts and dropping a boundary term reads $H=H_0+H_1$ with
\begin{equation}\label{eq:appH:H0}
  \begin{aligned}
    H_0 = \half \int_{-\infty}^\infty& \dd x \bigg\{ c [\partial_t\phi]^2 
      + \frac{1}{l}[\partial_x \phi]^2 \\
      &+ \frac{1}{\omega_P^2 l} [\partial_x\partial_t \phi]^2 \bigg\}
  \end{aligned}
\end{equation}
and
\begin{equation}\label{eq:appH:H1}
  \begin{aligned}
    H_1 = - \frac{\gamma}{2} \int_0^z \dd x [\partial_x \phi]^4.
  \end{aligned}
\end{equation}
Note that although we express the Hamiltonian density in terms of $\partial_t \phi$ at this stage, this should be considered a function of $\pi$.

Quantization follows by the usual prescription of promoting the fields to operators $\phi(x,t) \to \hphi(x,t)$, $\pi(x,t) \to \hpi(x,t)$ and imposing the canonical commutation relation
\begin{equation}\label{eq:appH:comrel}
  [\hphi(x,t),\hpi(x',t)]=i\hbar\delta(x-x').
\end{equation}
We attack the problem by first considering only the linear Hamiltonian $H_0$, and diagonalizing this part by expanding $\hphi(x,t)$ in a set of mode functions and inserting into the Hamiltonian, not taking the interaction $H_1$ into account for the moment. We will subsequently use these modes to construct the full-nonlinear Hamiltonian.

We follow closely the treatment of a dielectric with an interface given by Santos and Loudon in Ref.~\cite{Santos95} and write for the flux
\begin{equation}\label{eq:appH:phianzats}
  \begin{aligned}
    \hphi(x,t) ={}& \sum_{\nu={\rm L,R}} \int_{0}^\infty \dd\omega \sqrt{\frac{\hbar}{2 c(x) \omega}} g_{\nu\omega}(x)
    \ha_{\nu\omega} \e^{-i\omega t} \\
    &+ \text{H.c.},
  \end{aligned}
\end{equation}
where 
$[\ha_{\nu\omega},\ha^\dagger_{\mu\omega'}]=\delta_{\nu\mu}\delta(\omega-\omega')$, and mode functions
\begin{align}\label{eq:modeanzats}
  g_{\nu\omega}(x) ={}& \sqrt{\frac{1}{2\pi \eta_{\omega}(x)v(x)}} \e^{\pm i k_{\omega}(x) x},
\end{align}
where the $+$ sign is for $\nu={\rm R}$ and the $-$ sign is for $\nu={\rm L}$, with $k_\omega(x) = \eta_\omega(x)\omega/v(x)$ the wavevector. The refractive index, $\eta_\omega(x)$, has a different value inside and outside the JTWPA section, according to~\cref{eq:appH:disprel}.
Note that the refractive index should be real and symmetric, as we are not including absorption in the theory~\cite{Santos95}.
Using~\cref{eq:appH:pi,eq:appH:phianzats} this gives for the  canonical momentum $\hpi(x,t)$,
\begin{equation}\label{eq:appH:pianzats}
  \begin{aligned}
    \hpi(x,t) ={}& -i \sum_{\nu={\rm L,R}} \int_{0}^\infty \dd \omega \sqrt{\frac{\hbar c(x) \omega}{2}} \\
    &\times \varepsilon_{\omega}(x)  g_{\nu\omega}(x) 
    \ha_{\nu\omega} \e^{-i\omega t}
    + \text{H.c.},
  \end{aligned}
\end{equation}
where we have defined the dielectric function
\begin{equation}
  \begin{aligned}
    \varepsilon_{\omega}(x)
    ={}&  1 + \frac{\omega^2}{\omega_P^2(x)}\eta_{\omega}(x)^2.
  \end{aligned}
\end{equation}
Using~\cref{eq:appH:disprel} this can also be expressed as
\begin{equation}
  \begin{aligned}
    \varepsilon_{\omega}(x)
    = \eta_\omega(x)^2 = \frac{1}{1-\omega^2/\omega_P^2}.
  \end{aligned}
\end{equation}
Note that the canonical commutation relation, \cref{eq:appH:comrel}, implies the following condition on the mode functions
\begin{equation}\label{eq:commrelcond}
  \begin{aligned}
  \half \sum_\nu & \int_{0}^\infty \dd\omega \varepsilon_\omega(x)\left[g_{\nu\omega}(x')g_{\nu\omega}^*(x) + \text{c.c}\right]  \\
  &= i\delta(x-x').
  \end{aligned}
\end{equation}
This equality was proven to hold true for canonical fields of the form~\cref{eq:appH:phianzats,eq:appH:pi} in Ref.~\cite{Santos95} for real and symmetric $\eta_\omega(x)$ taking, as in~\cref{eq:appH:disprel}, different constant values in different sections of a dielectric with interfaces. As was pointed out by these authors, however, the theory is not entirely satisfactory as traveling wave solutions do not exist for all frequencies. As a result the integration over all frequencies is not strictly valid~\cite{Santos95}. A more careful analysis has to take absorption into account leading to localized excitations of the junction degrees of freedom.

Using the mode decomposition of $\hphi(x,t)$ and $\hpi(x,t)$, we can diagonalize the linear Hamiltonian $\hH_0$. This follows directly from the results from Ref.~\cite{Santos95}. 
To that end, we first note that from~\cref{eq:appH:H0} we can write the differential equation
\begin{equation}
  \begin{aligned}
  \frac{\partial \hH_0}{\partial t} ={}& \half \int_{-\infty}^\infty \dd x \bigg\{ c \partial_t \hphi\, \partial_t^2 \hphi + c \partial_t^2 \hphi\, \partial_t \hphi \\
  & + \frac{1}{\omega_P^2l} \left[ \partial_x \partial_t^2 \hphi\, \partial_x \partial_t \hphi + \partial_x \partial_t \hphi\, \partial_x \partial_t^2 \hphi \right] \\
& + \frac{1}{l} \partial_x \hphi\, \partial_x \partial_t \hphi + \frac{1}{l} \partial_x \partial_t \hphi\, \partial_x \hphi \bigg\}.
  \end{aligned}
\end{equation}
After an integration by parts and dropping boundary terms, this reduces to
\begin{equation}\label{eq:appH:dH0}
  \begin{aligned}
  \frac{\partial \hH_0}{\partial t} ={}& \half \int_{-\infty}^\infty \dd x \bigg\{ \partial_t \hphi\, \partial_t \hpi + \partial_t \hpi\, \partial_t \hphi \\
& + \frac{1}{l} \partial_x \hphi\, \partial_x \partial_t \hphi + \frac{1}{l} \partial_x \partial_t \hphi\, \partial_x \hphi \bigg\},
  \end{aligned}
\end{equation}
which is exactly Equation (3.3) of Ref.~\cite{Santos95} (with the identifications $\hat{E} \to -\partial_t \hphi$, $\hat{B} \to \partial_x \hphi$, $\hat{D} \to -\hpi$, $\mu_0 \to l$). It immediately follows from the results there [Eqs. (3.3) to (3.6)] that, after performing the integration over space, the linear Hamiltonian becomes
\begin{equation}\label{eq:appH:H0diag}
  \begin{aligned}
    \hH_0 ={}& \sum_{\nu} \int_{0}^\infty \dd \omega \hbar \omega
    \ha_{\nu\omega}^\dagger\ha_{\nu\omega},
  \end{aligned}
\end{equation}
where we have omitted the zero-point energy.

\subsection{With Resonances}

\begin{figure}
  \centering
  \includegraphics{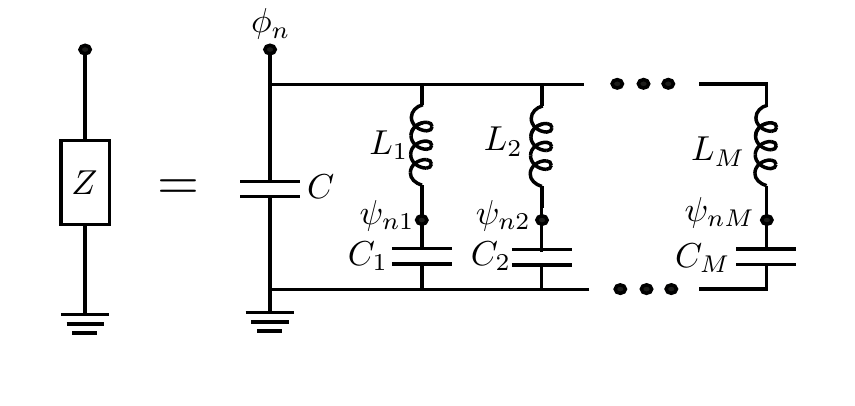}
  \caption{\label{fig:admittance}Representation of a general lossless admittance, $Z^{-1}(\omega)$, in terms of a set of modes, $\psi_{nq}$, with resonance frequencies $\omega_q=1/\sqrt{L_q C_q}$. Note that we assume the presence of a zero-frequency component represented by the capacitance $C$.}
\end{figure}
We now consider a general impedance $Z(\omega)$ describing a reactive, dissipationless element, coupling the node fluxes to ground in each unit cell. In general, we can represent this impedance with a circuit as illustrated in~\cref{fig:admittance}, and write for the inverse impedance (the admittance)~\cite{Devoret95,Nigg12},
\begin{equation}\label{eq:appH:Zinv}
  Z^{-1}(\omega) = i\omega C + \sum_q \left(\frac{1}{i\omega C_q} + i\omega L_q\right)^{-1},
\end{equation}
where we include a zero-frequency component represented by the capacitance $C$. Following identical steps as before, the linear part of the continuum Lagrangian is now modified to
\begin{equation}\label{eq:appH:L2}
  \begin{aligned}
    L_0 ={}& \half \int_{-\infty}^\infty \dd x \bigg\{ c[\partial_t\phi]^2
      - \frac{1}{l}[\partial_x \phi]^2 \\
      &+ \frac{1}{\omega_P^2 l} [\partial_x\partial_t \phi]^2 \\
    &+ \sum_q \left[ c_q[\partial_t\psi_q]^2 - \frac{1}{l_q} \left(\phi-\psi_q\right)^2 \right] \bigg\},
  \end{aligned}
\end{equation}
where $c_q(x) = C_q/a$ and $l_q(x) = L_q/a$ for $0 < x < z$ are the capacitance and inductance per unit cell associated to the $q^{\rm th}$ resonance, and $\psi_{nq}(t) \to \psi_q(x,t)$ is the continuum limit field for the resonance with frequency $\omega_q$. We take $c_q(x)=0$, $l_q(x)=\infty$ outside the JTWPA section. Similarly the linear Hamiltonian now reads after an integration by parts and dropping a boundary term
\begin{equation}\label{eq:appH:H02}
  \begin{aligned}
    H_0 ={}& \half \int_{-\infty}^\infty \dd x \bigg\{ c [\partial_t\phi]^2
      + \frac{1}{l}[\partial_x \phi]^2 \\
      &+ \frac{1}{\omega_P^2 l} [\partial_x\partial_t \phi]^2 \\
    &+ \sum_q \left[c_q [\partial_t\psi_q]^2 + \frac{1}{l_q}\left(\phi - \psi_q\right)^2 \right] \bigg\}.
  \end{aligned}
\end{equation}

The field $\psi_q(x,t)$ plays the role of a ``matter field,'' with a single resonance at $\omega_q = 1/\sqrt{L_q C_q}$ coupling linearly to the ``photonic field'' $\phi(x,t)$. This type of light-matter coupling is a standard microscopic model, based on the Hopfield model~\cite{Hopfield58}, for dispersion in linear media~\cite{Huttner91,Huttner92,Drummond14}. 

The Euler-Lagrange equations for the fields are
\begin{subequations}\label{eq:appH:EL2}
\begin{align}
  \left( c \partial_t^2 - \frac{1}{l} \partial_x^2 - \frac{1}{\omega_P^2 l} \partial_x^2 \partial_t^2  \right) \phi ={}& \sum_q \frac{1}{l_q} ( \psi_q - \phi),\\
  c_q \partial_t^2\psi_q ={}&  \frac{1}{l_q} (\phi-\psi_q).
\end{align}
\end{subequations}
As before we seek stationary solutions of the form $\phi(x,t) = A \e^{-i\omega t + i k_\omega x}$ and $\psi_q(x,t) = B_q \e^{-i\omega t + i k_\omega x}$. The Euler-Lagrange equations can be used to find the dispersion relation of the medium, but also to relate the frequency components of the fields $\psi_q(x,t)$ to those of the field $\phi(x,t)$~\cite{Drummond14}. Indeed, solutions are found for~\cref{eq:appH:EL2} with 
\begin{equation}\label{eq:appH:B_q}
B_q = (1-\omega^2/\omega_q^2)^{-1} A,
\end{equation}
and
\begin{equation}\label{eq:appH:disprel2}
  k_\omega^2 = \frac{\omega^2}{v_1^2} \frac{1+\sum_q\frac{c_q/c}{1-\omega^2/\omega_q^2}}{1-\omega^2/\omega_P^2} = \frac{-i\omega z^{-1}(\omega) l}{1-\omega^2/\omega_P^2},
\end{equation}
inside the JTWPA section and $k_\omega^2 = \omega^2/v_0^2$ elsewhere, as before. $z^{-1}(\omega) = Z^{-1}(\omega)/a$ is here the admittance to ground per unit cell.
Using~\cref{eq:appH:phianzats} for $\phi(x,t)$ and~\cref{eq:appH:B_q} to relate the frequency components of the traveling wave solutions for the matter fields to those of the photonic field, we can write
\begin{equation}
  \begin{aligned}
    \psi_q(x,t) ={}& \sum_{\nu={\rm L}}^{\rm R} \int_{0}^\infty \dd\omega \sqrt{\frac{\hbar}{2 c(x) \omega}} \frac{1}{1-\omega^2/\omega_q^2(x)} \\
    &\times g_{\nu\omega}(x) a_{\nu\omega} \e^{-i\omega t}
    + \text{H.c.}
  \end{aligned}
\end{equation}

Performing steps analogous to those in the previous section again leads to a diagonal linear Hamiltonian, $H_0$, as given in~\cref{eq:appH:H0diag}. First we find a differential equation for $\hH_0$,
\begin{equation}
  \begin{aligned}
    \frac{\partial H_0}{\partial t} ={}& \half \int_{-\infty}^\infty \dd x \bigg\{ c \partial_t \phi\, \partial_t^2 \phi + c \partial_t^2 \phi\, \partial_t \phi \\
  & - \frac{1}{\omega_P^2l} \left[ \partial_x^2 \partial_t^2 \phi\, \partial_t \phi + \partial_t \phi\, \partial_x^2 \partial_t^2 \phi \right] \\
    & + \frac{1}{l} \partial_x \phi\, \partial_x \partial_t \phi + \frac{1}{l} \partial_x \partial_t \phi\, \partial_x \phi \\
    & + \sum_q \frac{1}{l_q} \Big[
    \partial_t \phi (\phi - \psi_q)
    +(\phi - \psi_q) \partial_t\phi
  \Big]
\bigg\},
  \end{aligned}
\end{equation}
where we have used~\cref{eq:appH:EL2} and performed a partial integration, dropping a boundary term. This is of the same form as~\cref{eq:appH:dH0} if we define $\pi(x,t)$ as in~\cref{eq:appH:pianzats} with a dielectric function
\begin{equation}
  \begin{aligned}
    \varepsilon_{\omega}(x)
    ={}&  1 + \frac{\omega^2}{\omega_P^2(x)}\eta_{\omega}(x)^2 + \sum_q \frac{c_q/c}{1-\omega^2/\omega_q^2},
  \end{aligned}
\end{equation}
which from~\cref{eq:appH:disprel2} reads
\begin{equation}\label{eq:appH:eta2}
  \begin{aligned}
    \varepsilon_{\omega}(x)
    = \eta_\omega(x)^2 = \frac{1+\sum_q \frac{c_q/c}{1-\omega^2/\omega_q^2}}{1-\omega^2/\omega_P^2}.
  \end{aligned}
\end{equation}
With this, we can promote the fields to operators, and the canonical commutation relation as well as the diagonal form of $\hH_0$ follows, again, from the results of Ref.~\cite{Santos95}.

Note that the numerator of the refractive index given by~\cref{eq:appH:eta2}
has the standard form of a Sellmeir expansion that one expects for a dispersive medium with material resonances at $\omega_q$~\cite{Drummond14}. The plasma oscillations of the junctions, however, play a somewhat different role from the ``material'' resonances due to the admittance $Z^{-1}(\omega)$. Indeed, a term of the type $(\partial_x \partial_t \phi)^2$, as appearing in our Hamiltonian, is used in macroscopic models for dispersive dielectrics~\cite{Drummond14}. It is therefore interesting to note that we here have two distinct ``types'' of dispersion present at the same time---a term $(\partial_x \partial_t \phi)^2$ in the Hamiltonian, stemming from junction plasma oscillations, \emph{and} resonances associated to harmonic oscillators coupled linearly to the field $\hphi(x,t)$---and both effects arise from a microscopic theory in our case.

It is of course important to keep in mind that traveling wave solutions only exists for frequencies such that the right hand side of~\cref{eq:appH:disprel2} is positive.
Introducing resonances as illustrated in~\cref{fig:impedances} introduces band gaps in the dispersion relation, as illustrated in~\cref{fig:bandgap}. The theory developed here is valid away from any bandgap. A more general treatment would require keeping the ``matter field'' degrees of freedom and subsequently diagonalizing the linear Hamiltonian by introducing polariton fields~\cite{Huttner91,Huttner92}.

\subsection{Nonlinear Hamiltonian}

Having diagonalized the linear Hamiltonian $H_0$ we now consider the nonlinear contribution
\begin{equation}
  \hH_1 = -\frac{\gamma}{2} \int_{0}^{z} \dd x [\partial_x\hphi(x,t)]^4.
\end{equation}
We use the expansion of $\hphi(x,t)$ in frequency modes $\ha_{\nu\omega}$ introduced above, and assume that the fields $\ha_{\nu\omega}$ are small except for close to the single pump frequency. More precisely, we assume a strong right moving, classical pump centered at a frequency $\Omega_{p}$ and corresponding wave number $k_p$, and take $\ha_{R\omega} \to \ha_{R\omega} + b(\omega)$, with $b(\omega)$ a c-number. 

After dropping fast-rotating terms, the highly phase-mismatched left-moving field, and neglecting terms smaller than $\bigO[b(\omega)^2]$ we arrive at~\cref{eq:H_1,eq:H_CPM,eq:H_SQ} in~\cref{sect:inout}. Note that there is also a contribution from $\hH_1$ involving only the classical pump,
\begin{equation}\label{eq:appH:H_SPM}
  \begin{aligned}
    H_{\rm SPM} =&-\frac{\hbar}{4\pi} \int_0^\infty \dd\omega \dd\omega' \dd\Omega \dd\Omega' \sqrt{k_\omega k_{\omega'}} \\
      &\times \beta^*(\Omega)\beta(\Omega') \Phi(\omega,\omega',\Omega,\Omega') b^*(\omega) b(\omega') \\
    &+ \text{H.c.},
  \end{aligned}
\end{equation}
which describes self-phase modulation.
Here, the dimensionless pump amplitude is defined as
\begin{equation}
\beta(\Omega) = \sqrt{3 \gamma l} k_p B(\Omega),
\end{equation}
where
\begin{equation}
  B(\Omega) = \sqrt{\frac{\hbar Z_0}{4\pi \eta_{\Omega}\Omega}} b(\Omega).
\end{equation}
and we drop the $x$-argument on variables when there is no danger of confusion, as $\gamma \neq 0$ only inside the JTWPA section. The dimensionless pump amplitude can also be written 
in terms of the ratio of the pump current to the Josephson junction critical current,
\begin{equation}
  \beta(\Omega) = \frac{I_p(\Omega)}{4 I_c},
\end{equation}
where $I_c = (2\pi/\Phi_0) E_J$ is the critical current,
and we have related
the pump amplitude to the pump current through $B(\Omega) = I_p(\Omega) Z_0/\Omega$ where $Z_0 = \sqrt{l/c}$ is the characteristic impedance.

As explained in~\cref{sect:inout}, by defining an interaction picture evolution operator and taking the initial and final times to minus and plus infinity respectively, we can define an asymptotic evolution operator (to first order in $\hH_1$)
\begin{equation}\label{eq:appH:U}
  \hU \equiv \hU(-\infty,\infty) = \e^{-\frac{i}{\hbar} \hK_1 },
\end{equation}
where
\begin{equation}\label{eq:K_1}
  \hK_1 = \hK_{\rm CPM} + \hK_{\rm SQ} + K_{SPM}.
\end{equation}
and
\begin{equation}
  \begin{aligned}
    \hK_{\rm CPM}& = -\hbar \int_0^\infty \dd \omega \dd\Omega \dd\Omega' \sqrt{k_\omega k_{\omega+\Omega-\Omega'}} \\
    &\times \beta^*(\Omega)\beta(\Omega')
    \Phi(\omega,\omega + \Omega-\Omega',\Omega,\Omega') \\
    &\times \ha^\dagger_{R\omega} \ha_{R (\omega + \Omega-\Omega')}
    + \text{H.c.},
  \end{aligned}
\end{equation}
describes cross-phase modulation due to the pump,
\begin{equation}
  \begin{aligned}
    \hK_{\rm SQ}& = -\hbar \int_0^\infty \dd \omega \dd\Omega \dd\Omega' \sqrt{k_\omega k_{\Omega+\Omega'-\omega}} \\
    &\times \beta(\Omega)\beta(\Omega')
    \Phi(\omega,\Omega,\Omega+\Omega'-\omega,\Omega') \\
    &\times \ha^\dagger_{R\omega} \ha^\dagger_{R(\Omega+\Omega'-\omega)}
    + \text{H.c.},
  \end{aligned}
\end{equation}
describes broadband squeezing, and
\begin{equation}
  \begin{aligned}
    K_{\rm SPM} =&-\frac{\hbar}{2} \int_0^\infty \dd\omega \dd\Omega \dd\Omega' \sqrt{k_\omega k_{\omega_\Omega-\Omega'}} \\
      &\times \beta^*(\Omega)\beta(\Omega') \Phi(\omega,\omega+\Omega-\Omega',\Omega,\Omega') \\
      &\times b^*(\omega) b(\omega')
      + \text{H.c.},
  \end{aligned}
\end{equation}
describes pump self-phase modulation. Finally, taking the monochromatic pump limit, $b(\omega) \to b_p \delta(\omega-\Omega_p)$, with $b_p$ a c-number, we arrive at~\cref{eq:K_CPM,eq:K_SQ}, while the classical pump Hamiltonian is simply
$K_{SPM} = -\hbar z |\beta|^2 k_p  b^*_p b_p$.

\section{\label{app:inout}Asymptotic Input-Output Equations}

We define Heisenberg picture asymptotic output fields $\aR{\omega} = \hU \ha_{R\omega}\hU^\dagger$, where $\hU$ is defined in~\cref{eq:appH:U}.
To make a connection with the spatial equations of motion derived in Refs.~\cite{Yaakobi13,OBrien14,White15} we note that $\aR{\omega}$ is the solution to the following differential equation
\begin{equation}\label{eq:EOM}
  \begin{aligned}
    \partial_z \ha_{R\omega} ={}& \frac{i}{\hbar} \left[ \frac{\dd}{\dd z} \hK_1, \ha_{R\omega} \right] \\
    = 2i& |\beta|^2 k_\omega \ha_{R\omega} + i \lambda(\omega) \e^{i \Delta k_L(\omega) z} \ha^\dagger_{R(2\Omega_p-\omega)},
  \end{aligned}
\end{equation}
where we recall that $z$ is the length of the JTWPA section,
while the pump similarly is the solution to the classical equation of motion
\begin{equation}\label{eq:EOM_pump}
  \begin{aligned}
    \partial_z b_p ={}& - \left\{ \frac{\dd}{\dd z} K_{\rm SPM}, b_p \right\}
    = i |\beta|^2 k_p b_p,
  \end{aligned}
\end{equation}
where $\{\,,\}$ denotes the Poisson bracket and we have neglected any back-action onto the pump from the quantized frequency components.
\cref{eq:EOM,eq:EOM_pump} are formally identical, up to a frequency-dependent normalization of the wave amplitudes, to the classical equations of motion derived in Refs.~\cite{Yaakobi13,OBrien14,White15}.
With the anzats $\ha_{R\omega}(z) = \ta_{R\omega}(z) \e^{2i|\beta|^2 k_\omega z}$ and $b_p(z)=\tb_p(z) \e^{i|\beta|^2 k_p z}$, we have the differential equations
\begin{equation}\label{eq:EOM_tilde}
  \begin{aligned}
    \partial_z \ta_{R\omega} ={}& i \lambda(\omega) \e^{i \Delta k(\omega) z} \ta^\dagger_{R(2\Omega_p-\omega)},
  \end{aligned}
\end{equation}
and $\partial_z \tb_p = 0$,
where
\begin{equation}
  \Delta k(\omega) = \Delta k_L(\omega) + 2|\beta|^2(k_p-k_{2\Omega_p-\omega}-k_\omega).
\end{equation}
\cref{eq:EOM_tilde} can be solved exactly to give~\cite{OBrien14}
\begin{equation}\label{eq:appinout:modesol}
  \begin{aligned}
    &\ta_{R\omega}(z) = \e^{i\Delta k(\omega) z/2}\\
    &\times \bigg[ u(\omega,z) \ta_{R\omega}(0)
    + i v(\omega,z) \ta_{R(2\Omega_p-\omega)}^\dagger(0)  \bigg],
  \end{aligned}
\end{equation}
where
\begin{align}
  u(\omega,z) =\,& \cosh [g(\omega) z] - \frac{i\Delta k(\omega)}{2g(\omega)} \sinh [g(\omega) z],\\
  v(\omega,z) =\,& \frac{\lambda(\omega)}{g(\omega)} \sinh [g(\omega) z],\\
  g(\omega) =\,& \sqrt{|\lambda(\omega)|^2 - \left(\frac{\Delta k(\omega)}{2}\right)^2}.
\end{align}
This leads to~\cref{eq:modesol}.
A straightforward calculation shows that $|u(\omega,z)|^2-|v(\omega,z)|^2 = 1$, and that the modes satisfy the commutation relation
\begin{equation}
  [\aR{\omega},\aRd{\omega'}] = \delta(\omega-\omega'),
\end{equation}
for any $z$, as they should.

From~\cref{eq:appinout:modesol} it is straight forward to find the non-zero output field correlation functions for an incoming vacuum field,
\begin{subequations}\label{eq:appinout:corr}
\begin{align}
  &\begin{aligned}
    \braket{\aRd{\omega} \aR{\omega'}} 
    \equiv{}& N_R(\omega,z)\delta(\omega-\omega') \\
    ={}& |v(\omega,z)|^2\delta(\omega-\omega'),
  \end{aligned}\\
  &\begin{aligned}
    \braket{\aR{\omega} \aRd{\omega'}} 
    \equiv{}& [N_R(\omega,z)+1]\delta(\omega-\omega') \\
    ={}& |u(\omega,z)|^2\delta(\omega-\omega'),
  \end{aligned}\\
  &\begin{aligned}
    &\braket{\aR{\omega} \aR{\omega'}} 
    \equiv{} M(\omega,z)\delta(2\Omega_p-\omega-\omega')\\
    &= i u(\omega,z) v(\omega,z) \e^{i\Delta k(\omega) z} \delta(2\Omega_p-\omega-\omega'),
  \end{aligned}
\end{align}
\end{subequations}
from which the squeezing spectrum can be computed easily.

The squeezing spectrum is typically probed in experiments by heterodyne measurement of filtered field quadratures~\cite{Eichler11,Flurin12,Eichler14,Forgues15}. We here give some more details on how the squeezing spectrum defined in~\cref{eq:S} can be measured in this way.
We first define filtered output fields~\cite{Eichler11,Grimsmo16}
\begin{equation}
  \hb_{\omega_0}(t) = \int_{-\infty}^\infty \dd\omega f[\omega-\omega_0] \e^{-i(\omega-\omega_0)t}\aR{\omega},
\end{equation}
where $f[\omega]$ satisfies $f[-\omega]=f^*[\omega]$ and refers to a narrowband low-pass filter capturing the experimental bandwidth~\cite{Clerk10}. 
We also define filtered quadratures $\hX_{\omega_0}(t) = \hb_{\omega_0}^\dagger(t)+\hb_{\omega_0}(t)$, $\hY_{\omega_0}(t) = i\left[\hb_{\omega_0}^\dagger(t)-\hb_{\omega_0}(t)\right]$ and
two-mode quadratures 
\begin{align}
  \hU^{\theta\pm}_{\omega_0}(t) ={}& \frac{1}{\sqrt{2}} \left(\hY_{\omega_0}^\theta \pm \hY_{2\Omega_p-\omega_0}^\theta\right),\\
  \hV^{\theta\pm}_{\omega_0}(t) ={}& \frac{1}{\sqrt{2}} \left(\hX_{\omega_0}^\theta \pm \hX_{2\Omega_p-\omega_0}^\theta\right).
\end{align}
Two-mode squeezing refers to the fluctuations in some of these two-mode quadratures being below the vacuum level. In the present case we find squeezing in the fluctuations
\begin{equation}\label{eq:DeltaU}
  \begin{aligned}
    \Delta U_{\omega_0}^+ = \overline{\braket{\Delta \hU_{\omega_0}^+(t)\Delta \hU_{\omega_0}^+(t)}} = \overline{\braket{\Delta \hV_{\omega_0}^-(t)\Delta \hV_{\omega_0}^-(t)}},
  \end{aligned}
\end{equation}
where the overline denotes time-averaging.
For a vacuum input field we find that the two-mode fluctuations are
\begin{equation}
  \begin{aligned}
    \Delta U_{\omega_0}^+ ={}& \half \int_0^\infty \dd\omega \Big\{ |f[\omega-\omega_0]|^2 S_R(\omega,z)  \\
      & + |f[2\Omega_p-\omega -\omega_0]|^2 S_R(\omega,z),
  \end{aligned}
\end{equation}
giving a filtered version of the squeezing spectrum defined in~\cref{eq:S}.

\section{\label{app:bath}Squeezed Bath Engineering}

In this appendix we derive a master equation for a collection of quantum systems interacting with multiple multimode squeezed baths, and give detailed results for the case of two qubits placed at the output of a JTWPA pumped by a single classical pump.

\subsection{Master Equation for Multiple Systems and Multiple Squeezed Baths}

We consider a number of quantum systems distributed along linear sections of a set of transmission lines. We label the transmission lines by the index $i$ and the quantum systems coupled to the $i$th line by an index $m_i$. The systems couple to the transmission line voltages, $\hV_i(x_i,t) = \partial_t \hphi_i(x_i,t)$, at some position $x_i$, via system operators $\hc_{m_i}$. These system operators are furthermore assumed to rotate with frequencies $\omega_{m_i}$ in the interaction picture.
The interaction Hamiltonian in the interaction picture is then of the form
\begin{align}
  \hH_I(t) =&\,\hbar \sum_{i, m_i} \hB_i(t) \sqrt{\frac{\kappa_{m_i}}{2\pi}} \left(\hc_{m_i}\e^{-i\omega_{m_i} t} + \text{H.c.}\right),\\
  \hB_i(t) =& i \int_{0}^\infty \dd \omega \left( \hb_{i\omega}^\dagger\e^{i\omega t} - \text{H.c.} \right),
\end{align}
where $[\hb_{i\omega},\hb_{j\omega'}^\dagger] = \delta_{ij}\delta(\omega-\omega')$ are bath operators and $\kappa_{m_i}$ describes the coupling strength of system $m_i$ to the $i$th bath.

We do not consider a situation where the systems are cascaded~\cite{Carmichael93,Gardiner93}, \emph{i.e.}, none of the systems are driven by the output of any of the other. The different baths can, however, be correlated, with a defining set of correlation functions
\begin{subequations}\label{eq:appbath:corr}
\begin{align}
  \braket{\hb_{j\omega}^\dagger \hb_{i\omega'}} ={}& N_{ji}(\omega)\delta(\omega-\omega'),\\
  \braket{\hb_{j\omega} \hb_{i\omega'}^\dagger} ={}& [N_{ji}(\omega)+1]\delta(\omega-\omega'),\\
  \braket{\hb_{j\omega} \hb_{i\omega'}} ={}& \sum_k M_{ji}^k(\omega) \delta(\Omega_k-\omega-\omega'),
\end{align}
\end{subequations}
where $N_{ji}(\omega) = N_{ij}(\omega)$ and $M_{ji}^k(\omega) = M_{ij}^k(\omega)$. This generalizes~\cref{eq:appinout:corr} and includes, for example, setups of the type illustrated in~\cref{fig:cluster_D1,fig:cluster_D2} where the output field of multiple JTWPAs are combined on beam splitters before being incident on the quantum systems.

Following the standard approach~\cite{Gardiner04}, we find the following master equation for the reduced system density matrix under the usual Born-Markov approximation
\begin{widetext}
\begin{equation}\label{eq:appbath:ME}
  \begin{aligned}
    \dot{\rho}(t) =\,& 
    \sum_{ij} \sum_{m_i n_j} \frac{\sqrt{\kappa_{m_i} \kappa_{n_j}}}{2\pi}
    \Big\{
      \e^{-i(\omega_{m_i}-\omega_{n_j}) t} S^-_{ji}(t,\omega_{n_j}) \left( c_{m_i} \rho c_{n_j}^\dagger - \rho c_{n_j}^\dagger c_{m_i} \right)
      +\e^{i(\omega_{m_i}-\omega_{n_j}) t} S^+_{ji}(t,\omega_{n_j}) \left( c_{m_i}^\dagger \rho c_{n_j} - \rho c_{n_j} c_{m_i}^\dagger \right) \\
      &+\e^{i(\omega_{m_i}+\omega_{n_j}) t} S^-_{ji}(t,\omega_{n_j}) \left( c_{m_i}^\dagger \rho c_{n_j}^\dagger - \rho c_{n_j}^\dagger c_{m_i}^\dagger \right)
      +\e^{-i(\omega_{m_i}+\omega_{n_j}) t} S^+_{ji}(t,\omega_{n_j}) \left( c_{m_i}\rho c_{n_j} - \rho c_{n_j} c_{m_i} \right)
    +\text{H.c.}
    \Big\},
  \end{aligned}
\end{equation}
\end{widetext}
where we have defined bath correlation functions
\begin{equation}
  \begin{aligned}
    S^\pm_{ji}(t,\omega) =\,& \int_0^\infty \dd\tau \e^{\pm i\omega\tau} \braket{\hB_j(t-\tau)\hB_i(t)},
  \end{aligned}
\end{equation}
with $\omega > 0$ and the brackets refer to an expectation value with respect to the bath density matrix $\rho_B$.
At this stage we can invoke a rotating wave approximation (RWA). For example, we can approximate
\begin{equation}\label{eq:appbath:RWA1}
  \begin{aligned}
    \e^{i(\omega_m+\omega_n) t}& S^-_{ji}(t,\omega_n) \simeq{} - \int_0^\infty \dd\tau \int_{-\infty}^\infty \dd\omega\dd\omega' \e^{-i(\omega_n-\omega) \tau}\\
    &\times \braket{\hb_{j\omega}\hb_{i\omega'}}_B \e^{i(\omega_m+\omega_n-\omega-\omega')t} \\
    ={}& -\pi \sum_k M_{ij}^k(\omega_n) \e^{i(\omega_m+\omega_n-2\Omega_k)t},
  \end{aligned}
\end{equation}
where we have used
\begin{equation}\label{eq:delta_principal}
  \begin{aligned}
    \int_0^\infty \dd\tau \e^{i(\omega+\omega')\tau} {=} \pi\delta(\omega+\omega')
    + iP\left(\frac{1}{\omega+\omega'}\right),
  \end{aligned}
\end{equation}
and dropped the small principal part.
We find similar expressions for the other terms in~\cref{eq:appbath:ME}, leading to the interaction picture Master equation
\begin{widetext}
\begin{equation}\label{eq:appbath:ME2}
  \begin{aligned}
    \dot{\rho}(t) =
    \sum_{ijk} \sum_{m_i n_j} \frac{\sqrt{\kappa_{m_i} \kappa_{n_j}}}{2}
    \Big\{&
      \e^{-i(\omega_{m_i}-\omega_{n_j}) t} [N_{ji}(\omega_{n_j})+1] \left( c_{m_i} \rho c_{n_j}^\dagger  -\rho c_{n_j}^\dagger c_{m_i} \right) \\
      +&\e^{i(\omega_{m_i}-\omega_{n_j}) t} N_{ji}(\omega_{n_j}) \left( c_{m_i}^\dagger \rho c_{n_j} - \rho c_{n_j} c_{m_i}^\dagger\right) \\
      -& \e^{i(\omega_{m_i}+\omega_{n_j} - 2\Omega_k) t} M^k_{ji}(\omega_{n_j}) \left( c_{m_i}^\dagger \rho c_{n_j}^\dagger - \rho c_{n_j}^\dagger c_{m_i}^\dagger \right) \\
      -&\e^{-i(\omega_{m_i}+\omega_{n_j}-2\Omega_k) t} M^{k*}_{ji}(\omega_{n_j}) \left( c_{m_i}\rho c_{n_j} - \rho c_{n_j} c_{m_i} \right)
    +\text{H.c.}
      \Big\}.
  \end{aligned}
\end{equation}
\end{widetext}
Note that this master equation includes dissipative interactions between systems even if they are not coupled to the same bath, $i\neq j$, if the baths are correlated according to~\cref{eq:appbath:corr}. As we will see in~\cref{app:cluster}, this is the type of interactions that give rise to the diagonal edges in the cluster state graph in~\cref{fig:cluster_D1}. But first, we focus in the next subsection on the simple case of two qubits coupled to the left- and right-moving fields of a single transmission line.

\subsection{Two Qubits}

In this section we consider two qubits coupled to the left- and right moving field of a transmission line (in general both fields can be squeezed). We take both fields to have correlation functions as in~\cref{eq:appinout:corr} without any cross correlations between the two fields ($N_{ji}, M_{ji} \propto \delta_{ji}$ in~\cref{eq:appbath:corr}).

The qubits have free Hamiltonians $\hH_m = \omega_m\hs_z^{(m)}/2$, $m=1,2$, and we take $c_{m_i} \to \sigma_-^{(m)}$ in~\cref{eq:appbath:ME2}. We also assume that the qubits are off-resonant, $\omega_1\neq\omega_2$. The resulting master equation is
\begin{align}
  \mathcal{L} ={}&
    \sum_{\substack{\nu={\rm L,R} \\ m=1,2}}\left\{
    \frac{\gamma_m}{2}(N_{m,\nu}+1) \mathcal{D}[\hs_-^{(m)}] + \frac{\gamma_m}{2} N_{m,\nu} \mathcal{D}[\hs_+^{(m)}] \right\} \nonumber \\
    &-\frac{\sqrt{\gamma_1\gamma_2}}{2} \Big\{ \e^{i(\omega_1+\omega_2-2\Omega_\nu) t}M_\nu\mathcal{C}[\hs_+^{(1)},\hs_+^{(2)}] \label{eq:apptwoqubits:ME} \\
    &\qquad\qquad\,\,\,\,+\e^{-i(\omega_1+\omega_2-2\Omega_\nu) t}M_\nu^*\mathcal{C}[\hs_-^{(1)},\hs_-^{(2)}] \Big\}, \nonumber
\end{align}
where we have dropped off-resonant terms rotating at $\omega_1-\omega_2$, defined
\begin{equation}
  \mathcal{C}[A,B]\rho = \left(A\rho B + B\rho A - \{AB,\rho\}\right),
\end{equation}
and $\gamma_m$ the decay rate of qubit $m$, which has contributions from both the left- and the right-moving field.
$N_{m,\nu}=N_\nu(\omega_m)$ and $M_{\nu}=\left[M_\nu(\omega_1)+M_\nu(\omega_2)\right]/2$ are, respectively, thermal photon number and squeezing parameters for each field.

If the qubits are not tuned into resonance with the squeezing interaction induced by the pump fields, $\omega_1+\omega_2 \not\simeq 2\Omega_\nu$, the last two lines of~\cref{eq:apptwoqubits:ME} can be dropped in an RWA. In this case, the two qubits are effectively interacting with independent thermal baths, and will reach an uncorrelated thermal steady state, $\rho_{\rm ss} = \rho_1 \otimes \rho_2$, where
\begin{equation}
  \rho_{m} = 
  \left(
  \begin{array}{cc}
    \frac{N_{m,{\rm tot}}}{2N_{m,{\rm tot}}+2} & 0 \\
    0 & \frac{N_{m,{\rm tot}}+2}{2N_{m,{\rm tot}}+2}
  \end{array}
  \right),
\end{equation}
where $N_{m,{\rm tot}} = N_{m,{\rm R}} + N_{m,{\rm L}}$ is the total thermal photon number at the frequency of qubit $m$. For a left moving field in the vacuum state, $N_{\rm{L},m}=0$, this gives the atomic inversion quoted in~\cref{sect:twoqubits}.

As discussed in~\cref{sect:twoqubits}, when the qubits are tuned into resonance with the squeezing interaction, they become entangled and encode information about the squeezing spectrum. This can be exploited to use the qubits as a spectroscopic probe. Assuming a left moving field in the vacuum state, and taking $\omega_1+\omega_2=2\Omega_R$, \cref{eq:apptwoqubits:ME} reduces to
\begin{align}
  \mathcal{L} ={}&
    \sum_{m=1,2}\left\{
    \frac{\gamma_m}{2}(N_{m}+2) \mathcal{D}[\hs_-^{(m)}] + \frac{\gamma_m}{2} N_{m} \mathcal{D}[\hs_+^{(m)}] \right\} \nonumber \\
    &-\frac{\sqrt{\gamma_1\gamma_2}}{2} \mathcal{S}_{M_{\rm R}}[\hs_+^{(1)},\hs_+^{(2)}], \label{eq:apptwoqubits:ME2}
\end{align}
where $N_m = N_{m,{\rm R}}$, $M = \left[M_{\rm R}(\omega_1)+M_{\rm R}(\omega_2)\right]/2$ and
\begin{equation}
  \mathcal{S}_{M}[A,B]\rho = M \left(A\rho B + B\rho A - \{AB,\rho\}\right) + \text{H.c.}
\end{equation}
The steady state of~\cref{eq:apptwoqubits:ME2} can be found analytically if we assume that the qubits start in the ground state. We find (in the basis $\{\ket{ee},\ket{eg},\ket{ge},\ket{gg}\}$)
\begin{widetext}
\begin{equation}
  \rho_{\rm ss} = 
  \left(
  \begin{array}{cccc}
    \frac{N_1N_2}{A} + C & 0 & 0 & \sqrt{\gamma_1\gamma_2} B\, M \\
    0 & \frac{N_1(N_2+2)}{A} - C & 0 & 0 \\
    0 & 0 & \frac{(N_1+2) N_2}{A} - C& 0 \\
    \sqrt{\gamma_1\gamma_2} B\, M & 0 & 0 & \frac{(N_1+2)(N_2+2)}{A} + C\\
  \end{array}
  \right),
\end{equation}
\end{widetext}
where
\begin{align}
  A ={}& 4(N_1+1)(N_2+1),\\
  B ={}& \gamma_1(N_1+1)+\gamma_2(N_2+2),\\
  C ={}& \frac{1}{A} \frac{4 \gamma_1\gamma_2 |M|^2}{B^2-4\gamma_1\gamma_2 |M|^2}.
\end{align}
This gives steady state expectation values
\begin{align}
  \braket{\hs_z^{(m)}} ={}& -\frac{1}{N_m+1},\\
  \braket{\hs_x^{(1)}\hs_x^{(2)}} ={}& -\braket{\hs_y^{(1)}\hs_y^{(2)}} = \frac{1}{A}\frac{2\sqrt{\gamma_1\gamma_2}\,\Re M}{B^2-4\gamma_1\gamma_2|M|^2},\\
  \braket{\hs_x^{(1)}\hs_y^{(2)}} ={}& -\frac{1}{A}\frac{2\sqrt{\gamma_1\gamma_2}\,\Im M}{B^2-4\gamma_1\gamma_2|M|^2},
\end{align}
from which both $N_m$ and $M_m$, and thus the squeezing spectrum of the source, can be extracted. For the simplifying symmetric assumptions $\gamma_1=\gamma_2\equiv\gamma$ and $N_1=N_2\equiv N$, this leads to~\cref{eq:twoqubits:xx_corr,eq:twoqubits:xy_corr}.

Another interesting case is when both the left- and right moving fields are squeezed or, equivalently, the qubit couples to a single squeezed field. This could be realized, \emph{e.g.}, by ending the transmission line with a mirror as indicated in~\cref{fig:opmodes} (c). This gives higher degrees of entanglement between the qubits, as the vacuum noise of the left moving field otherwise reduces correlations between them. In particular, taking $N_{m,\nu} = N/2$, $M_{\nu} = M/2$ and $\gamma_1=\gamma_2$ in~\cref{eq:apptwoqubits:ME}, and furthermore assuming $|M|=\sqrt{N(N+1)}$, we find that the steady state is given by the pure state~\cref{eq:twoqubits:Psi}.

\section{\label{app:cluster}Generating cluster states}

In this appendix we give more details regarding the dissipative generation of CV cluster states. Recall that the cluster state $\ket{\phi_G}$ with respect to a graph $G$ is the unique pure state satisfying
\begin{equation}\label{eq:appcluster:csdef1}
  \Big( \hy_v - \sum_{w\in \mathcal{N}(v)} a_{vw} \hx_w \Big)\ket{\phi_G} = 0 \quad\forall v \in V,
\end{equation}
where $-1 \le a_{vw} \le 1$ is the weight of the (undirected) edge $\{v,w\}$ and $\mathcal{N}(v)$ denotes the neighborhood of $v$. We can define an adjacency matrix, $A = [a_{vw}]$ (with $a_{vw}=0$ if there is no edge $\{v,w\}$ in the graph), and we use the graph $G$ and its adjacency matrix $A$ interchangeably when referring to the graph in the following.

The general question of dissipatively preparing pure Gaussian states was addressed by Koga and Yamamoto in Ref.~\cite{Koga12} and, in particular, they found a Lindblad master equation $\dot{\rho} = \mathcal{L}_G(\varepsilon)\rho$ whose unique steady state approaces~\cref{eq:appcluster:csdef1}, in the limit of infinite squeezing. A Lindbladian that achieves this is
\begin{equation}\label{eq:appcluster:L_G}
  \mathcal{L}_G(\varepsilon)/\kappa = \sum_v \mathcal{D}[L_v(\varepsilon)],
\end{equation}
where $\kappa$ is a decay rate setting the overall time-scale and the Lindblad operators are
\begin{equation}
  L_v(\varepsilon) = \hy_v - \sum_{w\in \mathcal{N}(v)} a_{vw} \hx_w -i\varepsilon \hx_v.
\end{equation}
The steady state approaches~\cref{eq:appcluster:csdef1} in the limit $\varepsilon \to 0^+$.

In the special case of a bicolorable graph, meaning that every vertex can be given one out of two colors in such a way that every edge connects vertices of different colors, the Lindbladian~\cref{eq:appcluster:L_G} is to first order in $\varepsilon$
\begin{equation}\label{eq:cv:L_G2}
  \begin{aligned}
    &\mathcal{L}_G(\varepsilon)/\kappa =\half \sum_{v \in V} \Big\{ 
      (1+D_v+2\varepsilon)\mathcal{D} [\hc_v] \\
      &+ (1+D_v-2\varepsilon)\mathcal{D} [\hc_v^\dagger]
      + (D_v-1)\mathcal{S}_{1}[\hc_v^\dagger,\hc_v^\dagger]
  \Big\} \\
  &- \sum_{\{v,w\} \in E} a_{vw} \mathcal{S}_{i}[\hc_v^\dagger,\hc_w^\dagger]\\
  &+ \sum_{\{w,w'\} \in E_2} \frac{D_{w,w'}}{2}
    \Big\{ 
    \mathcal{S}_1[\hc_{w}^\dagger,\hc_{w'}^\dagger] + \mathcal{D}_1 [\hc_{w},\hc_{w'}] \Big\}.
  \end{aligned}
\end{equation}
Recall that $E$ is the set of all pairs of vertices $\{v,w\}$ connected by an edge. Furthermore, we have defined $E_2$ as the set of all pairs of distinct vertices $\{w,w'\}$ such that there exists some $v \neq w,w'$ such that $w,w'\in \mathcal{N}(v)$. In other words, $E_2$ is the set of all next-nearest neighbors. Note that the sum over edges in $E$ is a sum over \emph{undirected} edges, \emph{i.e.}, there is \emph{one} term in the sum for every \emph{pair} of vertices connected by an edge, and similarly for the sum over $E_2$. We have furthermore defined
\begin{align}
  &D_v = \sum_{w\in \mathcal{N}(v)} a_{v,w}^2,\\
  &D_{w,w'} = \sum_{\substack{v \text{ s.t. } \\ w,w' \in \mathcal{N}(v)}} a_{v,w}a_{v,w'}.
\end{align}
The former can be recognized as the sum of the weights of all ``2-paths'' starting and ending at $v$, and the latter as the sum of the weights of all ``2-paths'' connecting the two vertices $\{w,w'\}$. The weight of a 2-path is here defined to be the product of the weights of the two respective edges.

\cref{eq:cv:L_G2} takes a particularly simple form if the following two geometric conditions for the graph $G$ are satisfied:
\begin{enumerate}
  \item The sum of all 2-paths starting and ending at the same vertex add up to one: $D_v = 1$.
  \item The sum of all 2-paths connecting two different vertices cancel out: $D_{w,w'} = 0$.
\end{enumerate}
Note that these two criteria are equivalent to saying that the adjacency matrix defining the graph is self-inverse, $A=A^{-1}$~\cite{Flammia09}.
In this case,
\begin{equation}\label{eq:appcluster:L_Gselfsimilar}
  \begin{aligned}
    \mathcal{L}_G(\varepsilon) ={}& \sum_{v \in V} \Big\{ 
      \kappa(1+\varepsilon)\mathcal{D} [\hc_v]
      + \kappa(1-\varepsilon)\mathcal{D} [\hc_v^\dagger]  \Big\} \\
      - &\sum_{\{v,w\} \in E}\kappa a_{vw} \mathcal{S}_{i}[\hc_v^\dagger,\hc_w^\dagger].
  \end{aligned}
\end{equation}
Or, by defining $|M|=1/2\varepsilon$, and $N$ through $|M| = \sqrt{N(N+1)}$ we can write to first order in $\varepsilon$
\begin{equation}\label{eq:appcluster:L_squeezing}
  \begin{aligned}
    |M|\mathcal{L}_G(\varepsilon) ={}&
    \sum_{v \in V} \Big\{ 
      \kappa (N+1)\mathcal{D} [\hc_v]
      + \kappa N \mathcal{D} [\hc_v^\dagger]  \Big\} \\
      - &\sum_{\{v,w\} \in E} \kappa |M| a_{vw} \mathcal{S}_{i}[\hc_v^\dagger,\hc_w^\dagger].
  \end{aligned}
\end{equation}

Next we show how~\cref{eq:appcluster:L_squeezing} can be generated through squeezed bath engineering. Remarkably, the three criteria we have assumed for the graph $G$, that it is bicolorable and satisfies the two conditions listed above, are exactly the criteria first studied by Menicucci, Flammia and Pfister in Refs.~\cite{Menicucci08,Flammia09}. These authors showed that graphs satisfying these criteria corresponding to cluster states that are universal for quantum computing exists, and moreover how such cluster states can be prepared using a single optical parametric oscillator (OPO). Later, Wang and coworkers~\cite{Wang14} presented an alternative implementation using a network of OPOs and beam splitters. 

We adopt the scheme from Ref.~\cite{Wang14} and use JTWPAs as squeezing sources in place of OPOs. The main difference between what we propose here and previous proposals is the dissipative nature of the interactions. While in Ref.~\cite{Wang14} and previous work~\cite{Menicucci08,Flammia09} the schemes were based on Hamiltonian interactions between internal cavity modes of OPOs, we are using dissipative interactions between resonator modes that are \emph{external} to the squeezing sources. The CV modes of the cluster state never interact directly, but rather become entangled through absorption and stimulated emission of correlated photons from their environment. Even so, the graph constructions from Ref.~\cite{Wang14} can be adopted almost directly for our purposes as well.

Following Ref.~\cite{Wang14} the modes of the cluster states will be resonator modes with equally spaced frequencies $\omega_m = \omega_0+m\Delta$, where $\omega_0$ is some frequency offset and $\Delta$ the frequency separation. We require a number of degenerate modes for each frequency $\omega_m$: to create a $D$-dimensional cluster state requires a $2\times D$-fold degeneracy per frequency. Each mode is a vertex in the cluster state graph, and as will become clear below, a set of degenerate modes can be thought of as a graph ``macronode''~\cite{Wang14}. It is convenient to relabel the frequencies $\omega_m$ with a ``macronode index'' $\mathcal{M} = (-1)^m m$.

A simple implementation is to embody the modes of the cluster state in $2\times D$ identical multi-mode resonators. Each resonator is assumed to couple to a single squeezed bath, which we label by an index $i$. We use the same number of squeezing sources, and each source is pumped by a frequency $\Omega_i = \omega_0 + i\Delta/2$ where $i=m+n$ for some choice of frequencies $\omega_{m}\neq\omega_{n}$.
The key to creating useful cluster states is to correlate the different baths by combining the output fields of broadband squeezing sources using interferometers. 

\begin{figure}
  \centering
  \includegraphics{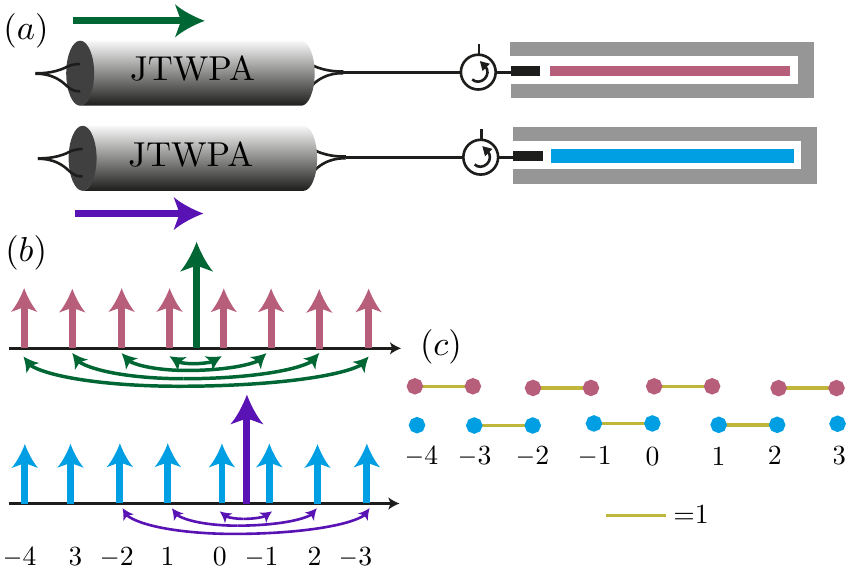}
  \caption{\label{fig:appcluster:resonators1}Independent broadband squeezing sources with a single pump frequency generates trivial cluster state graphs consisting of disjoint pairs of connected vertices. $(a)$ Experimental setup with JTWPAs whose output fields are incident on independent resonators. $(b)$ placement of pump frequencies with respect to the frequencies of the resonator modes. The numbers indicate the macronode index $\mathcal{M}$. Note that the two resonators have identical spectral profiles. $(c)$ The graph generated by the dissipative interactions. To generate non-trivial graphs with edges between modes belonging to different resonators, a beam splitter enacting a Hadamard transformation can be used, as in~\cref{fig:cluster_D1}.}
\end{figure}

We first consider the simplest situation where each resonator is coupled to the output field of a single squeezing source, as illustrated in~\cref{fig:appcluster:resonators1} for $D=1$.
The output field of each squeezing source has correlation functions of the form given in~\cref{eq:appinout:corr}
\begin{subequations}\label{eq:appcluster:corr1}
\begin{align}
  &\begin{aligned}
    \braket{\hb^\dagger_{j\omega} \hb_{i\omega'}} 
    ={}& \delta_{ji}N_j(\omega)\delta(\omega-\omega'),
  \end{aligned}\\
  &\begin{aligned}
    \braket{\hb_{j\omega} \hb^\dagger_{i\omega'}} 
    ={}& \delta_{ji}[N_j(\omega)+1]\delta(\omega-\omega'),
  \end{aligned}\\
  &\begin{aligned}
    &\braket{\hb_{j\omega} \hb_{i\omega'}} 
    ={} i\delta_{ji}M_j(\omega)\delta(2\Omega_j-\omega-\omega'),
  \end{aligned}
\end{align}
\end{subequations}
where we assume that $M_j(\omega)>0$ is real. With a JTPWA as a squeezing source, as long as the phase mismatch is small $\Delta k(\omega) \simeq 0$, the squeezing angle is frequency independent and $M(\omega)$ can thus be assumed real without loss of generality since any complex phase can be absorbed in a redefinition of the system operators $\hc_v$.
Using this in~\cref{eq:appbath:ME2} and doing an RWA leaves us with the Lindbladian
\begin{equation}\label{eq:appcluster:LM1}
  \begin{aligned}
    \mathcal{L}_M ={}&
    \sum_{v} \kappa_v
    \Big\{
      (N_{v}+1) \mathcal{D}[c_v]
      +N_{v} \mathcal{D}[c_v^\dagger] \\
      &- \sum_{vw} \sqrt{\kappa_v\kappa_w} M_{vw}\mathcal{S}_{i}[c_{v}^\dagger,c_{w}^\dagger]
    \Big\},
  \end{aligned}
\end{equation}
where we have labeled the resonator modes by a vertex index $v$, $N_{v} = N_j(\omega_v)$, with $\omega_v$ the frequency of mode $v$ of the $j$th resonator, and the matrix $M_{vw}$ is non-zero with value $M_{vw}=M_j(\omega_v)$ only if $\omega_v+\omega_w=2\Omega_j$ and $v$ and $w$ both are associated to the $j$th resonator. This is of the form~\cref{eq:appcluster:L_squeezing} if we assume equal decay rates, $\kappa_v\equiv\kappa$ and  $N_v \equiv N$ and we write $M_{vw} = |M| a_{vw}$ with $|M| = \sqrt{N(N+1)}$ and $-1 \le a_{vw} \le 1$.

The graph generated by~\cref{eq:appcluster:LM1}, defined by the adjacency matrix $A \sim M= [M_{vw}]$, is however not useful for measurement based quantum computing: it consists of a set of disjoint edges, as illustrated in~\cref{fig:appcluster:resonators1}~$(c)$. Note that for the adjacency matrix of this graph to be self-inverse, $A=A^{-1}$, requires edges with weights $a_{vw} = 0,\pm 1$, meaning a flat, quantum limited squeezing spectrum such that $|M_j(\omega)| = |M_i(\omega')| = |M| = \sqrt{N(N+1)}$ within the relevant bandwidth. We assume in the following that $M(\omega) = M >0$, such that the graph consists of a collection of identical two-mode cluster states with egdge weights $+1$.

Note that~\cref{eq:appcluster:LM1} can be brought into Lindblad form,
\begin{equation}
  \mathcal{L}_M = \sum_v \mathcal{D}[L_v],
\end{equation}
withLindblad operators
\begin{equation}
  L_v = \cosh(r) \ha_v + i\sinh(r) \sum_{w\in \mathcal{N}(v)} a_{vw} \ha_w^\dagger,
\end{equation}
where we have used that $A=A^{-1}$ and that we can write $N = \sinh(r)^2$, $M=\sinh(r)\cosh(r)$.

There are, of course, no edges between vertices associated to modes embodied in different resonators at this stage, as they interact with independent, uncorrelated baths. The key to generate non-trivial cluster states is to combine the output fields of the squeezing sources on beam splitters before being incident on the resonators, as illustrated for the $D=1$ case in~\cref{fig:cluster_D1} in~\cref{sect:cluster}.

A general two-port beam splitter transformation for a pair of fields,
\begin{equation}\label{eq:appcluster:Hadamard}
  R_{ij}(\omega) =
  \left(\begin{array}{cc}
      t_i(\omega) & r_i(\omega) \\
      r_j(\omega) & t_j(\omega)
  \end{array}\right),
\end{equation}
gives $\hb_{i\omega} \to t_i(\omega) \hb_{i\omega} + r_i(\omega) \hb_{j\omega}$, where $t_i(\omega)$ and $r_i(\omega)$ are the transmittance and reflectance coefficients. The correlation functions in~\cref{eq:appcluster:corr1} thus transform to
\begin{subequations}
\begin{align}
  &\begin{aligned}
    \braket{\hb^\dagger_{j\omega} \hb_{i\omega'}} 
    ={}& \delta_{ij} N_j \delta(\omega-\omega'),
  \end{aligned}\\
  &\begin{aligned}
    \braket{\hb_{j\omega} \hb^\dagger_{i\omega'}} 
    ={}& \delta_{ij} (N_j+1)\delta(\omega-\omega'),
  \end{aligned}\\
  &\begin{aligned}
    \braket{\hb_{j\omega} \hb_{i\omega'}} 
    ={}& it_jr_i M_j\delta(2\Omega_j-\omega-\omega') \\
    &+ ir_jt_i M_i\delta(2\Omega_i-\omega-\omega')].
  \end{aligned}
\end{align}
\end{subequations}
where we have suppressed the $\omega$ argument on the functions on the right hand side for notational simplicity,
and used that unitarity of the beam splitter requires $t_j^*r_i + r_j^*t_i =0$. We see that the form of the Lindbladian~\cref{eq:appcluster:L_squeezing} is preserved, with a transformed squeezing matrix, $M=[M_{vw}] \to R M R^T$, where $R = \bigoplus R_{ij}$ represents the beam splitter matrix $R_{ij}$ acting on all pairs of degenerate modes belonging to resonators $i$ and $j$.

Following Ref.~\cite{Wang14} we take the beam splitter to be a Hadamard transformation
\begin{equation}\label{eq:appcluster:Hadamard}
  H = \frac{1}{\sqrt{2}}
  \left(\begin{array}{cc}
      1 & 1 \\
      1 & -1
  \end{array}\right),
\end{equation}
such that the squeezing matrix
\begin{equation}
  M \to R M R^T,
\end{equation}
where $R=\bigoplus H$ represents a Hadamard matrix acting on pairs of degenerate modes. Since $M\sim A$, the graph of course transforms in the same way. For the $D=1$ case this gives the graph illustrated in~\cref{fig:cluster_D1}. Note that the transformed adjacency matrix $R A R^T$ is still self-inverse since $R$ is orthogonal~\cite{Wang14}.

This can be generalized in a straight forward manner to combine multiple squeezing sources. For $2\times D$ sources, we can construct a $2D \times 2D$ Hadamard matrix by composing together $2\times 2$ Hadamard matrices, $H_D = H^{\otimes D}$~\cite{Wang14}. In practice this is done by combining the output fields pairwise on two-port beam splitters in a cascaded fashion, as illustrated for the case $D=2$ in~\cref{fig:cluster_D2} and explained in more detail in Ref.~\cite{Wang14}. 
The total transformation on the squeezing matrix is $R = \bigoplus_{\mathcal{M}} H_D$, representing a $2D \times 2D$ Hadamard transformation on each \emph{macronode} $\mathcal{M}$, \emph{i.e.}, each set of $2\times D$ degenerate modes.

As shown in Ref.~\cite{Wang14}, graphs defined by the adjacency matrix $A \sim M$ corresponding to $D$-dimensional cluster states can be generated in this way, where in particular the $D=2$ case is universal for measurement based quantum computing.

\bibliography{refs}

\begin{thebibliography}{95}
\expandafter\ifx\csname natexlab\endcsname\relax\def\natexlab#1{#1}\fi
\expandafter\ifx\csname bibnamefont\endcsname\relax
  \def\bibnamefont#1{#1}\fi
\expandafter\ifx\csname bibfnamefont\endcsname\relax
  \def\bibfnamefont#1{#1}\fi
\expandafter\ifx\csname citenamefont\endcsname\relax
  \def\citenamefont#1{#1}\fi
\expandafter\ifx\csname url\endcsname\relax
  \def\url#1{\texttt{#1}}\fi
\expandafter\ifx\csname urlprefix\endcsname\relax\def\urlprefix{URL }\fi
\providecommand{\bibinfo}[2]{#2}
\providecommand{\eprint}[2][]{\url{#2}}

\bibitem[{\citenamefont{Wallraff et~al.}(2004)\citenamefont{Wallraff, Schuster,
  Blais, Frunzio, Huang, Majer, Kumar, Girvin, and Schoelkopf}}]{Wallraff09}
\bibinfo{author}{\bibfnamefont{A.}~\bibnamefont{Wallraff}},
  \bibinfo{author}{\bibfnamefont{D.~I.} \bibnamefont{Schuster}},
  \bibinfo{author}{\bibfnamefont{A.}~\bibnamefont{Blais}},
  \bibinfo{author}{\bibfnamefont{L.}~\bibnamefont{Frunzio}},
  \bibinfo{author}{\bibfnamefont{R.-S.} \bibnamefont{Huang}},
  \bibinfo{author}{\bibfnamefont{J.}~\bibnamefont{Majer}},
  \bibinfo{author}{\bibfnamefont{S.}~\bibnamefont{Kumar}},
  \bibinfo{author}{\bibfnamefont{S.~M.} \bibnamefont{Girvin}},
  \bibnamefont{and} \bibinfo{author}{\bibfnamefont{R.~J.}
  \bibnamefont{Schoelkopf}}, \bibinfo{journal}{Nature}
  \textbf{\bibinfo{volume}{431}}, \bibinfo{pages}{162} (\bibinfo{year}{2004}).

\bibitem[{\citenamefont{Van~Loo et~al.}(2013)\citenamefont{Van~Loo, Fedorov,
  Lalumi{\`e}re, Sanders, Blais, and Wallraff}}]{Van13}
\bibinfo{author}{\bibfnamefont{A.~F.} \bibnamefont{Van~Loo}},
  \bibinfo{author}{\bibfnamefont{A.}~\bibnamefont{Fedorov}},
  \bibinfo{author}{\bibfnamefont{K.}~\bibnamefont{Lalumi{\`e}re}},
  \bibinfo{author}{\bibfnamefont{B.~C.} \bibnamefont{Sanders}},
  \bibinfo{author}{\bibfnamefont{A.}~\bibnamefont{Blais}}, \bibnamefont{and}
  \bibinfo{author}{\bibfnamefont{A.}~\bibnamefont{Wallraff}},
  \bibinfo{journal}{Science} \textbf{\bibinfo{volume}{342}},
  \bibinfo{pages}{1494} (\bibinfo{year}{2013}).

\bibitem[{\citenamefont{Haroche and Raimond}(2006)}]{Haroche06}
\bibinfo{author}{\bibfnamefont{S.}~\bibnamefont{Haroche}} \bibnamefont{and}
  \bibinfo{author}{\bibfnamefont{J.-M.} \bibnamefont{Raimond}},
  \emph{\bibinfo{title}{Exploring the quantum: atoms, cavities, and photons}}
  (\bibinfo{publisher}{Oxford university press}, \bibinfo{year}{2006}).

\bibitem[{\citenamefont{Vetsch et~al.}(2010)\citenamefont{Vetsch, Reitz,
  Sagu{\'e}, Schmidt, Dawkins, and Rauschenbeutel}}]{Vetsch10}
\bibinfo{author}{\bibfnamefont{E.}~\bibnamefont{Vetsch}},
  \bibinfo{author}{\bibfnamefont{D.}~\bibnamefont{Reitz}},
  \bibinfo{author}{\bibfnamefont{G.}~\bibnamefont{Sagu{\'e}}},
  \bibinfo{author}{\bibfnamefont{R.}~\bibnamefont{Schmidt}},
  \bibinfo{author}{\bibfnamefont{S.}~\bibnamefont{Dawkins}}, \bibnamefont{and}
  \bibinfo{author}{\bibfnamefont{A.}~\bibnamefont{Rauschenbeutel}},
  \bibinfo{journal}{Physical review letters} \textbf{\bibinfo{volume}{104}},
  \bibinfo{pages}{203603} (\bibinfo{year}{2010}).

\bibitem[{\citenamefont{Goban et~al.}(2014)\citenamefont{Goban, Hung, Yu, Hood,
  Muniz, Lee, Martin, McClung, Choi, Chang et~al.}}]{Goban14}
\bibinfo{author}{\bibfnamefont{A.}~\bibnamefont{Goban}},
  \bibinfo{author}{\bibfnamefont{C.-L.} \bibnamefont{Hung}},
  \bibinfo{author}{\bibfnamefont{S.-P.} \bibnamefont{Yu}},
  \bibinfo{author}{\bibfnamefont{J.}~\bibnamefont{Hood}},
  \bibinfo{author}{\bibfnamefont{J.}~\bibnamefont{Muniz}},
  \bibinfo{author}{\bibfnamefont{J.}~\bibnamefont{Lee}},
  \bibinfo{author}{\bibfnamefont{M.}~\bibnamefont{Martin}},
  \bibinfo{author}{\bibfnamefont{A.}~\bibnamefont{McClung}},
  \bibinfo{author}{\bibfnamefont{K.}~\bibnamefont{Choi}},
  \bibinfo{author}{\bibfnamefont{D.}~\bibnamefont{Chang}},
  \bibnamefont{et~al.}, \bibinfo{journal}{Nature communications}
  \textbf{\bibinfo{volume}{5}} (\bibinfo{year}{2014}).

\bibitem[{\citenamefont{Niemczyk et~al.}(2010)\citenamefont{Niemczyk, Deppe,
  Huebl, Menzel, Hocke, Schwarz, Garcia-Ripoll, Zueco, H{\"u}mmer, Solano
  et~al.}}]{Niemczyk10}
\bibinfo{author}{\bibfnamefont{T.}~\bibnamefont{Niemczyk}},
  \bibinfo{author}{\bibfnamefont{F.}~\bibnamefont{Deppe}},
  \bibinfo{author}{\bibfnamefont{H.}~\bibnamefont{Huebl}},
  \bibinfo{author}{\bibfnamefont{E.}~\bibnamefont{Menzel}},
  \bibinfo{author}{\bibfnamefont{F.}~\bibnamefont{Hocke}},
  \bibinfo{author}{\bibfnamefont{M.}~\bibnamefont{Schwarz}},
  \bibinfo{author}{\bibfnamefont{J.}~\bibnamefont{Garcia-Ripoll}},
  \bibinfo{author}{\bibfnamefont{D.}~\bibnamefont{Zueco}},
  \bibinfo{author}{\bibfnamefont{T.}~\bibnamefont{H{\"u}mmer}},
  \bibinfo{author}{\bibfnamefont{E.}~\bibnamefont{Solano}},
  \bibnamefont{et~al.}, \bibinfo{journal}{Nature Physics}
  \textbf{\bibinfo{volume}{6}}, \bibinfo{pages}{772} (\bibinfo{year}{2010}).

\bibitem[{\citenamefont{Wilson et~al.}(2011)\citenamefont{Wilson, Johansson,
  Pourkabirian, Simoen, Johansson, Duty, Nori, and Delsing}}]{Wilson11}
\bibinfo{author}{\bibfnamefont{C.}~\bibnamefont{Wilson}},
  \bibinfo{author}{\bibfnamefont{G.}~\bibnamefont{Johansson}},
  \bibinfo{author}{\bibfnamefont{A.}~\bibnamefont{Pourkabirian}},
  \bibinfo{author}{\bibfnamefont{M.}~\bibnamefont{Simoen}},
  \bibinfo{author}{\bibfnamefont{J.}~\bibnamefont{Johansson}},
  \bibinfo{author}{\bibfnamefont{T.}~\bibnamefont{Duty}},
  \bibinfo{author}{\bibfnamefont{F.}~\bibnamefont{Nori}}, \bibnamefont{and}
  \bibinfo{author}{\bibfnamefont{P.}~\bibnamefont{Delsing}},
  \bibinfo{journal}{Nature} \textbf{\bibinfo{volume}{479}},
  \bibinfo{pages}{376} (\bibinfo{year}{2011}).

\bibitem[{\citenamefont{O'Brien et~al.}(2014)\citenamefont{O'Brien, Macklin,
  Siddiqi, and Zhang}}]{OBrien14}
\bibinfo{author}{\bibfnamefont{K.}~\bibnamefont{O'Brien}},
  \bibinfo{author}{\bibfnamefont{C.}~\bibnamefont{Macklin}},
  \bibinfo{author}{\bibfnamefont{I.}~\bibnamefont{Siddiqi}}, \bibnamefont{and}
  \bibinfo{author}{\bibfnamefont{X.}~\bibnamefont{Zhang}},
  \bibinfo{journal}{Phys. Rev. Lett.} \textbf{\bibinfo{volume}{113}},
  \bibinfo{pages}{157001} (\bibinfo{year}{2014}).

\bibitem[{\citenamefont{Macklin et~al.}(2015)\citenamefont{Macklin, O’Brien,
  Hover, Schwartz, Bolkhovsky, Zhang, Oliver, and Siddiqi}}]{Macklin15}
\bibinfo{author}{\bibfnamefont{C.}~\bibnamefont{Macklin}},
  \bibinfo{author}{\bibfnamefont{K.}~\bibnamefont{O’Brien}},
  \bibinfo{author}{\bibfnamefont{D.}~\bibnamefont{Hover}},
  \bibinfo{author}{\bibfnamefont{M.}~\bibnamefont{Schwartz}},
  \bibinfo{author}{\bibfnamefont{V.}~\bibnamefont{Bolkhovsky}},
  \bibinfo{author}{\bibfnamefont{X.}~\bibnamefont{Zhang}},
  \bibinfo{author}{\bibfnamefont{W.}~\bibnamefont{Oliver}}, \bibnamefont{and}
  \bibinfo{author}{\bibfnamefont{I.}~\bibnamefont{Siddiqi}},
  \bibinfo{journal}{Science} \textbf{\bibinfo{volume}{350}},
  \bibinfo{pages}{307} (\bibinfo{year}{2015}).

\bibitem[{\citenamefont{White et~al.}(2015)\citenamefont{White, Mutus, Hoi,
  Barends, Campbell, Chen, Chen, Chiaro, Dunsworth, Jeffrey et~al.}}]{White15}
\bibinfo{author}{\bibfnamefont{T.}~\bibnamefont{White}},
  \bibinfo{author}{\bibfnamefont{J.}~\bibnamefont{Mutus}},
  \bibinfo{author}{\bibfnamefont{I.-C.} \bibnamefont{Hoi}},
  \bibinfo{author}{\bibfnamefont{R.}~\bibnamefont{Barends}},
  \bibinfo{author}{\bibfnamefont{B.}~\bibnamefont{Campbell}},
  \bibinfo{author}{\bibfnamefont{Y.}~\bibnamefont{Chen}},
  \bibinfo{author}{\bibfnamefont{Z.}~\bibnamefont{Chen}},
  \bibinfo{author}{\bibfnamefont{B.}~\bibnamefont{Chiaro}},
  \bibinfo{author}{\bibfnamefont{A.}~\bibnamefont{Dunsworth}},
  \bibinfo{author}{\bibfnamefont{E.}~\bibnamefont{Jeffrey}},
  \bibnamefont{et~al.}, \bibinfo{journal}{Applied Physics Letters}
  \textbf{\bibinfo{volume}{106}}, \bibinfo{pages}{242601}
  (\bibinfo{year}{2015}).

\bibitem[{\citenamefont{Hillery}(2004)}]{Hillery04}
\bibinfo{author}{\bibfnamefont{M.}~\bibnamefont{Hillery}},
  \emph{\bibinfo{title}{Quantum squeezing}} (\bibinfo{publisher}{Springer},
  \bibinfo{year}{2004}), chap.~\bibinfo{chapter}{2}.

\bibitem[{\citenamefont{Caves}(1982)}]{Caves82}
\bibinfo{author}{\bibfnamefont{C.~M.} \bibnamefont{Caves}},
  \bibinfo{journal}{Phys. Rev. D} \textbf{\bibinfo{volume}{26}},
  \bibinfo{pages}{1817} (\bibinfo{year}{1982}).

\bibitem[{\citenamefont{Jeffrey
  et~al.}(2014{\natexlab{a}})\citenamefont{Jeffrey, Sank, Mutus, White, Kelly,
  Barends, Chen, Chen, Chiaro, Dunsworth et~al.}}]{Jeffrey14}
\bibinfo{author}{\bibfnamefont{E.}~\bibnamefont{Jeffrey}},
  \bibinfo{author}{\bibfnamefont{D.}~\bibnamefont{Sank}},
  \bibinfo{author}{\bibfnamefont{J.}~\bibnamefont{Mutus}},
  \bibinfo{author}{\bibfnamefont{T.}~\bibnamefont{White}},
  \bibinfo{author}{\bibfnamefont{J.}~\bibnamefont{Kelly}},
  \bibinfo{author}{\bibfnamefont{R.}~\bibnamefont{Barends}},
  \bibinfo{author}{\bibfnamefont{Y.}~\bibnamefont{Chen}},
  \bibinfo{author}{\bibfnamefont{Z.}~\bibnamefont{Chen}},
  \bibinfo{author}{\bibfnamefont{B.}~\bibnamefont{Chiaro}},
  \bibinfo{author}{\bibfnamefont{A.}~\bibnamefont{Dunsworth}},
  \bibnamefont{et~al.}, \bibinfo{journal}{Phys. Rev. Lett.}
  \textbf{\bibinfo{volume}{112}}, \bibinfo{pages}{190504}
  (\bibinfo{year}{2014}{\natexlab{a}}).

\bibitem[{\citenamefont{Castellanos-Beltran and Lehnert}(2007)}]{Castellanos07}
\bibinfo{author}{\bibfnamefont{M.}~\bibnamefont{Castellanos-Beltran}}
  \bibnamefont{and} \bibinfo{author}{\bibfnamefont{K.}~\bibnamefont{Lehnert}},
  \bibinfo{journal}{Applied Physics Letters} p. \bibinfo{pages}{083509}
  (\bibinfo{year}{2007}).

\bibitem[{\citenamefont{Bergeal et~al.}(2010)\citenamefont{Bergeal, Schackert,
  Metcalfe, Vijay, Manucharyan, Frunzio, Prober, Schoelkopf, Girvin, and
  Devoret}}]{Bergeal10}
\bibinfo{author}{\bibfnamefont{N.}~\bibnamefont{Bergeal}},
  \bibinfo{author}{\bibfnamefont{F.}~\bibnamefont{Schackert}},
  \bibinfo{author}{\bibfnamefont{M.}~\bibnamefont{Metcalfe}},
  \bibinfo{author}{\bibfnamefont{R.}~\bibnamefont{Vijay}},
  \bibinfo{author}{\bibfnamefont{V.}~\bibnamefont{Manucharyan}},
  \bibinfo{author}{\bibfnamefont{L.}~\bibnamefont{Frunzio}},
  \bibinfo{author}{\bibfnamefont{D.}~\bibnamefont{Prober}},
  \bibinfo{author}{\bibfnamefont{R.}~\bibnamefont{Schoelkopf}},
  \bibinfo{author}{\bibfnamefont{S.}~\bibnamefont{Girvin}}, \bibnamefont{and}
  \bibinfo{author}{\bibfnamefont{M.}~\bibnamefont{Devoret}},
  \bibinfo{journal}{Nature} \textbf{\bibinfo{volume}{465}}, \bibinfo{pages}{64}
  (\bibinfo{year}{2010}).

\bibitem[{\citenamefont{Hatridge et~al.}(2011)\citenamefont{Hatridge, Vijay,
  Slichter, Clarke, and Siddiqi}}]{Hatridge11}
\bibinfo{author}{\bibfnamefont{M.}~\bibnamefont{Hatridge}},
  \bibinfo{author}{\bibfnamefont{R.}~\bibnamefont{Vijay}},
  \bibinfo{author}{\bibfnamefont{D.}~\bibnamefont{Slichter}},
  \bibinfo{author}{\bibfnamefont{J.}~\bibnamefont{Clarke}}, \bibnamefont{and}
  \bibinfo{author}{\bibfnamefont{I.}~\bibnamefont{Siddiqi}},
  \bibinfo{journal}{Physical Review B} \textbf{\bibinfo{volume}{83}},
  \bibinfo{pages}{134501} (\bibinfo{year}{2011}).

\bibitem[{\citenamefont{Caves et~al.}(2012)\citenamefont{Caves, Combes, Jiang,
  and Pandey}}]{Caves12}
\bibinfo{author}{\bibfnamefont{C.~M.} \bibnamefont{Caves}},
  \bibinfo{author}{\bibfnamefont{J.}~\bibnamefont{Combes}},
  \bibinfo{author}{\bibfnamefont{Z.}~\bibnamefont{Jiang}}, \bibnamefont{and}
  \bibinfo{author}{\bibfnamefont{S.}~\bibnamefont{Pandey}},
  \bibinfo{journal}{Phys. Rev. A} \textbf{\bibinfo{volume}{86}},
  \bibinfo{pages}{063802} (\bibinfo{year}{2012}).

\bibitem[{\citenamefont{Ong et~al.}(2013)\citenamefont{Ong, Boissonneault,
  Mallet, Doherty, Blais, Vion, Esteve, and Bertet}}]{Ong13}
\bibinfo{author}{\bibfnamefont{F.~R.} \bibnamefont{Ong}},
  \bibinfo{author}{\bibfnamefont{M.}~\bibnamefont{Boissonneault}},
  \bibinfo{author}{\bibfnamefont{F.}~\bibnamefont{Mallet}},
  \bibinfo{author}{\bibfnamefont{A.~C.} \bibnamefont{Doherty}},
  \bibinfo{author}{\bibfnamefont{A.}~\bibnamefont{Blais}},
  \bibinfo{author}{\bibfnamefont{D.}~\bibnamefont{Vion}},
  \bibinfo{author}{\bibfnamefont{D.}~\bibnamefont{Esteve}}, \bibnamefont{and}
  \bibinfo{author}{\bibfnamefont{P.}~\bibnamefont{Bertet}},
  \bibinfo{journal}{Phys. Rev. Lett.} \textbf{\bibinfo{volume}{110}},
  \bibinfo{pages}{047001} (\bibinfo{year}{2013}).

\bibitem[{\citenamefont{Raussendorf and Briegel}(2001)}]{Raussendorf01}
\bibinfo{author}{\bibfnamefont{R.}~\bibnamefont{Raussendorf}} \bibnamefont{and}
  \bibinfo{author}{\bibfnamefont{H.~J.} \bibnamefont{Briegel}},
  \bibinfo{journal}{Physical Review Letters} \textbf{\bibinfo{volume}{86}},
  \bibinfo{pages}{5188} (\bibinfo{year}{2001}).

\bibitem[{\citenamefont{Menicucci et~al.}(2006)\citenamefont{Menicucci, van
  Loock, Gu, Weedbrook, Ralph, and Nielsen}}]{Menicucci06}
\bibinfo{author}{\bibfnamefont{N.~C.} \bibnamefont{Menicucci}},
  \bibinfo{author}{\bibfnamefont{P.}~\bibnamefont{van Loock}},
  \bibinfo{author}{\bibfnamefont{M.}~\bibnamefont{Gu}},
  \bibinfo{author}{\bibfnamefont{C.}~\bibnamefont{Weedbrook}},
  \bibinfo{author}{\bibfnamefont{T.~C.} \bibnamefont{Ralph}}, \bibnamefont{and}
  \bibinfo{author}{\bibfnamefont{M.~A.} \bibnamefont{Nielsen}},
  \bibinfo{journal}{Phys. Rev. Lett.} \textbf{\bibinfo{volume}{97}},
  \bibinfo{pages}{110501} (\bibinfo{year}{2006}).

\bibitem[{\citenamefont{Aharonov et~al.}(2008)\citenamefont{Aharonov, Van~Dam,
  Kempe, Landau, Lloyd, and Regev}}]{Aharonov08}
\bibinfo{author}{\bibfnamefont{D.}~\bibnamefont{Aharonov}},
  \bibinfo{author}{\bibfnamefont{W.}~\bibnamefont{Van~Dam}},
  \bibinfo{author}{\bibfnamefont{J.}~\bibnamefont{Kempe}},
  \bibinfo{author}{\bibfnamefont{Z.}~\bibnamefont{Landau}},
  \bibinfo{author}{\bibfnamefont{S.}~\bibnamefont{Lloyd}}, \bibnamefont{and}
  \bibinfo{author}{\bibfnamefont{O.}~\bibnamefont{Regev}},
  \bibinfo{journal}{SIAM review} \textbf{\bibinfo{volume}{50}},
  \bibinfo{pages}{755} (\bibinfo{year}{2008}).

\bibitem[{\citenamefont{Nielsen and Chuang}(2010)}]{Nielsen10}
\bibinfo{author}{\bibfnamefont{M.~A.} \bibnamefont{Nielsen}} \bibnamefont{and}
  \bibinfo{author}{\bibfnamefont{I.~L.} \bibnamefont{Chuang}},
  \emph{\bibinfo{title}{Quantum computation and quantum information}}
  (\bibinfo{publisher}{Cambridge university press}, \bibinfo{year}{2010}).

\bibitem[{\citenamefont{Verstraete et~al.}(2009)\citenamefont{Verstraete, Wolf,
  and Cirac}}]{Verstraete09}
\bibinfo{author}{\bibfnamefont{F.}~\bibnamefont{Verstraete}},
  \bibinfo{author}{\bibfnamefont{M.~M.} \bibnamefont{Wolf}}, \bibnamefont{and}
  \bibinfo{author}{\bibfnamefont{J.~I.} \bibnamefont{Cirac}},
  \bibinfo{journal}{Nature physics} \textbf{\bibinfo{volume}{5}},
  \bibinfo{pages}{633} (\bibinfo{year}{2009}).

\bibitem[{\citenamefont{Kraus et~al.}(2008)\citenamefont{Kraus, B\"uchler,
  Diehl, Kantian, Micheli, and Zoller}}]{Kraus08}
\bibinfo{author}{\bibfnamefont{B.}~\bibnamefont{Kraus}},
  \bibinfo{author}{\bibfnamefont{H.~P.} \bibnamefont{B\"uchler}},
  \bibinfo{author}{\bibfnamefont{S.}~\bibnamefont{Diehl}},
  \bibinfo{author}{\bibfnamefont{A.}~\bibnamefont{Kantian}},
  \bibinfo{author}{\bibfnamefont{A.}~\bibnamefont{Micheli}}, \bibnamefont{and}
  \bibinfo{author}{\bibfnamefont{P.}~\bibnamefont{Zoller}},
  \bibinfo{journal}{Phys. Rev. A} \textbf{\bibinfo{volume}{78}},
  \bibinfo{pages}{042307} (\bibinfo{year}{2008}).

\bibitem[{\citenamefont{Diehl et~al.}(2008)\citenamefont{Diehl, Micheli,
  Kantian, Kraus, B{\"u}chler, and Zoller}}]{Diehl08}
\bibinfo{author}{\bibfnamefont{S.}~\bibnamefont{Diehl}},
  \bibinfo{author}{\bibfnamefont{A.}~\bibnamefont{Micheli}},
  \bibinfo{author}{\bibfnamefont{A.}~\bibnamefont{Kantian}},
  \bibinfo{author}{\bibfnamefont{B.}~\bibnamefont{Kraus}},
  \bibinfo{author}{\bibfnamefont{H.}~\bibnamefont{B{\"u}chler}},
  \bibnamefont{and} \bibinfo{author}{\bibfnamefont{P.}~\bibnamefont{Zoller}},
  \bibinfo{journal}{Nature Physics} \textbf{\bibinfo{volume}{4}},
  \bibinfo{pages}{878} (\bibinfo{year}{2008}).

\bibitem[{\citenamefont{Krauter et~al.}(2011)\citenamefont{Krauter, Muschik,
  Jensen, Wasilewski, Petersen, Cirac, and Polzik}}]{Krauter11}
\bibinfo{author}{\bibfnamefont{H.}~\bibnamefont{Krauter}},
  \bibinfo{author}{\bibfnamefont{C.~A.} \bibnamefont{Muschik}},
  \bibinfo{author}{\bibfnamefont{K.}~\bibnamefont{Jensen}},
  \bibinfo{author}{\bibfnamefont{W.}~\bibnamefont{Wasilewski}},
  \bibinfo{author}{\bibfnamefont{J.~M.} \bibnamefont{Petersen}},
  \bibinfo{author}{\bibfnamefont{J.~I.} \bibnamefont{Cirac}}, \bibnamefont{and}
  \bibinfo{author}{\bibfnamefont{E.~S.} \bibnamefont{Polzik}},
  \bibinfo{journal}{Physical review letters} \textbf{\bibinfo{volume}{107}},
  \bibinfo{pages}{080503} (\bibinfo{year}{2011}).

\bibitem[{\citenamefont{Lin et~al.}(2013)\citenamefont{Lin, Gaebler, Reiter,
  Tan, Bowler, S{\o}rensen, Leibfried, and Wineland}}]{Lin13}
\bibinfo{author}{\bibfnamefont{Y.}~\bibnamefont{Lin}},
  \bibinfo{author}{\bibfnamefont{J.}~\bibnamefont{Gaebler}},
  \bibinfo{author}{\bibfnamefont{F.}~\bibnamefont{Reiter}},
  \bibinfo{author}{\bibfnamefont{T.}~\bibnamefont{Tan}},
  \bibinfo{author}{\bibfnamefont{R.}~\bibnamefont{Bowler}},
  \bibinfo{author}{\bibfnamefont{A.}~\bibnamefont{S{\o}rensen}},
  \bibinfo{author}{\bibfnamefont{D.}~\bibnamefont{Leibfried}},
  \bibnamefont{and} \bibinfo{author}{\bibfnamefont{D.}~\bibnamefont{Wineland}},
  \bibinfo{journal}{Nature} \textbf{\bibinfo{volume}{504}},
  \bibinfo{pages}{415} (\bibinfo{year}{2013}).

\bibitem[{\citenamefont{Shankar et~al.}(2013)\citenamefont{Shankar, Hatridge,
  Leghtas, Sliwa, Narla, Vool, Girvin, Frunzio, Mirrahimi, and
  Devoret}}]{Shankar13}
\bibinfo{author}{\bibfnamefont{S.}~\bibnamefont{Shankar}},
  \bibinfo{author}{\bibfnamefont{M.}~\bibnamefont{Hatridge}},
  \bibinfo{author}{\bibfnamefont{Z.}~\bibnamefont{Leghtas}},
  \bibinfo{author}{\bibfnamefont{K.}~\bibnamefont{Sliwa}},
  \bibinfo{author}{\bibfnamefont{A.}~\bibnamefont{Narla}},
  \bibinfo{author}{\bibfnamefont{U.}~\bibnamefont{Vool}},
  \bibinfo{author}{\bibfnamefont{S.~M.} \bibnamefont{Girvin}},
  \bibinfo{author}{\bibfnamefont{L.}~\bibnamefont{Frunzio}},
  \bibinfo{author}{\bibfnamefont{M.}~\bibnamefont{Mirrahimi}},
  \bibnamefont{and} \bibinfo{author}{\bibfnamefont{M.~H.}
  \bibnamefont{Devoret}}, \bibinfo{journal}{Nature}
  \textbf{\bibinfo{volume}{504}}, \bibinfo{pages}{419} (\bibinfo{year}{2013}).

\bibitem[{\citenamefont{Stannigel et~al.}(2012)\citenamefont{Stannigel, Rabl,
  and Zoller}}]{Stannigel12}
\bibinfo{author}{\bibfnamefont{K.}~\bibnamefont{Stannigel}},
  \bibinfo{author}{\bibfnamefont{P.}~\bibnamefont{Rabl}}, \bibnamefont{and}
  \bibinfo{author}{\bibfnamefont{P.}~\bibnamefont{Zoller}},
  \bibinfo{journal}{New Journal of Physics} \textbf{\bibinfo{volume}{14}},
  \bibinfo{pages}{063014} (\bibinfo{year}{2012}).

\bibitem[{\citenamefont{Bardyn et~al.}(2012)\citenamefont{Bardyn, Baranov,
  Rico, {\.I}mamo{\u{g}}lu, Zoller, and Diehl}}]{Bardyn12}
\bibinfo{author}{\bibfnamefont{C.-E.} \bibnamefont{Bardyn}},
  \bibinfo{author}{\bibfnamefont{M.}~\bibnamefont{Baranov}},
  \bibinfo{author}{\bibfnamefont{E.}~\bibnamefont{Rico}},
  \bibinfo{author}{\bibfnamefont{A.}~\bibnamefont{{\.I}mamo{\u{g}}lu}},
  \bibinfo{author}{\bibfnamefont{P.}~\bibnamefont{Zoller}}, \bibnamefont{and}
  \bibinfo{author}{\bibfnamefont{S.}~\bibnamefont{Diehl}},
  \bibinfo{journal}{Physical review letters} \textbf{\bibinfo{volume}{109}},
  \bibinfo{pages}{130402} (\bibinfo{year}{2012}).

\bibitem[{\citenamefont{Menicucci et~al.}(2008)\citenamefont{Menicucci,
  Flammia, and Pfister}}]{Menicucci08}
\bibinfo{author}{\bibfnamefont{N.~C.} \bibnamefont{Menicucci}},
  \bibinfo{author}{\bibfnamefont{S.~T.} \bibnamefont{Flammia}},
  \bibnamefont{and} \bibinfo{author}{\bibfnamefont{O.}~\bibnamefont{Pfister}},
  \bibinfo{journal}{Phys. Rev. Lett.} \textbf{\bibinfo{volume}{101}},
  \bibinfo{pages}{130501} (\bibinfo{year}{2008}).

\bibitem[{\citenamefont{Flammia et~al.}(2009)\citenamefont{Flammia, Menicucci,
  and Pfister}}]{Flammia09}
\bibinfo{author}{\bibfnamefont{S.~T.} \bibnamefont{Flammia}},
  \bibinfo{author}{\bibfnamefont{N.~C.} \bibnamefont{Menicucci}},
  \bibnamefont{and} \bibinfo{author}{\bibfnamefont{O.}~\bibnamefont{Pfister}},
  \bibinfo{journal}{Journal of Physics B: Atomic, Molecular and Optical
  Physics} \textbf{\bibinfo{volume}{42}}, \bibinfo{pages}{114009}
  (\bibinfo{year}{2009}).

\bibitem[{\citenamefont{Menicucci}(2011)}]{Menicucci11b}
\bibinfo{author}{\bibfnamefont{N.~C.} \bibnamefont{Menicucci}},
  \bibinfo{journal}{Phys. Rev. A} \textbf{\bibinfo{volume}{83}},
  \bibinfo{pages}{062314} (\bibinfo{year}{2011}).

\bibitem[{\citenamefont{Wang et~al.}(2014)\citenamefont{Wang, Chen, Menicucci,
  and Pfister}}]{Wang14}
\bibinfo{author}{\bibfnamefont{P.}~\bibnamefont{Wang}},
  \bibinfo{author}{\bibfnamefont{M.}~\bibnamefont{Chen}},
  \bibinfo{author}{\bibfnamefont{N.~C.} \bibnamefont{Menicucci}},
  \bibnamefont{and} \bibinfo{author}{\bibfnamefont{O.}~\bibnamefont{Pfister}},
  \bibinfo{journal}{Phys. Rev. A} \textbf{\bibinfo{volume}{90}},
  \bibinfo{pages}{032325} (\bibinfo{year}{2014}).

\bibitem[{\citenamefont{Chen et~al.}(2014)\citenamefont{Chen, Menicucci, and
  Pfister}}]{Chen14}
\bibinfo{author}{\bibfnamefont{M.}~\bibnamefont{Chen}},
  \bibinfo{author}{\bibfnamefont{N.~C.} \bibnamefont{Menicucci}},
  \bibnamefont{and} \bibinfo{author}{\bibfnamefont{O.}~\bibnamefont{Pfister}},
  \bibinfo{journal}{Phys. Rev. Lett.} \textbf{\bibinfo{volume}{112}},
  \bibinfo{pages}{120505} (\bibinfo{year}{2014}).

\bibitem[{\citenamefont{Alexander et~al.}(2015)\citenamefont{Alexander, Wang,
  Sridhar, Chen, Pfister, and Menicucci}}]{Alexander15}
\bibinfo{author}{\bibfnamefont{R.~N.} \bibnamefont{Alexander}},
  \bibinfo{author}{\bibfnamefont{P.}~\bibnamefont{Wang}},
  \bibinfo{author}{\bibfnamefont{N.}~\bibnamefont{Sridhar}},
  \bibinfo{author}{\bibfnamefont{M.}~\bibnamefont{Chen}},
  \bibinfo{author}{\bibfnamefont{O.}~\bibnamefont{Pfister}}, \bibnamefont{and}
  \bibinfo{author}{\bibfnamefont{N.~C.} \bibnamefont{Menicucci}},
  \bibinfo{journal}{arXiv:1509.00484}  (\bibinfo{year}{2015}).

\bibitem[{\citenamefont{Brecht et~al.}(2016)\citenamefont{Brecht, Pfaff, Wang,
  Chu, Frunzio, Devoret, and Schoelkopf}}]{Brecht:16a}
\bibinfo{author}{\bibfnamefont{T.}~\bibnamefont{Brecht}},
  \bibinfo{author}{\bibfnamefont{W.}~\bibnamefont{Pfaff}},
  \bibinfo{author}{\bibfnamefont{C.}~\bibnamefont{Wang}},
  \bibinfo{author}{\bibfnamefont{Y.}~\bibnamefont{Chu}},
  \bibinfo{author}{\bibfnamefont{L.}~\bibnamefont{Frunzio}},
  \bibinfo{author}{\bibfnamefont{M.~H.} \bibnamefont{Devoret}},
  \bibnamefont{and} \bibinfo{author}{\bibfnamefont{R.~J.}
  \bibnamefont{Schoelkopf}}, \bibinfo{journal}{Npj Quantum Information}
  \textbf{\bibinfo{volume}{2}}, \bibinfo{pages}{16002 EP }
  (\bibinfo{year}{2016}).

\bibitem[{\citenamefont{Eom et~al.}(2012)\citenamefont{Eom, Day, LeDuc, and
  Zmuidzinas}}]{Eom12}
\bibinfo{author}{\bibfnamefont{B.~H.} \bibnamefont{Eom}},
  \bibinfo{author}{\bibfnamefont{P.~K.} \bibnamefont{Day}},
  \bibinfo{author}{\bibfnamefont{H.~G.} \bibnamefont{LeDuc}}, \bibnamefont{and}
  \bibinfo{author}{\bibfnamefont{J.}~\bibnamefont{Zmuidzinas}},
  \bibinfo{journal}{Nature Physics} \textbf{\bibinfo{volume}{8}},
  \bibinfo{pages}{623} (\bibinfo{year}{2012}).

\bibitem[{\citenamefont{Bockstiegel et~al.}(2014)\citenamefont{Bockstiegel,
  Gao, Vissers, Sandberg, Chaudhuri, Sanders, Vale, Irwin, and
  Pappas}}]{Bockstiegel14}
\bibinfo{author}{\bibfnamefont{C.}~\bibnamefont{Bockstiegel}},
  \bibinfo{author}{\bibfnamefont{J.}~\bibnamefont{Gao}},
  \bibinfo{author}{\bibfnamefont{M.}~\bibnamefont{Vissers}},
  \bibinfo{author}{\bibfnamefont{M.}~\bibnamefont{Sandberg}},
  \bibinfo{author}{\bibfnamefont{S.}~\bibnamefont{Chaudhuri}},
  \bibinfo{author}{\bibfnamefont{A.}~\bibnamefont{Sanders}},
  \bibinfo{author}{\bibfnamefont{L.}~\bibnamefont{Vale}},
  \bibinfo{author}{\bibfnamefont{K.}~\bibnamefont{Irwin}}, \bibnamefont{and}
  \bibinfo{author}{\bibfnamefont{D.}~\bibnamefont{Pappas}},
  \bibinfo{journal}{Journal of Low Temperature Physics}
  \textbf{\bibinfo{volume}{176}}, \bibinfo{pages}{476} (\bibinfo{year}{2014}).

\bibitem[{\citenamefont{Roy et~al.}(2015)\citenamefont{Roy, Kundu, Chand,
  Vadiraj, Ranadive, Nehra, Patankar, Aumentado, Clerk, and Vijay}}]{Roy15}
\bibinfo{author}{\bibfnamefont{T.}~\bibnamefont{Roy}},
  \bibinfo{author}{\bibfnamefont{S.}~\bibnamefont{Kundu}},
  \bibinfo{author}{\bibfnamefont{M.}~\bibnamefont{Chand}},
  \bibinfo{author}{\bibfnamefont{A.}~\bibnamefont{Vadiraj}},
  \bibinfo{author}{\bibfnamefont{A.}~\bibnamefont{Ranadive}},
  \bibinfo{author}{\bibfnamefont{N.}~\bibnamefont{Nehra}},
  \bibinfo{author}{\bibfnamefont{M.~P.} \bibnamefont{Patankar}},
  \bibinfo{author}{\bibfnamefont{J.}~\bibnamefont{Aumentado}},
  \bibinfo{author}{\bibfnamefont{A.}~\bibnamefont{Clerk}}, \bibnamefont{and}
  \bibinfo{author}{\bibfnamefont{R.}~\bibnamefont{Vijay}},
  \bibinfo{journal}{Applied Physics Letters} \textbf{\bibinfo{volume}{107}},
  \bibinfo{pages}{262601} (\bibinfo{year}{2015}).

\bibitem[{\citenamefont{Metelmann and Clerk}(2014)}]{Metelmann14}
\bibinfo{author}{\bibfnamefont{A.}~\bibnamefont{Metelmann}} \bibnamefont{and}
  \bibinfo{author}{\bibfnamefont{A.~A.} \bibnamefont{Clerk}},
  \bibinfo{journal}{Phys. Rev. Lett.} \textbf{\bibinfo{volume}{112}},
  \bibinfo{pages}{133904} (\bibinfo{year}{2014}).

\bibitem[{\citenamefont{Forgues et~al.}(2015)\citenamefont{Forgues, Lupien, and
  Reulet}}]{Forgues15}
\bibinfo{author}{\bibfnamefont{J.-C.} \bibnamefont{Forgues}},
  \bibinfo{author}{\bibfnamefont{C.}~\bibnamefont{Lupien}}, \bibnamefont{and}
  \bibinfo{author}{\bibfnamefont{B.}~\bibnamefont{Reulet}},
  \bibinfo{journal}{Phys. Rev. Lett.} \textbf{\bibinfo{volume}{114}},
  \bibinfo{pages}{130403} (\bibinfo{year}{2015}).

\bibitem[{\citenamefont{Grimsmo et~al.}(2016)\citenamefont{Grimsmo, Qassemi,
  Reulet, and Blais}}]{Grimsmo16}
\bibinfo{author}{\bibfnamefont{A.~L.} \bibnamefont{Grimsmo}},
  \bibinfo{author}{\bibfnamefont{F.}~\bibnamefont{Qassemi}},
  \bibinfo{author}{\bibfnamefont{B.}~\bibnamefont{Reulet}}, \bibnamefont{and}
  \bibinfo{author}{\bibfnamefont{A.}~\bibnamefont{Blais}},
  \bibinfo{journal}{Phys. Rev. Lett.} \textbf{\bibinfo{volume}{116}},
  \bibinfo{pages}{043602} (\bibinfo{year}{2016}).

\bibitem[{\citenamefont{Yaakobi et~al.}(2013)\citenamefont{Yaakobi, Friedland,
  Macklin, and Siddiqi}}]{Yaakobi13}
\bibinfo{author}{\bibfnamefont{O.}~\bibnamefont{Yaakobi}},
  \bibinfo{author}{\bibfnamefont{L.}~\bibnamefont{Friedland}},
  \bibinfo{author}{\bibfnamefont{C.}~\bibnamefont{Macklin}}, \bibnamefont{and}
  \bibinfo{author}{\bibfnamefont{I.}~\bibnamefont{Siddiqi}},
  \bibinfo{journal}{Phys. Rev. B} \textbf{\bibinfo{volume}{87}},
  \bibinfo{pages}{144301} (\bibinfo{year}{2013}).

\bibitem[{\citenamefont{Quesada and Sipe}(2014)}]{Quesada14}
\bibinfo{author}{\bibfnamefont{N.}~\bibnamefont{Quesada}} \bibnamefont{and}
  \bibinfo{author}{\bibfnamefont{J.~E.} \bibnamefont{Sipe}},
  \bibinfo{journal}{Phys. Rev. A} \textbf{\bibinfo{volume}{90}},
  \bibinfo{pages}{063840} (\bibinfo{year}{2014}).

\bibitem[{\citenamefont{Quesada and Sipe}(2015)}]{Quesada15}
\bibinfo{author}{\bibfnamefont{N.}~\bibnamefont{Quesada}} \bibnamefont{and}
  \bibinfo{author}{\bibfnamefont{J.~E.} \bibnamefont{Sipe}},
  \bibinfo{journal}{Phys. Rev. Lett.} \textbf{\bibinfo{volume}{114}},
  \bibinfo{pages}{093903} (\bibinfo{year}{2015}).

\bibitem[{\citenamefont{Breit and Bethe}(1954)}]{Breit54}
\bibinfo{author}{\bibfnamefont{G.}~\bibnamefont{Breit}} \bibnamefont{and}
  \bibinfo{author}{\bibfnamefont{H.~A.} \bibnamefont{Bethe}},
  \bibinfo{journal}{Phys. Rev.} \textbf{\bibinfo{volume}{93}},
  \bibinfo{pages}{888} (\bibinfo{year}{1954}).

\bibitem[{\citenamefont{Liscidini et~al.}(2012)\citenamefont{Liscidini, Helt,
  and Sipe}}]{Liscidini12}
\bibinfo{author}{\bibfnamefont{M.}~\bibnamefont{Liscidini}},
  \bibinfo{author}{\bibfnamefont{L.~G.} \bibnamefont{Helt}}, \bibnamefont{and}
  \bibinfo{author}{\bibfnamefont{J.~E.} \bibnamefont{Sipe}},
  \bibinfo{journal}{Phys. Rev. A} \textbf{\bibinfo{volume}{85}},
  \bibinfo{pages}{013833} (\bibinfo{year}{2012}).

\bibitem[{\citenamefont{Drummond and Hillery}(2014)}]{Drummond14}
\bibinfo{author}{\bibfnamefont{P.~D.} \bibnamefont{Drummond}} \bibnamefont{and}
  \bibinfo{author}{\bibfnamefont{M.}~\bibnamefont{Hillery}},
  \emph{\bibinfo{title}{The quantum theory of nonlinear optics}}
  (\bibinfo{publisher}{Cambridge University Press}, \bibinfo{year}{2014}).

\bibitem[{\citenamefont{Caves and Crouch}(1987)}]{Caves87}
\bibinfo{author}{\bibfnamefont{C.~M.} \bibnamefont{Caves}} \bibnamefont{and}
  \bibinfo{author}{\bibfnamefont{D.~D.} \bibnamefont{Crouch}},
  \bibinfo{journal}{JOSA B} \textbf{\bibinfo{volume}{4}}, \bibinfo{pages}{1535}
  (\bibinfo{year}{1987}).

\bibitem[{\citenamefont{Hillery and Mlodinow}(1984)}]{Hillery84}
\bibinfo{author}{\bibfnamefont{M.}~\bibnamefont{Hillery}} \bibnamefont{and}
  \bibinfo{author}{\bibfnamefont{L.~D.} \bibnamefont{Mlodinow}},
  \bibinfo{journal}{Phys. Rev. A} \textbf{\bibinfo{volume}{30}},
  \bibinfo{pages}{1860} (\bibinfo{year}{1984}).

\bibitem[{\citenamefont{Bell and Samolov}(2015)}]{Bell15}
\bibinfo{author}{\bibfnamefont{M.~T.} \bibnamefont{Bell}} \bibnamefont{and}
  \bibinfo{author}{\bibfnamefont{A.}~\bibnamefont{Samolov}},
  \bibinfo{journal}{Phys. Rev. Applied} \textbf{\bibinfo{volume}{4}},
  \bibinfo{pages}{024014} (\bibinfo{year}{2015}).

\bibitem[{\citenamefont{Zorin}(2016)}]{Zorin16}
\bibinfo{author}{\bibfnamefont{A.}~\bibnamefont{Zorin}},
  \bibinfo{journal}{arXiv preprint arXiv:1602.02650}  (\bibinfo{year}{2016}).

\bibitem[{\citenamefont{Yurke and Denker}(1984)}]{Yurke84b}
\bibinfo{author}{\bibfnamefont{B.}~\bibnamefont{Yurke}} \bibnamefont{and}
  \bibinfo{author}{\bibfnamefont{J.~S.} \bibnamefont{Denker}},
  \bibinfo{journal}{Phys. Rev. A} \textbf{\bibinfo{volume}{29}},
  \bibinfo{pages}{1419} (\bibinfo{year}{1984}).

\bibitem[{\citenamefont{Yurke}(2004)}]{yurke:2004a}
\bibinfo{author}{\bibfnamefont{B.}~\bibnamefont{Yurke}},
  \emph{\bibinfo{title}{Quantum squeezing}} (\bibinfo{publisher}{Springer},
  \bibinfo{year}{2004}), chap.~\bibinfo{chapter}{3}.

\bibitem[{\citenamefont{Huttner et~al.}(1991)\citenamefont{Huttner, Baumberg,
  and Barnett}}]{Huttner91}
\bibinfo{author}{\bibfnamefont{B.}~\bibnamefont{Huttner}},
  \bibinfo{author}{\bibfnamefont{J.}~\bibnamefont{Baumberg}}, \bibnamefont{and}
  \bibinfo{author}{\bibfnamefont{S.}~\bibnamefont{Barnett}},
  \bibinfo{journal}{EPL (Europhysics Letters)} \textbf{\bibinfo{volume}{16}},
  \bibinfo{pages}{177} (\bibinfo{year}{1991}).

\bibitem[{\citenamefont{Huttner and Barnett}(1992)}]{Huttner92}
\bibinfo{author}{\bibfnamefont{B.}~\bibnamefont{Huttner}} \bibnamefont{and}
  \bibinfo{author}{\bibfnamefont{S.~M.} \bibnamefont{Barnett}},
  \bibinfo{journal}{Phys. Rev. A} \textbf{\bibinfo{volume}{46}},
  \bibinfo{pages}{4306} (\bibinfo{year}{1992}).

\bibitem[{\citenamefont{Gardiner and Collett}(1985)}]{Gardiner85}
\bibinfo{author}{\bibfnamefont{C.}~\bibnamefont{Gardiner}} \bibnamefont{and}
  \bibinfo{author}{\bibfnamefont{M.}~\bibnamefont{Collett}},
  \bibinfo{journal}{Physical Review A} \textbf{\bibinfo{volume}{31}},
  \bibinfo{pages}{3761} (\bibinfo{year}{1985}).

\bibitem[{\citenamefont{Wustmann and Shumeiko}(2013)}]{Wustmann13}
\bibinfo{author}{\bibfnamefont{W.}~\bibnamefont{Wustmann}} \bibnamefont{and}
  \bibinfo{author}{\bibfnamefont{V.}~\bibnamefont{Shumeiko}},
  \bibinfo{journal}{Phys. Rev. B} \textbf{\bibinfo{volume}{87}},
  \bibinfo{pages}{184501} (\bibinfo{year}{2013}).

\bibitem[{\citenamefont{Eichler et~al.}(2011)\citenamefont{Eichler, Bozyigit,
  Lang, Baur, Steffen, Fink, Filipp, and Wallraff}}]{Eichler11}
\bibinfo{author}{\bibfnamefont{C.}~\bibnamefont{Eichler}},
  \bibinfo{author}{\bibfnamefont{D.}~\bibnamefont{Bozyigit}},
  \bibinfo{author}{\bibfnamefont{C.}~\bibnamefont{Lang}},
  \bibinfo{author}{\bibfnamefont{M.}~\bibnamefont{Baur}},
  \bibinfo{author}{\bibfnamefont{L.}~\bibnamefont{Steffen}},
  \bibinfo{author}{\bibfnamefont{J.~M.} \bibnamefont{Fink}},
  \bibinfo{author}{\bibfnamefont{S.}~\bibnamefont{Filipp}}, \bibnamefont{and}
  \bibinfo{author}{\bibfnamefont{A.}~\bibnamefont{Wallraff}},
  \bibinfo{journal}{Phys. Rev. Lett.} \textbf{\bibinfo{volume}{107}},
  \bibinfo{pages}{113601} (\bibinfo{year}{2011}).

\bibitem[{\citenamefont{Flurin et~al.}(2012)\citenamefont{Flurin, Roch, Mallet,
  Devoret, and Huard}}]{Flurin12}
\bibinfo{author}{\bibfnamefont{E.}~\bibnamefont{Flurin}},
  \bibinfo{author}{\bibfnamefont{N.}~\bibnamefont{Roch}},
  \bibinfo{author}{\bibfnamefont{F.}~\bibnamefont{Mallet}},
  \bibinfo{author}{\bibfnamefont{M.~H.} \bibnamefont{Devoret}},
  \bibnamefont{and} \bibinfo{author}{\bibfnamefont{B.}~\bibnamefont{Huard}},
  \bibinfo{journal}{Phys. Rev. Lett.} \textbf{\bibinfo{volume}{109}},
  \bibinfo{pages}{183901} (\bibinfo{year}{2012}).

\bibitem[{\citenamefont{Eichler et~al.}(2014)\citenamefont{Eichler, Salathe,
  Mlynek, Schmidt, and Wallraff}}]{Eichler14}
\bibinfo{author}{\bibfnamefont{C.}~\bibnamefont{Eichler}},
  \bibinfo{author}{\bibfnamefont{Y.}~\bibnamefont{Salathe}},
  \bibinfo{author}{\bibfnamefont{J.}~\bibnamefont{Mlynek}},
  \bibinfo{author}{\bibfnamefont{S.}~\bibnamefont{Schmidt}}, \bibnamefont{and}
  \bibinfo{author}{\bibfnamefont{A.}~\bibnamefont{Wallraff}},
  \bibinfo{journal}{Phys. Rev. Lett.} \textbf{\bibinfo{volume}{113}},
  \bibinfo{pages}{110502} (\bibinfo{year}{2014}).

\bibitem[{\citenamefont{Palma and Knight}(1989)}]{Palma89}
\bibinfo{author}{\bibfnamefont{G.~M.} \bibnamefont{Palma}} \bibnamefont{and}
  \bibinfo{author}{\bibfnamefont{P.~L.} \bibnamefont{Knight}},
  \bibinfo{journal}{Phys. Rev. A} \textbf{\bibinfo{volume}{39}},
  \bibinfo{pages}{1962} (\bibinfo{year}{1989}).

\bibitem[{\citenamefont{G{\'o}mez et~al.}(2015)\citenamefont{G{\'o}mez,
  Rodr{\'\i}guez, Quiroga, and Garc{\'\i}a-Ripoll}}]{Gomez15}
\bibinfo{author}{\bibfnamefont{A.~V.} \bibnamefont{G{\'o}mez}},
  \bibinfo{author}{\bibfnamefont{F.~J.} \bibnamefont{Rodr{\'\i}guez}},
  \bibinfo{author}{\bibfnamefont{L.}~\bibnamefont{Quiroga}}, \bibnamefont{and}
  \bibinfo{author}{\bibfnamefont{J.~J.} \bibnamefont{Garc{\'\i}a-Ripoll}},
  \bibinfo{journal}{arXiv:1512.00269}  (\bibinfo{year}{2015}).

\bibitem[{\citenamefont{Furusawa et~al.}(1998)\citenamefont{Furusawa,
  S{\o}rensen, Braunstein, Fuchs, Kimble, and Polzik}}]{Furusawa98}
\bibinfo{author}{\bibfnamefont{A.}~\bibnamefont{Furusawa}},
  \bibinfo{author}{\bibfnamefont{J.~L.} \bibnamefont{S{\o}rensen}},
  \bibinfo{author}{\bibfnamefont{S.~L.} \bibnamefont{Braunstein}},
  \bibinfo{author}{\bibfnamefont{C.~A.} \bibnamefont{Fuchs}},
  \bibinfo{author}{\bibfnamefont{H.~J.} \bibnamefont{Kimble}},
  \bibnamefont{and} \bibinfo{author}{\bibfnamefont{E.~S.}
  \bibnamefont{Polzik}}, \bibinfo{journal}{Science}
  \textbf{\bibinfo{volume}{282}}, \bibinfo{pages}{706} (\bibinfo{year}{1998}).

\bibitem[{\citenamefont{Yurke et~al.}(1986)\citenamefont{Yurke, McCall, and
  Klauder}}]{Yurke86}
\bibinfo{author}{\bibfnamefont{B.}~\bibnamefont{Yurke}},
  \bibinfo{author}{\bibfnamefont{S.~L.} \bibnamefont{McCall}},
  \bibnamefont{and} \bibinfo{author}{\bibfnamefont{J.~R.}
  \bibnamefont{Klauder}}, \bibinfo{journal}{Phys. Rev. A}
  \textbf{\bibinfo{volume}{33}}, \bibinfo{pages}{4033} (\bibinfo{year}{1986}).

\bibitem[{\citenamefont{Tsang and Caves}(2012)}]{Tsang12}
\bibinfo{author}{\bibfnamefont{M.}~\bibnamefont{Tsang}} \bibnamefont{and}
  \bibinfo{author}{\bibfnamefont{C.~M.} \bibnamefont{Caves}},
  \bibinfo{journal}{Phys. Rev. X} \textbf{\bibinfo{volume}{2}},
  \bibinfo{pages}{031016} (\bibinfo{year}{2012}).

\bibitem[{\citenamefont{Barzanjeh et~al.}(2014)\citenamefont{Barzanjeh,
  DiVincenzo, and Terhal}}]{Barzanjeh14}
\bibinfo{author}{\bibfnamefont{S.}~\bibnamefont{Barzanjeh}},
  \bibinfo{author}{\bibfnamefont{D.~P.} \bibnamefont{DiVincenzo}},
  \bibnamefont{and} \bibinfo{author}{\bibfnamefont{B.~M.}
  \bibnamefont{Terhal}}, \bibinfo{journal}{Phys. Rev. B}
  \textbf{\bibinfo{volume}{90}}, \bibinfo{pages}{134515}
  (\bibinfo{year}{2014}).

\bibitem[{\citenamefont{Didier et~al.}(2015)\citenamefont{Didier, Kamal,
  Oliver, Blais, and Clerk}}]{Didier15}
\bibinfo{author}{\bibfnamefont{N.}~\bibnamefont{Didier}},
  \bibinfo{author}{\bibfnamefont{A.}~\bibnamefont{Kamal}},
  \bibinfo{author}{\bibfnamefont{W.~D.} \bibnamefont{Oliver}},
  \bibinfo{author}{\bibfnamefont{A.}~\bibnamefont{Blais}}, \bibnamefont{and}
  \bibinfo{author}{\bibfnamefont{A.~A.} \bibnamefont{Clerk}},
  \bibinfo{journal}{Physical review letters} \textbf{\bibinfo{volume}{115}},
  \bibinfo{pages}{093604} (\bibinfo{year}{2015}).

\bibitem[{\citenamefont{Royer et~al.}(2016)\citenamefont{Royer, Grimsmo,
  Didier, and Blais}}]{Royer16}
\bibinfo{author}{\bibfnamefont{B.}~\bibnamefont{Royer}},
  \bibinfo{author}{\bibfnamefont{A.~L.} \bibnamefont{Grimsmo}},
  \bibinfo{author}{\bibfnamefont{N.}~\bibnamefont{Didier}}, \bibnamefont{and}
  \bibinfo{author}{\bibfnamefont{A.}~\bibnamefont{Blais}},
  \bibinfo{journal}{arXiv:1603.04424}  (\bibinfo{year}{2016}).

\bibitem[{\citenamefont{Mallet et~al.}(2011)\citenamefont{Mallet,
  Castellanos-Beltran, Ku, Glancy, Knill, Irwin, Hilton, Vale, and
  Lehnert}}]{Mallet11}
\bibinfo{author}{\bibfnamefont{F.}~\bibnamefont{Mallet}},
  \bibinfo{author}{\bibfnamefont{M.~A.} \bibnamefont{Castellanos-Beltran}},
  \bibinfo{author}{\bibfnamefont{H.~S.} \bibnamefont{Ku}},
  \bibinfo{author}{\bibfnamefont{S.}~\bibnamefont{Glancy}},
  \bibinfo{author}{\bibfnamefont{E.}~\bibnamefont{Knill}},
  \bibinfo{author}{\bibfnamefont{K.~D.} \bibnamefont{Irwin}},
  \bibinfo{author}{\bibfnamefont{G.~C.} \bibnamefont{Hilton}},
  \bibinfo{author}{\bibfnamefont{L.~R.} \bibnamefont{Vale}}, \bibnamefont{and}
  \bibinfo{author}{\bibfnamefont{K.~W.} \bibnamefont{Lehnert}},
  \bibinfo{journal}{Phys. Rev. Lett.} \textbf{\bibinfo{volume}{106}},
  \bibinfo{pages}{220502} (\bibinfo{year}{2011}).

\bibitem[{\citenamefont{Blais et~al.}(2004)\citenamefont{Blais, Huang,
  Wallraff, Girvin, and Schoelkopf}}]{Blais04}
\bibinfo{author}{\bibfnamefont{A.}~\bibnamefont{Blais}},
  \bibinfo{author}{\bibfnamefont{R.-S.} \bibnamefont{Huang}},
  \bibinfo{author}{\bibfnamefont{A.}~\bibnamefont{Wallraff}},
  \bibinfo{author}{\bibfnamefont{S.~M.} \bibnamefont{Girvin}},
  \bibnamefont{and} \bibinfo{author}{\bibfnamefont{R.~J.}
  \bibnamefont{Schoelkopf}}, \bibinfo{journal}{Phys. Rev. A}
  \textbf{\bibinfo{volume}{69}}, \bibinfo{pages}{062320}
  (\bibinfo{year}{2004}).

\bibitem[{\citenamefont{Vijay et~al.}(2011)\citenamefont{Vijay, Slichter, and
  Siddiqi}}]{Vijay11}
\bibinfo{author}{\bibfnamefont{R.}~\bibnamefont{Vijay}},
  \bibinfo{author}{\bibfnamefont{D.~H.} \bibnamefont{Slichter}},
  \bibnamefont{and} \bibinfo{author}{\bibfnamefont{I.}~\bibnamefont{Siddiqi}},
  \bibinfo{journal}{Phys. Rev. Lett.} \textbf{\bibinfo{volume}{106}},
  \bibinfo{pages}{110502} (\bibinfo{year}{2011}).

\bibitem[{\citenamefont{Jeffrey
  et~al.}(2014{\natexlab{b}})\citenamefont{Jeffrey, Sank, Mutus, White, Kelly,
  Barends, Chen, Chen, Chiaro, Dunsworth et~al.}}]{Evan14}
\bibinfo{author}{\bibfnamefont{E.}~\bibnamefont{Jeffrey}},
  \bibinfo{author}{\bibfnamefont{D.}~\bibnamefont{Sank}},
  \bibinfo{author}{\bibfnamefont{J.~Y.} \bibnamefont{Mutus}},
  \bibinfo{author}{\bibfnamefont{T.~C.} \bibnamefont{White}},
  \bibinfo{author}{\bibfnamefont{J.}~\bibnamefont{Kelly}},
  \bibinfo{author}{\bibfnamefont{R.}~\bibnamefont{Barends}},
  \bibinfo{author}{\bibfnamefont{Y.}~\bibnamefont{Chen}},
  \bibinfo{author}{\bibfnamefont{Z.}~\bibnamefont{Chen}},
  \bibinfo{author}{\bibfnamefont{B.}~\bibnamefont{Chiaro}},
  \bibinfo{author}{\bibfnamefont{A.}~\bibnamefont{Dunsworth}},
  \bibnamefont{et~al.}, \bibinfo{journal}{Phys. Rev. Lett.}
  \textbf{\bibinfo{volume}{112}}, \bibinfo{pages}{190504}
  (\bibinfo{year}{2014}{\natexlab{b}}).

\bibitem[{\citenamefont{Wootters}(2001)}]{Wootters01}
\bibinfo{author}{\bibfnamefont{W.~K.} \bibnamefont{Wootters}},
  \bibinfo{journal}{Quantum Information \& Computation}
  \textbf{\bibinfo{volume}{1}}, \bibinfo{pages}{27} (\bibinfo{year}{2001}).

\bibitem[{\citenamefont{Gardiner}(1986)}]{Gardiner86}
\bibinfo{author}{\bibfnamefont{C.~W.} \bibnamefont{Gardiner}},
  \bibinfo{journal}{Phys. Rev. Lett.} \textbf{\bibinfo{volume}{56}},
  \bibinfo{pages}{1917} (\bibinfo{year}{1986}).

\bibitem[{\citenamefont{Carmichael et~al.}(1987)\citenamefont{Carmichael, Lane,
  and Walls}}]{Carmichael87}
\bibinfo{author}{\bibfnamefont{H.~J.} \bibnamefont{Carmichael}},
  \bibinfo{author}{\bibfnamefont{A.~S.} \bibnamefont{Lane}}, \bibnamefont{and}
  \bibinfo{author}{\bibfnamefont{D.~F.} \bibnamefont{Walls}},
  \bibinfo{journal}{Phys. Rev. Lett.} \textbf{\bibinfo{volume}{58}},
  \bibinfo{pages}{2539} (\bibinfo{year}{1987}).

\bibitem[{\citenamefont{Toyli et~al.}(2016)\citenamefont{Toyli, Eddins, Boutin,
  Puri, Hover, Bolkhovsky, Oliver, Blais, and Siddiqi}}]{Toyli16}
\bibinfo{author}{\bibfnamefont{D.}~\bibnamefont{Toyli}},
  \bibinfo{author}{\bibfnamefont{A.}~\bibnamefont{Eddins}},
  \bibinfo{author}{\bibfnamefont{S.}~\bibnamefont{Boutin}},
  \bibinfo{author}{\bibfnamefont{S.}~\bibnamefont{Puri}},
  \bibinfo{author}{\bibfnamefont{D.}~\bibnamefont{Hover}},
  \bibinfo{author}{\bibfnamefont{V.}~\bibnamefont{Bolkhovsky}},
  \bibinfo{author}{\bibfnamefont{W.}~\bibnamefont{Oliver}},
  \bibinfo{author}{\bibfnamefont{A.}~\bibnamefont{Blais}}, \bibnamefont{and}
  \bibinfo{author}{\bibfnamefont{I.}~\bibnamefont{Siddiqi}},
  \bibinfo{journal}{arXiv:1602.03240}  (\bibinfo{year}{2016}).

\bibitem[{\citenamefont{Raussendorf et~al.}(2003)\citenamefont{Raussendorf,
  Browne, and Briegel}}]{Raussendorf03}
\bibinfo{author}{\bibfnamefont{R.}~\bibnamefont{Raussendorf}},
  \bibinfo{author}{\bibfnamefont{D.~E.} \bibnamefont{Browne}},
  \bibnamefont{and} \bibinfo{author}{\bibfnamefont{H.~J.}
  \bibnamefont{Briegel}}, \bibinfo{journal}{Physical review A}
  \textbf{\bibinfo{volume}{68}}, \bibinfo{pages}{022312}
  (\bibinfo{year}{2003}).

\bibitem[{\citenamefont{Briegel et~al.}(2009)\citenamefont{Briegel, Browne,
  D{\"u}r, Raussendorf, and Van~den Nest}}]{Briegel09}
\bibinfo{author}{\bibfnamefont{H.~J.} \bibnamefont{Briegel}},
  \bibinfo{author}{\bibfnamefont{D.~E.} \bibnamefont{Browne}},
  \bibinfo{author}{\bibfnamefont{W.}~\bibnamefont{D{\"u}r}},
  \bibinfo{author}{\bibfnamefont{R.}~\bibnamefont{Raussendorf}},
  \bibnamefont{and} \bibinfo{author}{\bibfnamefont{M.}~\bibnamefont{Van~den
  Nest}}, \bibinfo{journal}{Nature Physics} \textbf{\bibinfo{volume}{5}},
  \bibinfo{pages}{19} (\bibinfo{year}{2009}).

\bibitem[{\citenamefont{Gu et~al.}(2009)\citenamefont{Gu, Weedbrook, Menicucci,
  Ralph, and van Loock}}]{Gu09}
\bibinfo{author}{\bibfnamefont{M.}~\bibnamefont{Gu}},
  \bibinfo{author}{\bibfnamefont{C.}~\bibnamefont{Weedbrook}},
  \bibinfo{author}{\bibfnamefont{N.~C.} \bibnamefont{Menicucci}},
  \bibinfo{author}{\bibfnamefont{T.~C.} \bibnamefont{Ralph}}, \bibnamefont{and}
  \bibinfo{author}{\bibfnamefont{P.}~\bibnamefont{van Loock}},
  \bibinfo{journal}{Physical Review A} \textbf{\bibinfo{volume}{79}},
  \bibinfo{pages}{062318} (\bibinfo{year}{2009}).

\bibitem[{\citenamefont{Menicucci et~al.}(2011)\citenamefont{Menicucci,
  Flammia, and van Loock}}]{Menicucci11}
\bibinfo{author}{\bibfnamefont{N.~C.} \bibnamefont{Menicucci}},
  \bibinfo{author}{\bibfnamefont{S.~T.} \bibnamefont{Flammia}},
  \bibnamefont{and} \bibinfo{author}{\bibfnamefont{P.}~\bibnamefont{van
  Loock}}, \bibinfo{journal}{Phys. Rev. A} \textbf{\bibinfo{volume}{83}},
  \bibinfo{pages}{042335} (\bibinfo{year}{2011}).

\bibitem[{\citenamefont{Zhang et~al.}(2008)\citenamefont{Zhang, Xie, Peng, and
  van Loock}}]{Zhang08}
\bibinfo{author}{\bibfnamefont{J.}~\bibnamefont{Zhang}},
  \bibinfo{author}{\bibfnamefont{C.}~\bibnamefont{Xie}},
  \bibinfo{author}{\bibfnamefont{K.}~\bibnamefont{Peng}}, \bibnamefont{and}
  \bibinfo{author}{\bibfnamefont{P.}~\bibnamefont{van Loock}},
  \bibinfo{journal}{Phys. Rev. A} \textbf{\bibinfo{volume}{78}},
  \bibinfo{pages}{052121} (\bibinfo{year}{2008}).

\bibitem[{\citenamefont{Dennis et~al.}(2002)\citenamefont{Dennis, Kitaev,
  Landahl, and Preskill}}]{Dennis02}
\bibinfo{author}{\bibfnamefont{E.}~\bibnamefont{Dennis}},
  \bibinfo{author}{\bibfnamefont{A.}~\bibnamefont{Kitaev}},
  \bibinfo{author}{\bibfnamefont{A.}~\bibnamefont{Landahl}}, \bibnamefont{and}
  \bibinfo{author}{\bibfnamefont{J.}~\bibnamefont{Preskill}},
  \bibinfo{journal}{Journal of Mathematical Physics}
  \textbf{\bibinfo{volume}{43}}, \bibinfo{pages}{4452} (\bibinfo{year}{2002}).

\bibitem[{\citenamefont{Schuster et~al.}(2007)\citenamefont{Schuster, Houck,
  Schreier, Wallraff, Gambetta, Blais, Frunzio, Majer, Johnson, Devoret
  et~al.}}]{Schuster07}
\bibinfo{author}{\bibfnamefont{D.}~\bibnamefont{Schuster}},
  \bibinfo{author}{\bibfnamefont{A.}~\bibnamefont{Houck}},
  \bibinfo{author}{\bibfnamefont{J.}~\bibnamefont{Schreier}},
  \bibinfo{author}{\bibfnamefont{A.}~\bibnamefont{Wallraff}},
  \bibinfo{author}{\bibfnamefont{J.}~\bibnamefont{Gambetta}},
  \bibinfo{author}{\bibfnamefont{A.}~\bibnamefont{Blais}},
  \bibinfo{author}{\bibfnamefont{L.}~\bibnamefont{Frunzio}},
  \bibinfo{author}{\bibfnamefont{J.}~\bibnamefont{Majer}},
  \bibinfo{author}{\bibfnamefont{B.}~\bibnamefont{Johnson}},
  \bibinfo{author}{\bibfnamefont{M.}~\bibnamefont{Devoret}},
  \bibnamefont{et~al.}, \bibinfo{journal}{Nature}
  \textbf{\bibinfo{volume}{445}}, \bibinfo{pages}{515} (\bibinfo{year}{2007}).

\bibitem[{\citenamefont{Devoret}(1995)}]{Devoret95}
\bibinfo{author}{\bibfnamefont{M.~H.} \bibnamefont{Devoret}},
  \bibinfo{journal}{Les Houches Lecture Notes}  (\bibinfo{year}{1995}).

\bibitem[{\citenamefont{Bhat and Sipe}(2006)}]{Bhat06}
\bibinfo{author}{\bibfnamefont{N.~A.~R.} \bibnamefont{Bhat}} \bibnamefont{and}
  \bibinfo{author}{\bibfnamefont{J.~E.} \bibnamefont{Sipe}},
  \bibinfo{journal}{Phys. Rev. A} \textbf{\bibinfo{volume}{73}},
  \bibinfo{pages}{063808} (\bibinfo{year}{2006}).

\bibitem[{\citenamefont{Santos and Loudon}(1995)}]{Santos95}
\bibinfo{author}{\bibfnamefont{D.~J.} \bibnamefont{Santos}} \bibnamefont{and}
  \bibinfo{author}{\bibfnamefont{R.}~\bibnamefont{Loudon}},
  \bibinfo{journal}{Phys. Rev. A} \textbf{\bibinfo{volume}{52}},
  \bibinfo{pages}{1538} (\bibinfo{year}{1995}).

\bibitem[{\citenamefont{Nigg et~al.}(2012)\citenamefont{Nigg, Paik, Vlastakis,
  Kirchmair, Shankar, Frunzio, Devoret, Schoelkopf, and Girvin}}]{Nigg12}
\bibinfo{author}{\bibfnamefont{S.~E.} \bibnamefont{Nigg}},
  \bibinfo{author}{\bibfnamefont{H.}~\bibnamefont{Paik}},
  \bibinfo{author}{\bibfnamefont{B.}~\bibnamefont{Vlastakis}},
  \bibinfo{author}{\bibfnamefont{G.}~\bibnamefont{Kirchmair}},
  \bibinfo{author}{\bibfnamefont{S.}~\bibnamefont{Shankar}},
  \bibinfo{author}{\bibfnamefont{L.}~\bibnamefont{Frunzio}},
  \bibinfo{author}{\bibfnamefont{M.}~\bibnamefont{Devoret}},
  \bibinfo{author}{\bibfnamefont{R.}~\bibnamefont{Schoelkopf}},
  \bibnamefont{and} \bibinfo{author}{\bibfnamefont{S.}~\bibnamefont{Girvin}},
  \bibinfo{journal}{Physical review letters} \textbf{\bibinfo{volume}{108}},
  \bibinfo{pages}{240502} (\bibinfo{year}{2012}).

\bibitem[{\citenamefont{Hopfield}(1958)}]{Hopfield58}
\bibinfo{author}{\bibfnamefont{J.}~\bibnamefont{Hopfield}},
  \bibinfo{journal}{Physical Review} \textbf{\bibinfo{volume}{112}},
  \bibinfo{pages}{1555} (\bibinfo{year}{1958}).

\bibitem[{\citenamefont{Clerk et~al.}(2010)\citenamefont{Clerk, Devoret,
  Girvin, Marquardt, and Schoelkopf}}]{Clerk10}
\bibinfo{author}{\bibfnamefont{A.~A.} \bibnamefont{Clerk}},
  \bibinfo{author}{\bibfnamefont{M.~H.} \bibnamefont{Devoret}},
  \bibinfo{author}{\bibfnamefont{S.~M.} \bibnamefont{Girvin}},
  \bibinfo{author}{\bibfnamefont{F.}~\bibnamefont{Marquardt}},
  \bibnamefont{and} \bibinfo{author}{\bibfnamefont{R.~J.}
  \bibnamefont{Schoelkopf}}, \bibinfo{journal}{Reviews of Modern Physics}
  \textbf{\bibinfo{volume}{82}}, \bibinfo{pages}{1155} (\bibinfo{year}{2010}).

\bibitem[{\citenamefont{Carmichael}(1993)}]{Carmichael93}
\bibinfo{author}{\bibfnamefont{H.~J.} \bibnamefont{Carmichael}},
  \bibinfo{journal}{Phys. Rev. Lett.} \textbf{\bibinfo{volume}{70}},
  \bibinfo{pages}{2273} (\bibinfo{year}{1993}).

\bibitem[{\citenamefont{Gardiner}(1993)}]{Gardiner93}
\bibinfo{author}{\bibfnamefont{C.~W.} \bibnamefont{Gardiner}},
  \bibinfo{journal}{Phys. Rev. Lett.} \textbf{\bibinfo{volume}{70}},
  \bibinfo{pages}{2269} (\bibinfo{year}{1993}).

\bibitem[{\citenamefont{Gardiner and Zoller}(2004)}]{Gardiner04}
\bibinfo{author}{\bibfnamefont{C.}~\bibnamefont{Gardiner}} \bibnamefont{and}
  \bibinfo{author}{\bibfnamefont{P.}~\bibnamefont{Zoller}},
  \emph{\bibinfo{title}{Quantum Noise: A Handbook of Markovian and
  Non-Markovian Quantum Stochastic Methods with Applications to Quantum
  Optics}} (\bibinfo{publisher}{Springer}, \bibinfo{address}{New York},
  \bibinfo{year}{2004}).

\bibitem[{\citenamefont{Koga and Yamamoto}(2012)}]{Koga12}
\bibinfo{author}{\bibfnamefont{K.}~\bibnamefont{Koga}} \bibnamefont{and}
  \bibinfo{author}{\bibfnamefont{N.}~\bibnamefont{Yamamoto}},
  \bibinfo{journal}{Physical Review A} \textbf{\bibinfo{volume}{85}},
  \bibinfo{pages}{022103} (\bibinfo{year}{2012}).

\end{thebibliography}

\end{document}